\documentclass[11pt]{article}

\usepackage[a4paper,left=2cm,right=2cm,top=2cm,bottom=2cm]{geometry}
\usepackage[utf8]{inputenc}
\usepackage{natbib}
\usepackage[leqno]{amsmath}
\usepackage{amsmath}
\usepackage{amsthm}
\usepackage{amssymb}
\usepackage[T1]{fontenc}
\usepackage{cite}
\usepackage{graphicx,subfigure}
\usepackage{chngcntr}
\usepackage{xcolor}
\usepackage{hyperref}
\usepackage[all]{hypcap} 

\counterwithin{equation}{section}
\hypersetup{
  colorlinks   = true, 
  urlcolor     = blue, 
  linkcolor    = blue, 
  citecolor   = blue 
}

\def\R{\mathbb{R}}
\def\C{\mathbb{C}}
\def\N{\mathbb{N}}
\def\v{\boldsymbol{v}}
\def\u{\boldsymbol{u}}
\def\U{\boldsymbol{U}}

\def\k{\boldsymbol{k}}

\def\a{\boldsymbol{a}}
\def\x{\boldsymbol{x}}
\def\2{\boldsymbol{e_2}}
\def\3{\boldsymbol{e_3}}

\def\p{\partial}
\def\b{\mathtt{b}}
\def\tb{\tilde{\b}}

\def\calL{\mathcal{L}}
\def\rmi{\mathrm{i}}
\def\transpose{^\mathsf{T}}

\def\sc{s}
\def\diag{\mathrm{diag}}
\def\calA{\mathcal{A}}
\def\calB{\mathcal{B}}
\def\Id{\mathrm{Id}}
\def\B{\boldsymbol{B}}
\def\LL{\mathcal{L}}
\def\Lb{\boldsymbol{L}}
\def\sgn{\mathrm{sgn}}
\def\pparagraph{\textbf}

\definecolor{colorJRred}{rgb}{1.,0.,0.} 

\begin{document}

\markboth{A.\ Prugger, J.\ Rademacher, J. Yang}{Geophysical flows, energy backscatter, unbounded growth}
\title{\textbf{\LARGE Geophysical fluid models with simple energy backscatter: explicit flows and unbounded exponential growth}}
\author{A.\ Prugger${\dag}$\thanks{Corresponding author. Email: a.prugger@uni-bremen.de}\,\, and J.D.M.\ Rademacher${\dag}$ and J.\ Yang${\ddag}$\vspace{18pt}\\\vspace{6pt}  
${\dag}$University of Bremen, Department 3 - Mathematics, 28359 Bremen, Germany\\ 
${\ddag}$Sun Yat-sen University, School of Mathematics (Zhuhai), 519082 Zhuhai, China}
\date{\today}

\maketitle

\begin{abstract}
Motivated by numerical schemes for large scale geophysical flow, we consider the rotating shallow water and Boussinesq equations on the whole space with horizontal kinetic energy backscatter source terms built from negative viscosity and stabilising hyperviscosity with constant parameters. We study the impact of this energy input through various explicit flows, which are simultaneously solving the nonlinear equations and the linear equations that arise upon dropping the transport nonlinearity, i.e. the linearisation in the zero state. These include barotropic, parallel and Kolmogorov flows as well as monochromatic inertia gravity waves. With focus on stable stratification we find that the backscatter generates numerous  solutions of this type that grow exponentially and unboundedly, also with vertical structure. This signifies the possibility of undesired energy concentration into specific modes due to the backscatter. Families of steady  state flows of this type arise as well and superposition principles in the nonlinear equations provide explicit sufficient conditions for instability of some of these. For certain steady barotropic flows of this type we provide numerical evidence of eigenmodes whose growth rates are proportional to the amplitude factor of the flow. For all other arising steady solutions we prove this is not possible.\\
\\
\textbf{Keywords:} explicit flows; instability; plane waves; rotating shallow water equations; rotating Boussinesq equations
\end{abstract}

\section{Introduction}\label{Introduction}
Spatial resolution of geophysical flows is limited not only in observational data, but also in numerical simulations for ocean and climate studies due to lack of computing power. Towards realistic simulations, this is compensated by so-called parameterisations of subgrid effects, i.e. subgrid models that are gauged a priori and simulate the influence of the missing small scale resolution on the resolved large scale flow. Moreover, these should account for numerical discretisation effects such as over-dissipation, and yet admit stable simulations of ocean and climate models. A practical solution to these problems that has come to frequent use are kinetic energy backscatter schemes that effectively introduce negative horizontal viscosity together with hyperviscosity, cf.\ e.g.\ \citep{Jansen2014, ZuritaEtAl2015,JansenEtAl2019, juricke2020kinematic,Perezhogin20}; we refer to \citep{M3} for a detailed discussion and relations to other approaches. Simulations with backscatter have been found to provide energy `at the right place', matching the result more closely to observations and high resolution comparisons.

Motivated by this, we consider the rotating shallow water and Boussinesq equations on the whole space with simplified kinetic energy backscatter source terms built from negative horizontal viscosity and stabilising hyperviscosity with constant parameters. The backscatter terms in the numerical scheme have non-constant coefficients from a coupled energy equation that aims to regulate energy consistency and thus these terms vanish for infinite resolution. Our simplified consideration on the continuum level is amenable to an analysis of qualitative features and energy distribution, so that the results point out potential issues and can guide further study of backscatter schemes. 

In particular, this idealisation admits a direct analytical study of the influence of backscatter through its impact on various explicit flows. Indeed, explicit flows are frequently used as a tool for benchmarking analytical and numerical studies in this and other contexts, e.g. \citep{Chai20,Drazin06,Dyck19,Majda03,MajdaWang06,Weinbaum67}, and also for turbulence studies, e.g. \citep{LeLongDunkerton1998,ghaemsaidi_mathur_2019,onuki_joubaud_dauxois_2021}.\newline
We start our investigations with the rotating shallow equations and then move to the rotating Boussinesq equations. In the spirit of \citep{prugger2020explicit}, we consider flows and waves, which are simultaneously solving the nonlinear equations and the linear equation that arise from dropping the transport nonlinearity, i.e. the linearisation in the zero state. These include flows that correspond to barotropic, parallel and Kolmogorov flows as well as monochromatic inertia gravity waves, cf.\ \citep{Achatz06,BalmforthYoung2005,prugger2020explicit,Yau04} for vanishing backscatter. More specifically, we identify these solutions from superpositions of plane waves with suitably selected wave vectors and wave directions. These are therefore specific eigenmodes of the linear operator $\calL$ arising from infinitesimal perturbations of the zero state and thus correspond to elements in the spectrum of $\calL$. Since these are equally solutions of the nonlinear equations we find subspaces of linear dynamics in the nonlinear system, where all these solutions come as a family with a free scaling factor. We confirm that the presence of backscatter implies unstable spectrum of $\calL$, i.e. spectrum with positive growth rates, also for stable stratification. Moreover, we identify explicit solutions of the nonlinear problem that grow exponentially and unboundedly, 
including solutions whose vertical velocity is growing unboundedly, despite the horizontal nature of the backscatter. This highlights the possibility of undesired concentration of energy due to backscatter, and is in contrast to the targeted energy redistribution. The solutions which have negative real spectrum likewise illustrate the possibility of ineffective energy input into certain scales. However, while solutions to consistent discretisations shadow continuum effects over finite times, the specific implications for numerical backscatter schemes require additional investigations beyond the present study. 

We refer to this unbounded exponential growth of linear type in the nonlinear system as \textit{unbounded instability} and note that it does not occur in generic damped-driven evolution equations, where unstable manifolds are nonlinear. 
Among these explicit flows there are numerous steady states and we identify admissible superpositions in the nonlinear equations, which provide explicit conditions for unbounded instability in many cases. For certain steady barotropic flows of this type we provide numerical evidence of eigenmodes whose growth rates are proportional to the amplitude factor of these steady flows, thus featuring arbitrarily strong local instability. For all other steady solutions of the type considered, we analyse the resulting large amplitude regime and prove that growth rates must scale sublinear with the amplitude. We next summarise and describe the results in more detail.

\medskip
We first consider the rotating shallow water equations with backscatter and flat bottom topography in the $f$-plane approximation given by
\begin{subequations}\label{eq: RSWB}
\begin{align}
\frac{\p\v}{\p t}+(\v\cdot\nabla)\v &= -f\v^{\perp}-g\nabla\eta - \left(\begin{array}{cc} d_1\Delta^2+b_1\Delta & 0 \\ 0 & d_2\Delta^2+b_2\Delta \end{array}\right)\v\label{eq: RSWBa}\\
\frac{\p\eta}{\p t}+(\v\cdot\nabla)\eta &= -(H_0+\eta)\text{div}(\v)\,,\label{eq: RSWBb}
\end{align}
\end{subequations}
where $\v=\v(t,\x)\in\R^2$ is the velocity field on the whole space $\x\in\R^2$ at time $t\geq 0$ and $\eta=\eta(t,\x)\in\R$ is the deviation of the fluid layer from the characteristic fluid depth $H_0>0$, giving $H_0+\eta$ as the fluid layer thickness. In addition, $f\neq0$ is the constant Coriolis parameter, $g>0$ gravity acceleration and the backscatter parameters are $b_1,\, b_2,\,  d_1,\, d_2>0$.\newline
The explicit solutions that we consider in this case derive from the plane wave ansatz 
\begin{align}\label{sol: RSWB1}
\v = \psi(t,\k\cdot\x)\k^{\perp}\,, \quad \eta = \frac{f}{g}\phi(\k\cdot\x)\,,
\end{align}
which implies vanishing nonlinear terms and, for single Fourier modes in $\psi$ and $\phi$, readily yields certain nonlinear relations of parameters in order to obtain solutions to \eqref{eq: RSWB}. These turn into geostrophically balanced Rossby waves when the backscatter parameters tend to zero. 
We consider the wave vectors $\k$ as the primary parameters, which (in this case) admits a direct comparison to the spectrum of $\calL$, where \textit{each} $\k\in\R^2$ corresponds to up to three eigenmodes. From this viewpoint, the set of $\k$ that admit plane waves as solutions to the nonlinear equations are more constrained, forming curves $\k(s)$, with $s\in\R$, in the $\k$-plane with corresponding growth rates $\lambda(s)\in\R$. We study the geometry of $\k(s)$ and find that, up to sign, at most three lie on the same line in the $\k$-plane through the origin. Due to a \textit{radial superposition principle} that allows to superpose plane wave flows whose wave vectors lie in the same line through the origin, this gives rise to several invariant subspace with linear dynamics of dimensions $n=1,\,2,\,3$. Regarding the backscatter coefficients, the latter requires \textit{anisotropy} in the sense that $n>1$ requires $b_1/d_1\neq b_2/d_2$. In addition to $\k$, an organising parameter in this analysis is the relative size of the amplitudes of $\v$ and $\eta$, which lies in a certain interval for existing steady solutions \eqref{sol: RSWB1} and determines those, which are unboundedly unstable.\newline 
Moreover, we show that when $\lambda(s)$ changes sign, the transition to unstable spectrum related to plane waves \eqref{sol: RSWB1} is a long-wavelength, modulational instability; this occurs in addition to unstable spectrum from other modes that we numerically find. The free amplitude parameter $a$ of steady plane flows naturally leads to the asymptotic regimes $|a|\ll 1$ and $|a|\gg 1$ of small and large amplitudes. In the small amplitude regime the unstable spectrum of the zero state generates unstable spectrum of the plane wave flows, implying positive growth rates in addition to the explicit ones, but these are not expected to create unbounded growth. In order to study the large amplitude regime, we consider a rescaled problem and show that the resulting operator possesses purely imaginary spectrum. Hence, for these cases the unstable growth rates must scale sublinear with respect to the amplitude factor $a$. Numerical computations suggest such unstable spectrum indeed occurs.

\medskip
We then turn to the rotating Boussinesq equations augmented with backscatter 
\begin{subequations}\label{eq: introB}
\begin{align}
\frac{\p\v}{\p t}+(\v\cdot\nabla)\v+f\3\times\v+\nabla p-\3 \b &= -\mathrm{diag}\bigl(d_1\Delta+b_1,d_2\Delta+b_2,-\nu\bigr)\Delta \v\label{eq: introBa}\\
\nabla\cdot\v&= 0\label{eq: introBb}\\
\frac{\p \b}{\p t}+(\v\cdot\nabla)\b+N^2v_3 &= \mu\Delta \b\,,\label{eq: introBc}
\end{align}
\end{subequations}
with the horizontal backscatter parameters $b_i,\, d_i>0$, $i=1,\,2$ and vertical viscosity $\nu\geq 0$ in the diagonal matrix operator, following the idealised kinetic energy backscatter. Other quantities in \eqref{eq: introB} are the velocity field $\v(t,\x)\in\R^3$ for $\x\in\R^3$, $t\geq 0$, pressure and buoyancy $p(t,\x),\, \b(t,\x)\in\R$, the nonzero Coriolis parameter $f\in\R\backslash\{0\}$, the vertical unit vector $\3$ and thermal diffusivity $\mu\geq0$. 
As usual, the buoyancy considered here is of the form $\b(t,\x)=-g(\rho(t,\x)-\overline{\rho}(z))/\rho_0\in\R$ with fluid density $\rho(t,\x)\in\R$ and reference density field $\overline{\rho}(z)$ depending on the vertical space direction $z$ only, characteristic density $\rho_0$ and gravitational acceleration $g$. Then $N^2=-(g/\rho_0)d\overline{\rho}/dz$ is the Brunt-V\"ais\"al\"a frequency with stable stratification $d\overline{\rho}/dz<0$. 

We focus on stable stratification so that destabilisation develops from backscatter in the horizontal directions only. Guided by \citep{prugger2020explicit}, we first find barotropic \textit{horizontal flows} of similar plane wave type as in the shallow water case, which arise from sinusoidal wave shapes by means of a \textit{radial superposition principle}, akin to the shallow water case, and in addition from an \textit{angular superposition principle} of plane waves with the same wave length. Each of these allows for infinite dimensional invariant subspaces with linear dynamics. In contrast to the shallow water case, here the nonlinear constraints admit a horizontal flow for any horizontal wave vector; the pressure can be determined explicitly in all cases. These flows are in general not fully geostrophically balanced since the pressure not only compensates the Coriolis force, but also the gradient part of the nonlinear term, which may occur with angular superposition. In addition, for anisotropic backscatter (similar to the shallow water case) the backscatter partially compensates the Coriolis force.\newline 
We determine the loci and stability properties of steady states related to these flows  -- in particular the unboundedly unstable ones -- and also the large amplitude regime. Unlike the shallow water plane wave flows, we numerically find that certain superposed horizontal flows have spectral growth rates that are proportional to the amplitude factor, i.e. arbitrarily strong instabilities. \newline
In the same spirit we investigate steady and unboundedly growing explicit flows with vertical structure and coupled buoyancy. These relate to known flows in absence of backscatter: parallel flow, Kolmogorov flow and monochromatic inertia gravity waves (MGWs). 
While the existence of parallel flows is unaffected by the (horizontal) backscatter, superposition with small wave number horizontal flows implies unbounded instability of any parallel flow. However, the situation is more subtle for the other flows. Notably, we find that in Kolmogorov flows and MGWs, the purely horizontal backscatter triggers unbounded growth in the vertical velocity. We also identify possible superpositions of Kolmogorov flows and MGWs. Regarding the large amplitude regime, for these flows the growth rates cannot be proportional to their amplitude factors.

\medskip
We note that the above analytical realisations of backscatter replace the usual molecular viscosity operator $\nu \Delta$ by operators of the form $-(d\Delta^2 + b\Delta)$, familiar from the scalar Kuramoto-Sivashinsky equations (KS),
 \[
 \frac{\p u}{\p t} + \frac 1 2 |\nabla u|^2 = -\Delta^2 u - \Delta u\,, \quad \x\in\R^n\,, \quad n\leq 3\,.
 \]
These have been derived in various contexts and in particular the one-dimensional case appears broadly, e.g. for interfacial layers \citep{Wei06}. 
Posed on tori, for solutions with globally bounded gradient the deviations from the spatial mean admit a finite dimensional global attractor \citep{NST85} and the one-dimensional KS is a paradigm for chaos in a partial differential equation, cf.\ e.g.\ \citep{NST85, Smyrlis91, Kalogirou15} and the references therein. 
Differentiating KS yields a system for $\v=\nabla u$ with the fluid transport nonlinearity, thus relating more closely to \eqref{eq: RSWB} and \eqref{eq: introB} in case all backscatter coefficients are equal, although this relation is clearly far from complete. Moreover, the solutions that we are focussing on, in particular the unboundedly growing ones, are all non-gradient and do not exist on one-dimensional space, therefore they are unrelated even on these levels.

\medskip
This paper is organised as follows. In \S\ref{Shallow Water} we discuss the horizontal flows of the rotating shallow water equations with backscatter. Section \ref{Rotating Boussinesq} is devoted to the rotating Boussinesq equations with backscatter and the analysis of existence, growth and unbounded instability properties of the aforementioned different types of flows.

\section{Rotating shallow water with backscatter}\label{Shallow Water}
In this section we consider the rotating shallow water equations with backscatter \eqref{eq: RSWB} and first identify certain explicit flows. The inviscid rotating shallow water equations without backscatter, i.e. $b_1=b_2= d_1=d_2 = 0$, possess the explicit plane wave steady solutions
\begin{align*}
\v = \phi'(\k\cdot\x)\k^{\perp}\,, \quad \eta = \frac{f}{g}\phi(\k\cdot\x)\,,
\end{align*}
for any wave vector $\k\in\R^2$ and sufficiently smooth wave shape $\phi$, e.g. \citep{prugger2020explicit}. These are also in geostrophic balance, corresponding to Rossby waves.

\begin{figure}
\begin{center}
\subfigure[$d_2=1.0$ and $\alpha_2=0.5$]{
\includegraphics[trim=5.1cm 9.3cm 5.6cm 9.7cm, clip, width=0.32\linewidth]{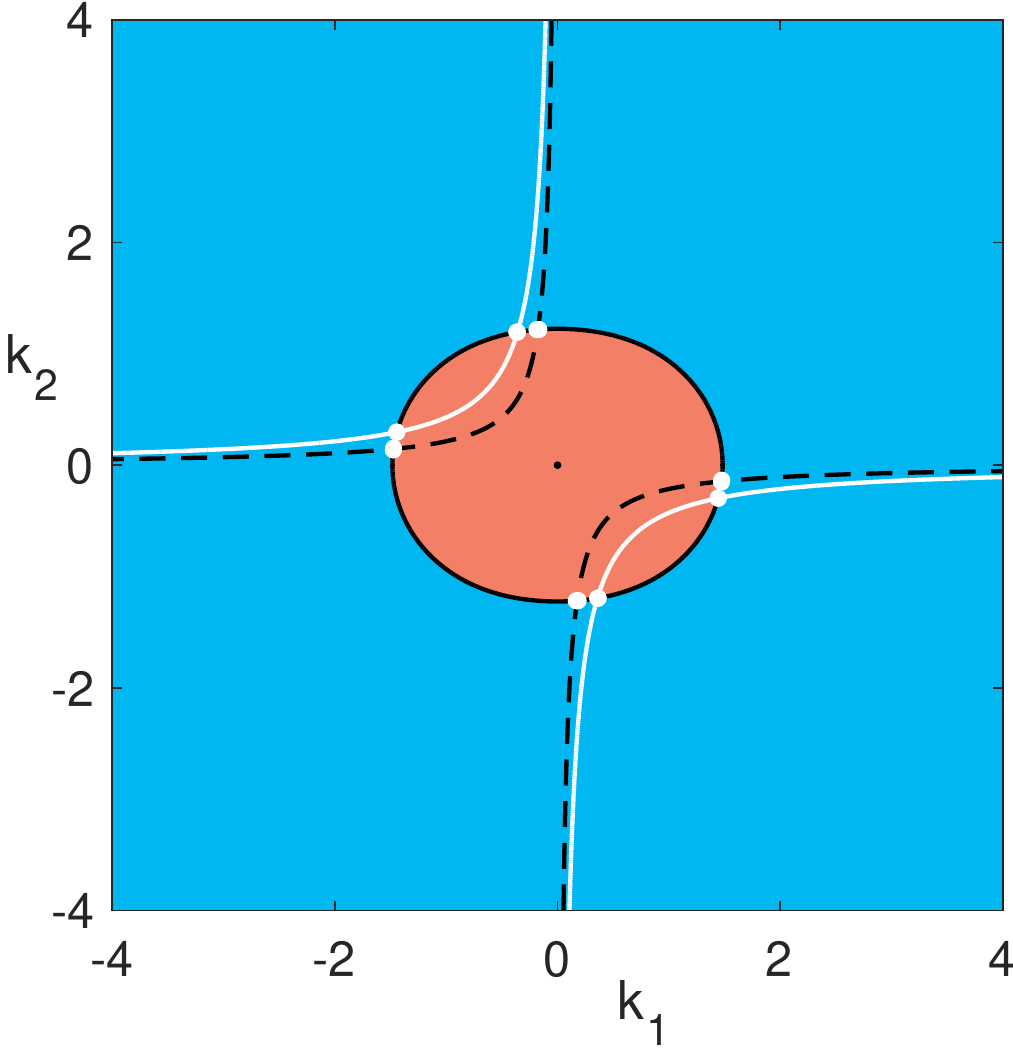}\label{Fig. 1a}}
\subfigure[$d_2=1.04$ and $\alpha_2= -0.5$]{
\includegraphics[trim=5.1cm 9.3cm 5.6cm 9.7cm, clip, width=0.32\linewidth]{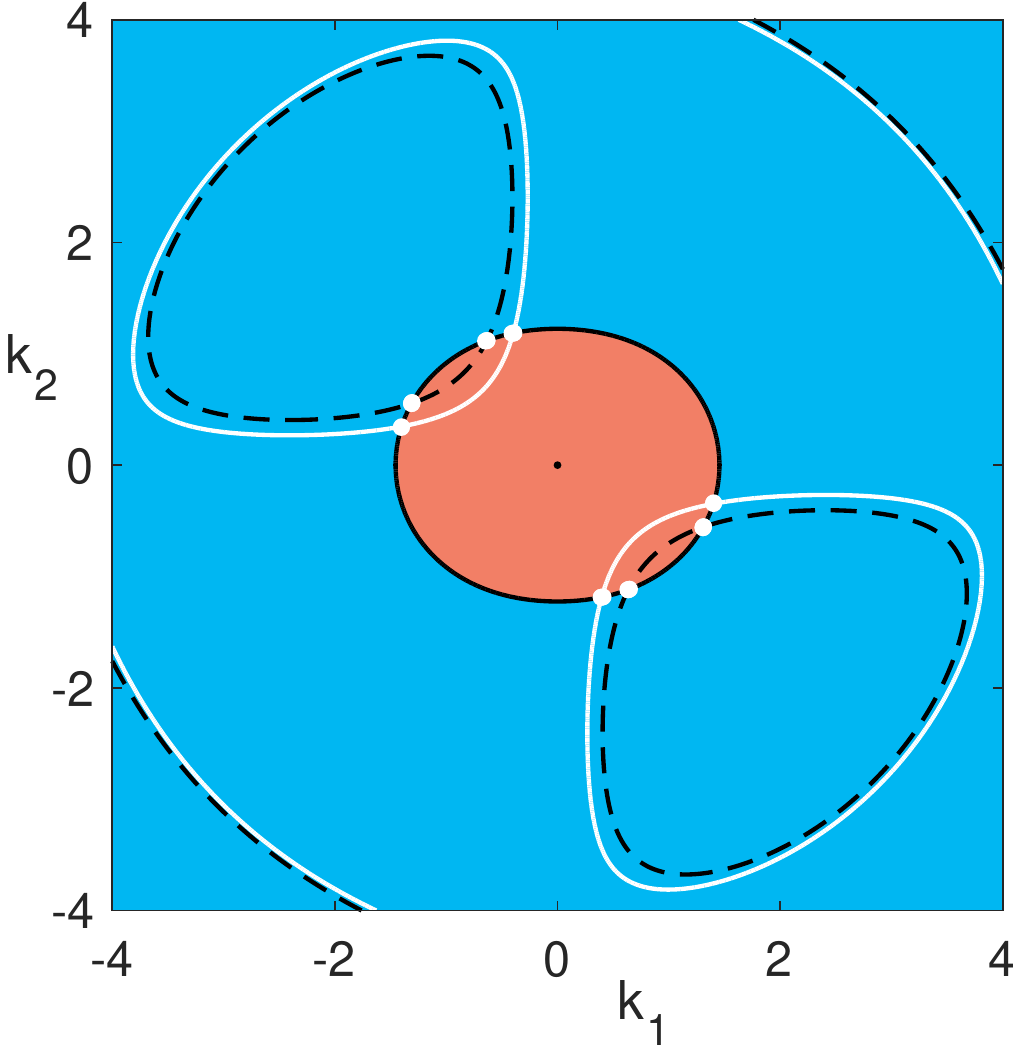}\label{Fig. 1b}}
\subfigure[$d_2=1.04$]{
\includegraphics[trim=5.1cm 9.3cm 5.6cm 9.7cm, clip, width=0.32\linewidth]{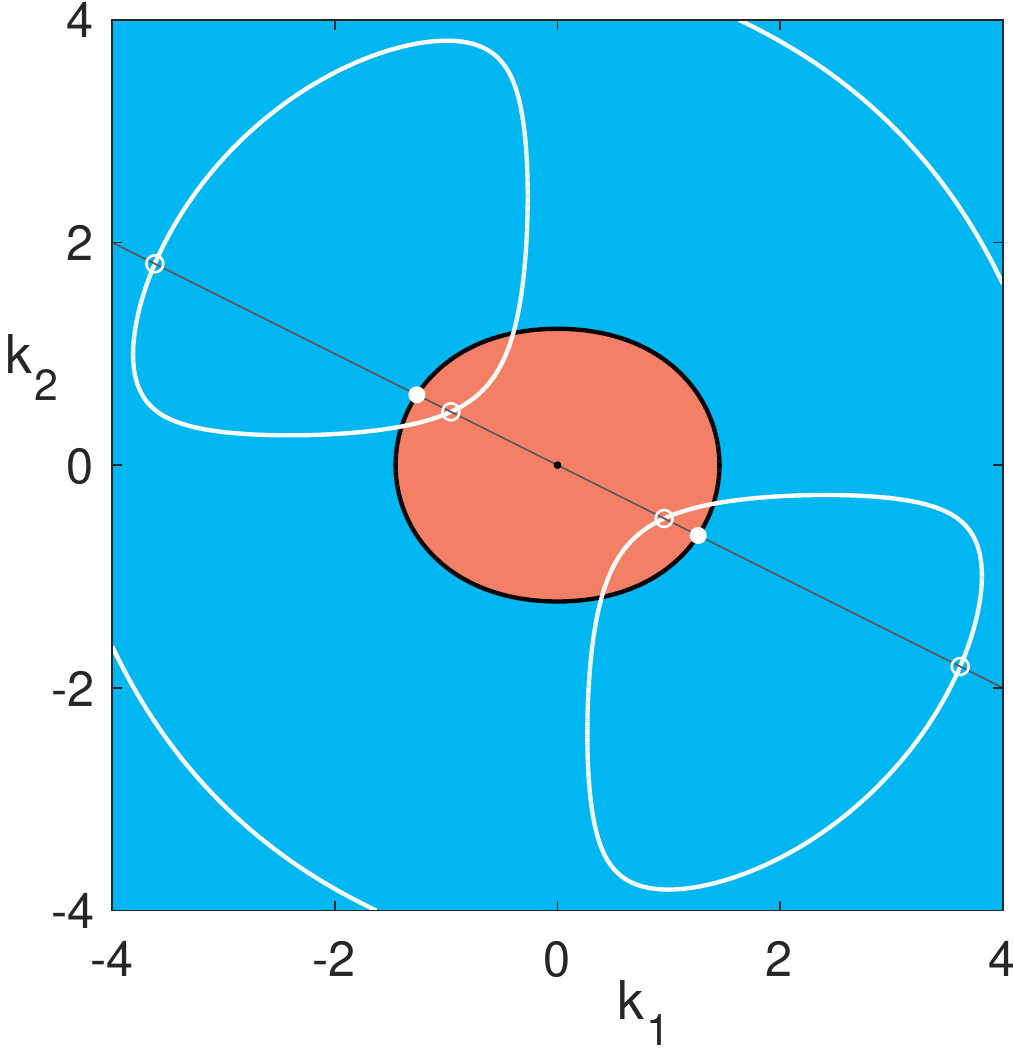}\label{Fig. 1c}}
\caption{\label{Fig. 1}We plot the occurrence of explicit solutions \eqref{sol: RSWB2} in the plane of wave vectors with fixed parameters $d_1=1.0,\, b_1=1.5,\, b_2=2.2,\, f=0.3,\, g = 9.8,\, H_0 = 0.1,\, \alpha_1=1.0$ and $d_2,\, \alpha_2$ as in the subcaptions. Red regions: $\lambda>0$, i.e. unbounded growth; blue regions: $\lambda<0$; black curves: $\lambda=0$, i.e. steady states. The white curves mark solutions with $\alpha_2=0$, the white bullets mark steady solutions with $\lambda=0$. The black dashed curves mark solutions of condition \eqref{cond: RSWB2b} only, for $\alpha_2\neq0$ with values as in the subcaptions, which also solve \eqref{cond: RSWB2a} at the white bullets ($\lambda=0$). In (c) we mark a line of wave vectors in a fixed direction (gray), and the growing or decaying solutions (white circles) on it, whose superpositions with or without the steady state on the gray line again yield explicit solutions.}
\end{center}
\end{figure}

For the case of backscatter in \eqref{eq: RSWB} we seek solutions of the similar form \eqref{sol: RSWB1} for any wave vector $\k=(k_1,k_2)\transpose\in\R^2$ and sufficiently smooth wave shapes $\psi$ and $\phi$. The time-independence of $\eta$ results from equation \eqref{eq: RSWBb}, since $\v$ is divergence free and the nonlinear terms vanish in this case. Inserting \eqref{sol: RSWB1} into \eqref{eq: RSWBa} yields the \textit{linear} equation
\begin{align*}
\frac{\p\psi}{\p t}\k^{\perp}+f\left(\frac{\p\phi}{\p\xi}-\psi\right)\k = -|\k|^2\left(\begin{array}{cc} d_1|\k|^2\frac{\p^4\psi}{\p\xi^4}+b_1\frac{\p^2\psi}{\p\xi^2} & 0 \\ 0 & d_2|\k|^2\frac{\p^4\psi}{\p\xi^4}+b_2\frac{\p^2\psi}{\p\xi^2} \end{array}\right)\k^{\perp}\,.
\end{align*}
Every vector in $\R^2$ on the right hand side has a unique representation by the orthogonal basis vectors on the left hand side. The scalar product with $\k^\perp$ and $\k$, respectively, gives
\begin{subequations}\label{cond: RSWB1}
\begin{align}
\frac{\p\psi}{\p t} &= -|\k|^2\left(d_1k_2^2+d_2k_1^2\right)\frac{\p^4\psi}{\p\xi^4}-\left(b_1k_2^2+b_2k_1^2\right)\frac{\p^2\psi}{\p\xi^2}\label{cond: RSWB1a} \\[2mm]
f\left(\frac{\p\phi}{\p\xi}-\psi\right) &= k_1k_2\left((d_1-d_2)|\k|^2\frac{\p^4\psi}{\p\xi^4}+(b_1-b_2)\frac{\p^2\psi}{\p\xi^2}\right)\,.\label{cond: RSWB1b}
\end{align}
\end{subequations}
We focus on monochromatic solutions, i.e. that contain a single Fourier mode, and later investigate possible superpositions. Equations \eqref{cond: RSWB1} restrict such solutions to the form
\begin{align}\label{sol: RSWB2}
\v = \alpha_1e^{\lambda t}\cos(\k\cdot\x + \tau)\k^{\perp}\,, \quad \eta = \alpha_2\frac{f}{g}\sin(\k\cdot\x + \tau) + s\,,
\end{align}
with arbitrary shifts $\tau,\, s\in\R$ and the rest of the real parameters must satisfy
\begin{subequations}\label{cond: RSWB2}
\begin{align}
\lambda &=  (b_1-d_1|\k|^2)k_2^2+(b_2-d_2|\k|^2)k_1^2 \label{cond: RSWB2a} \\
\frac{\alpha_2-\alpha_1}{\alpha_1}f &= k_1k_2 \bigl((d_1-d_2)|\k|^2+b_2-b_1\bigr)\label{cond: RSWB2b}\\
\alpha_2\cdot\lambda&=0\label{cond: RSWB2c}\,.
\end{align}
\end{subequations}
Specifically, \eqref{eq: RSWBa} possesses explicit solutions \eqref{sol: RSWB2} with parameters satisfying \eqref{cond: RSWB2a} and \eqref{cond: RSWB2b}, and the time-independence of $\eta$ coming from \eqref{eq: RSWBb} requires condition \eqref{cond: RSWB2c}, which means $\alpha_2$ or $\lambda$ is zero. In particular, \eqref{cond: RSWB2c} means that these explicit solutions with non-trivial depth variation $\eta$ are steady. Notably, solutions with $\lambda> 0$, whose existence is studied in \S\ref{s:directions+superpositions}, grow exponentially and unboundedly. As mentioned before, we refer to this as \textit{unbounded instability} of the zero state, and more generally of any other solution which admits superposition with such growing explicit solutions. In Figure~\ref{Fig. 1} we plot the loci of these different solutions. The blue and red regions show the sign of the growth rate $\lambda$, which characterise the exponentially decaying and growing explicit solutions. In fact, we show in \S\ref{s:SWE-lin_stability_trivial} that the red region describes a subset of real unstable eigenmodes of the linearisation of the zero state. 

We proceed as follows: in \S\ref{s:directions+superpositions} we will discuss the sets of solutions in terms of their wave vectors, the organising parameters and possible superpositions. In \S\ref{Stability} we then analyse the unbounded instability and the linear stability of explicit steady solutions.

\subsection{Sets of solutions and superpositions}\label{s:directions+superpositions}
In order to analyse the existence of solutions \eqref{sol: RSWB2} to \eqref{eq: RSWB} in more detail, it is convenient to write the conditions \eqref{cond: RSWB2a} and \eqref{cond: RSWB2b} in the form
\begin{subequations}\label{cond: sigma}
\begin{align}
\lambda &= -(d_1k_2^2+d_2k_1^2)|\k|^2+b_1k_2^2+b_2k_1^2 \label{cond: sigmaa} \\
0 &= \bigl((d_1-d_2)|\k|^2+b_2-b_1\bigr)k_1k_2+\sigma\,,\label{cond: sigmab}
\end{align}
\end{subequations}
with real parameter $\sigma=f\frac{\alpha_1-\alpha_2}{\alpha_1}$ describing the relative difference between the amplitudes of the velocity vector $\v$ and the fluid depth variation $\eta$. Steady solutions satisfy \eqref{cond: sigmaa} with $\lambda=0$ and it is then natural to view $\sigma$ as an adjustment, defined by \eqref{cond: sigmab}, of the relation between the amplitudes $\alpha_1,\,\alpha_2$ depending in particular on the wave vector $\k$.
For the time-dependent case $\lambda\neq 0$ we have $\sigma=f$, since $\alpha_2=0$ is required due to \eqref{cond: RSWB2c}, and viewing \eqref{cond: sigmaa} as a definition for $\lambda$. The natural free parameter is the wave vector $\k$. The existence and growth or decay properties of solutions of the form \eqref{sol: RSWB2}, as well as the locations of unboundedly unstable steady states of this kind, are strongly connected with the values of $\sigma$, which we therefore consider as an organising parameter.\newline

\subsubsection{Superpositions of explicit flows}\label{s:SWE-superpositions}
Before discussing existence conditions, we briefly note that superpositions of solutions of the form \eqref{sol: RSWB2} are also solutions, if all wave vectors $\k$ lie on the same line through the origin in the wave vector plane. We plot examples in Figure \ref{Fig. 1c}. The reason is, that for these superpositions the nonlinear terms in \eqref{eq: RSWB} still vanish due to the orthogonality of wave vectors and flow directions, and the remaining linear equations are satisfied by each superposed explicit solution. 
This \textit{radial superposition principle} of wave vectors gives non-trivial subspaces of initial data to \eqref{eq: RSWBa} in which the dynamics are linear. In the example of Figure \ref{Fig. 1c} this space is three dimensional, since the negated wave vectors give linearly dependent solutions, and this is the maximum possible as shown below. \newline

\subsubsection{Steady explicit solutions}\label{s:SWE-steady_solutions}
For steady states, we only need to investigate the wave vectors $\k$ satisfying \eqref{cond: sigmaa} with $\lambda=0$. These form a simple closed curve around the origin in wave vector space that is symmetric with respect to axis reflection, and whose interior is star shaped, i.e. all points of the set are connected with the origin through a direct line contained in the set, but it need not be convex. We plot an example in Figure \ref{Fig. 1}. To see this, note that for wave vectors $\k=r \k'$, $\k'$ with $|\k'|=1$ fixed, the right-hand side of equation \eqref{cond: sigmaa} is linear in the squared wave vector length $|\k|^2=r^2$ (after using $\k=r\k'$ with $|\k'|=1$ and division by $r^2$). Furthermore, for any fixed $\k'$ there is exactly one $r_0>0$ so that $\lambda=0$ for $r=r_0$, and $\lambda>0$ for $0<r<r_0$ as well as $\lambda<0$ for $r_0<r$. This means, that $\lambda$ is positive in the interior of the closed curve of steady solutions \eqref{sol: RSWB2} (red regions in Figure \ref{Fig. 1}), except for the origin, where $\lambda=0$, and $\lambda$ is negative outside (blue regions in Figure \ref{Fig. 1}).\newline 
In polar coordinates 
\begin{align*}
\k=r\left(\begin{array}{c} \cos(\varphi)\\\sin(\varphi) \end{array}\right)\,,
\end{align*}
with angle $\varphi\in[0,2\pi)$ and wave number $r\geq 0$, the curve for explicit non-trivial steady solutions \eqref{sol: RSWB2}, i.e. the wave vectors with $\lambda=0$, is parameterised by the angle $\varphi$ with the wave number given by
\begin{align}\label{cond: SteadyPhi}
r=\sqrt{\frac{b_1\sin(\varphi)^2+b_2\cos(\varphi)^2}{d_1\sin(\varphi)^2+d_2\cos(\varphi)^2}}\,.
\end{align} 
Generally, these steady solutions have different values of $\sigma$, the relative difference of amplitudes $\alpha_1$ and $\alpha_2$; recall that time-dependent explicit solutions ($\lambda\neq 0$) all have the same value $\sigma=f$.  In either case, the explicit solutions \eqref{sol: RSWB2} form a linear space since their amplitudes only enter into the ratio $(\alpha_2-\alpha_1)/\alpha_1$ (so into $\sigma$), and are therefore naturally parameterised by an arbitrary amplitude parameter $a\geq 0$ that is a common factor of both $\alpha_1$ and $\alpha_2$, and thus does not change the value of $\sigma$.\newline

In the following we further investigate the conditions \eqref{cond: sigma} for the existence of explicit solutions \eqref{sol: RSWB2}. First, we analyse the occurrence and shapes of the curves defined by \eqref{cond: sigmab} and use this to determine the time-dependent explicit solutions, i.e. $\alpha_2=0$, as well as the steady explicit solutions for which \eqref{cond: sigmaa} is satisfied with $\lambda=0$. Second, we discuss the values of $\sigma$, for which steady solutions exist; clearly any steady solution of the form \eqref{sol: RSWB2} has a corresponding value of $\sigma$. But not every $\sigma$ admits such a steady solution and the value of $\sigma$ for the time-dependent solutions ($\lambda\neq0$) is fixed at $\sigma=f$, since these solutions require $\alpha_2=0$.\newline

\subsubsection{Set of solutions}\label{s:SWE-solutions_set}
\begin{figure}
\begin{center}
\subfigure[$d_2=1.0,\, b_2=1.5,\, \alpha_2=1.0$ ]{
\includegraphics[trim=5.1cm 9.3cm 5.6cm 9.7cm, clip, width=0.32\linewidth]{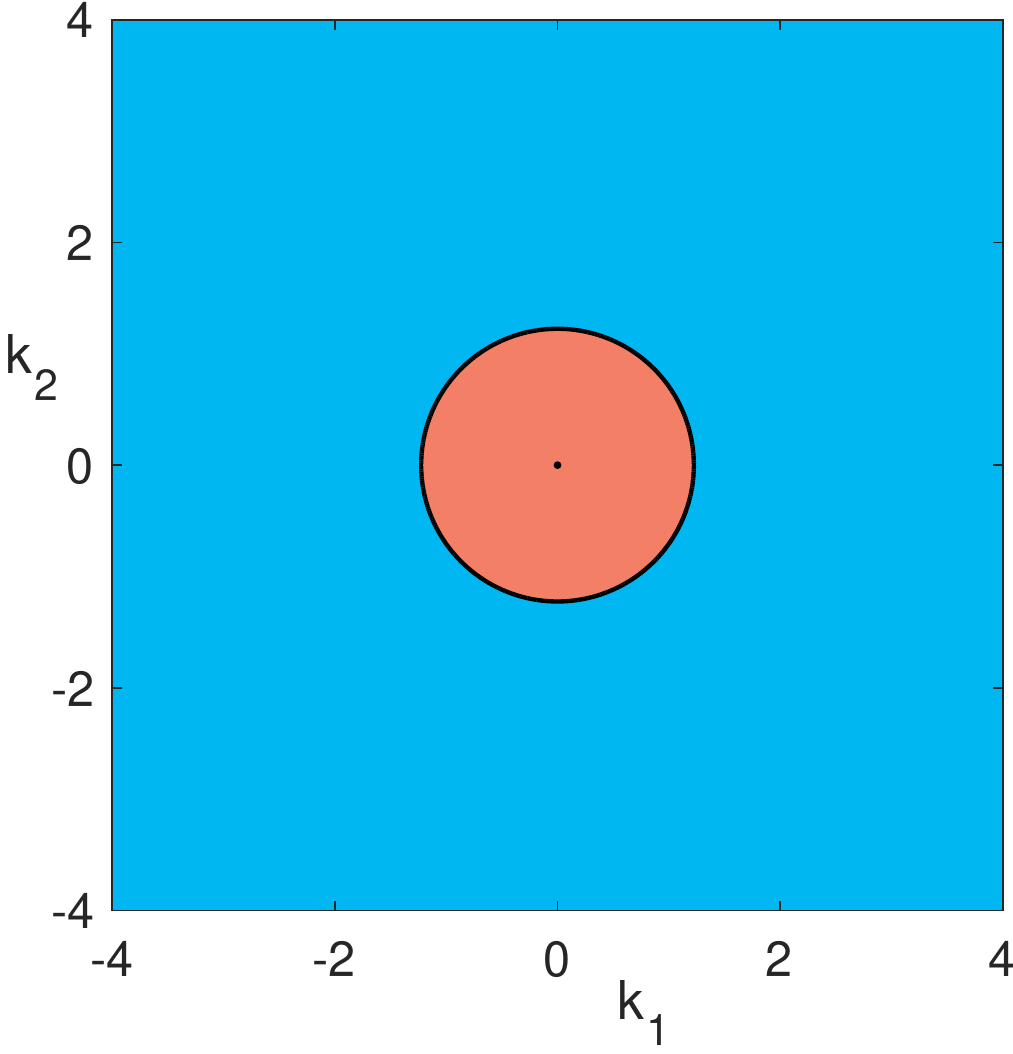}\label{Fig. 3a}}
\subfigure[$d_2=1.0,\, b_2=2.2,\, \alpha_2=3.0$]{
\includegraphics[trim=5.1cm 9.3cm 5.6cm 9.7cm, clip, width=0.32\linewidth]{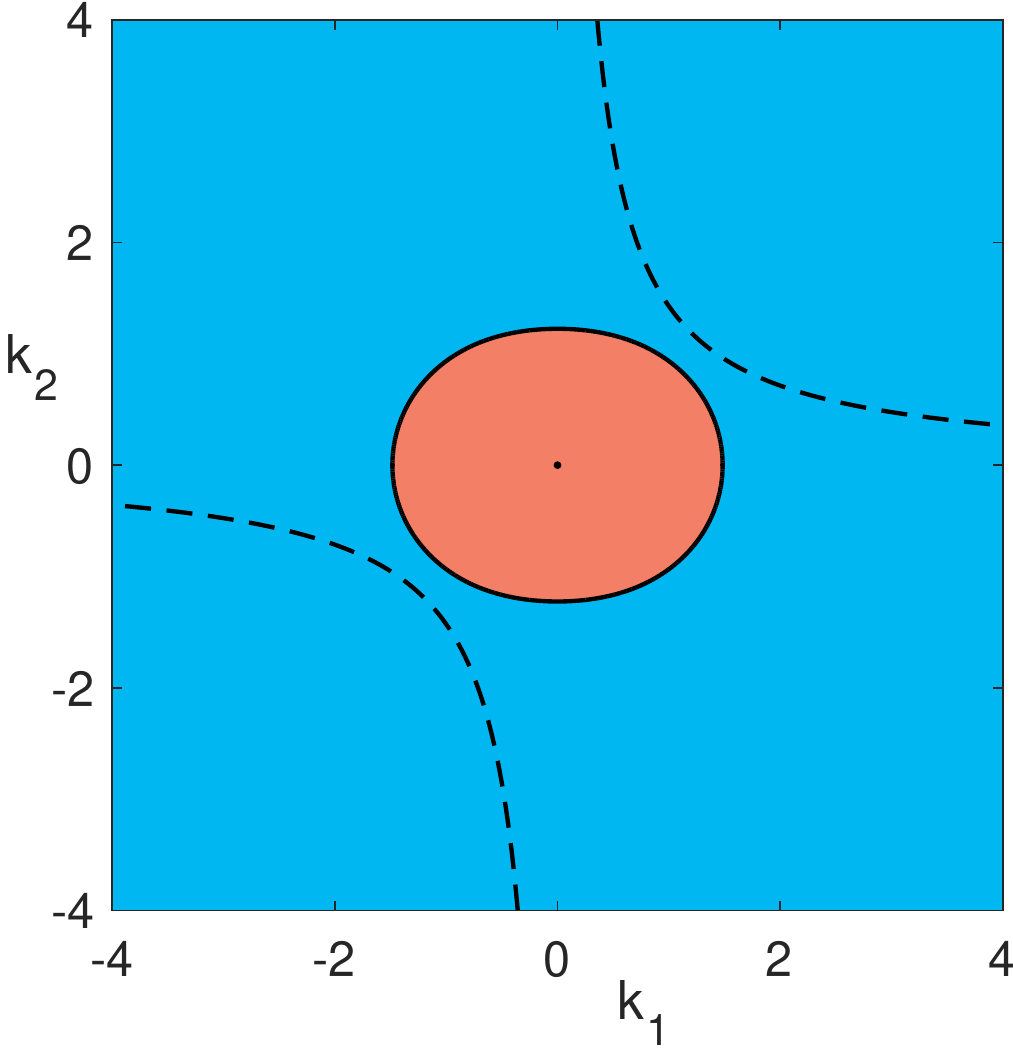}\label{Fig. 3b}}
\subfigure[$d_2=1.04,\, b_2=2.2,\, \alpha_2=3.0$]{
\includegraphics[trim=5.1cm 9.3cm 5.6cm 9.7cm, clip, width=0.32\linewidth]{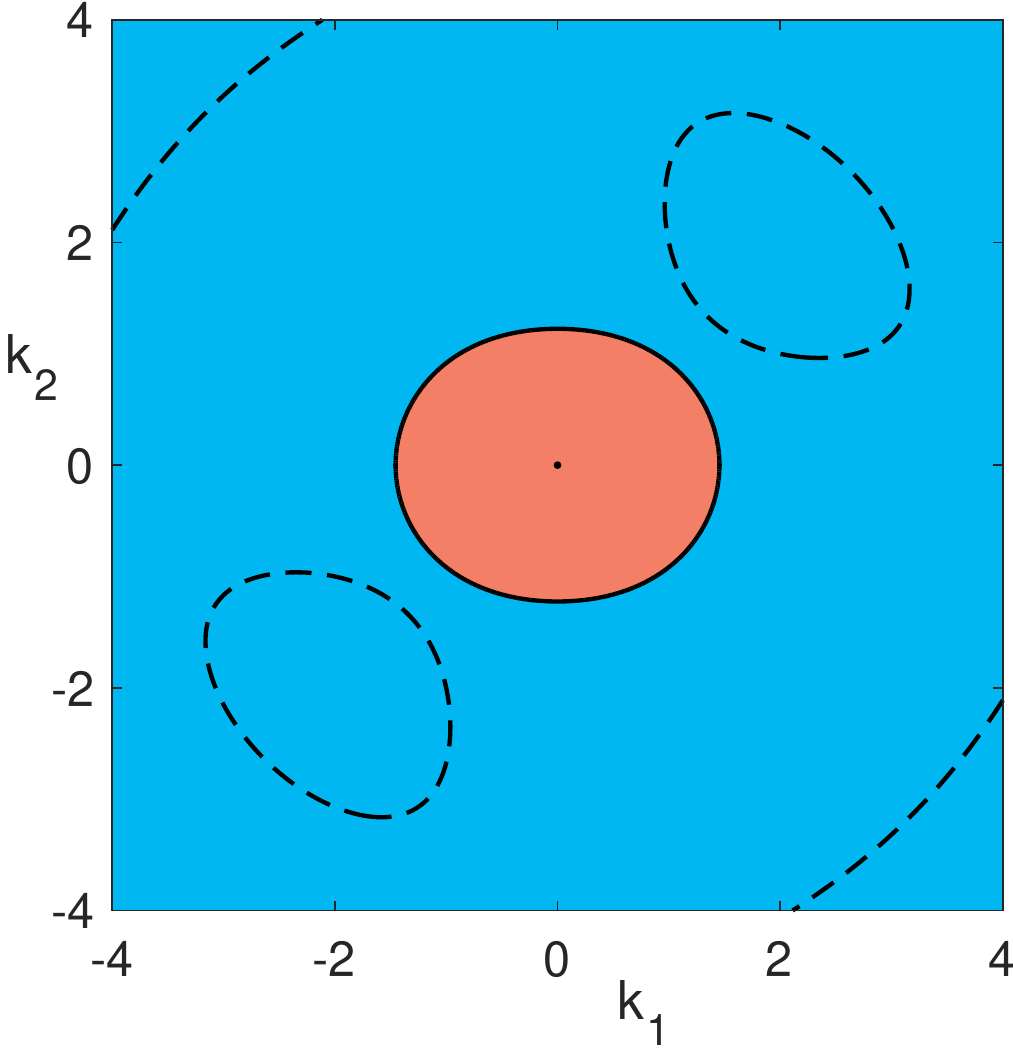}\label{Fig. 3c}}
\caption{\label{Fig. 3}Possible structures of solution curves of \eqref{cond: sigmab} analogous to Figure \ref{Fig. 1}. Fixed parameters: $d_1=1.0,\, b_1=1.5,\, f=0.5,\, g = 9.8,\, H_0 = 0.1,\, \alpha_1=1.0$; It is $\sigma=0$ in (a) and $\sigma=-1$ in (b) and (c).}
\end{center}
\end{figure}

In order to investigate the set of explicit solutions \eqref{sol: RSWB2}, primarily of the time-dependent ones with $\alpha_2=0$ and $\lambda\neq0$, we analyse the shapes of the curves defined by \eqref{cond: sigmab}. We start with two special cases: \newline
In the isotropic case $b_1=b_2$ and $d_1=d_2$, equation \eqref{cond: sigmab} requires $\sigma=0$, i.e. $\alpha_1=\alpha_2$, so that in this case all non-trivial solutions are steady, i.e. $\lambda=0$, and have $\k=0$ or $\k$ on the circle with radius $\sqrt{b_1/d_1}$ defined by \eqref{cond: sigmaa} (with $\lambda=0$), see Figure \ref{Fig. 3a}. Thus, non-steady solutions ($\lambda\neq 0$) of the form \eqref{sol: RSWB2} arise from anisotropy in the backscatter.\newline
There is also a special anisotropic case. If $d_1\neq d_2$, then \eqref{cond: sigmab} is satisfied in the origin and on the circle $|\k|^2=\frac{b_1-b_2}{d_1-d_2}$ with $\sigma=0$, and $\lambda$ defined by \eqref{cond: sigmaa} is always constant on that circle. If additionally $b_1/d_1=b_2/d_2$, then all solutions of \eqref{cond: sigmab} on the circle $|\k|^2=\frac{b_1-b_2}{d_1-d_2}$ also solve \eqref{cond: sigmaa} with $\lambda=0$, so all of these give explicit steady solutions, which have $\sigma=0$. In case $b_1/d_1\neq b_2/d_2$ the value of $\sigma$ for the steady states is not constant, as mentioned above. 

It remains to discuss the general anisotropic case, for which we consider the wave vectors in polar coordinates $\k=r(\cos(\varphi), \sin(\varphi))\transpose$ as above; by symmetry of \eqref{cond: sigmab} it suffices to take $\varphi\in[0,\pi]$. The special cases $\varphi\in\{0,\frac{\pi}{2},\pi\}$ requires $\sigma=0$, i.e. steady solutions (since then $\alpha_2\neq 0$), and the corresponding wave vectors are 
\begin{align*}
\k\in\left\{\left(\begin{array}{c} 0\\0 \end{array}\right), \pm\sqrt{\frac{b_1}{d_1}}\left(\begin{array}{c} 0\\1 \end{array}\right), \pm\sqrt{\frac{b_2}{d_2}}\left(\begin{array}{c} 1\\0 \end{array}\right)\right\}\,.
\end{align*} 
We now consider $\varphi\in(0,\pi)\backslash\{\frac{\pi}{2}\}$ only. In the case $d_1=d_2$ solutions of \eqref{cond: sigmab} are
\begin{align}\label{cond: polar1}
r=\pm\sqrt{\frac{2\sigma}{(b_1-b_2)\sin(2\varphi)}}\,,\quad \text{for} \quad \sgn(\sigma)=\sgn\bigl((b_1-b_2)\sin(2\varphi)\bigr)\,,
\end{align}
with $\sgn(\cdot)$ the sign function (see, e.g. Figure \ref{Fig. 3b}). For $d_1\neq d_2$ we get
\begin{align}\label{cond: polar2}
r=\sqrt{\frac{b_1-b_2}{2(d_1-d_2)}\pm\sqrt{\frac{(b_1-b_2)^2}{4(d_1-d_2)^2}-\frac{2\sigma}{(d_1-d_2)\sin(2\varphi)}}}\,,
\end{align}
which gives real solutions to \eqref{cond: sigmab}, if and only if the expressions in the square roots of \eqref{cond: polar2} are non-negative. This means that for fixed angle $\varphi$ we have two cases:\newline
(1) $b_1=b_2$ or $\sgn(b_1-b_2)=-\sgn(d_1-d_2)$ requires for at least one solution that
\begin{align}\label{cond: polar3}
\sgn(\sigma)=-\sgn\bigl((d_1-d_2)\sin(2\varphi)\bigr)\,.
\end{align}
(2) $b_1\neq b_2$ and $\sgn(b_1-b_2)=\sgn(d_1-d_2)$ requires for at least one solution that
\begin{subequations}\label{cond: polar4}
\begin{align}
\frac{(b_1-b_2)^2}{8(d_1-d_2)}\sin(2\varphi) &\geq \sigma\quad\text{for } \varphi\in\left(0,\frac{\pi}{2}\right)\left(\text{ if } d_1>d_2\text{, otherwise } \varphi\in\left(\frac{\pi}{2},\pi\right)\right)\,,\label{cond: polar4a} \\[2mm]
\frac{(b_1-b_2)^2}{8(d_1-d_2)}\sin(2\varphi) &\leq \sigma\quad\text{for } \varphi\in\left(\frac{\pi}{2},\pi\right)\left(\text{ if } d_1>d_2\text{, otherwise } \varphi\in\left(0,\frac{\pi}{2}\right)\right)\,.\label{cond: polar4b}
\end{align}
\end{subequations}
Two solutions for a fixed angle $\varphi$ occur if and only if the following three conditions are satisfied:
\begin{subequations}\label{cond: polar5}
\begin{align}
&\sgn(b_1-b_2)=\sgn(d_1-d_2)\,,\quad\text{so also } b_1\neq b_2\,,\label{cond: polar5a} \\[2mm]
&\sgn(\sigma)=\sgn\bigl((d_1-d_2)\sin(2\varphi)\bigr)\,,\label{cond: polar5b} \\[2mm]
&\frac{(b_1-b_2)^2}{8|d_1-d_2|}|\sin(2\varphi)| \geq |\sigma|\,.\label{cond: polar5c}
\end{align}
\end{subequations}
In Figure \ref{Fig. 3} we plot examples, where up to one (Fig.\ \ref{Fig. 3a} and Fig.\ \ref{Fig. 3b}) or up to two (Fig.\ \ref{Fig. 3c}) solutions of \eqref{cond: sigmab} for certain angles $\varphi$ arise for fixed $\sigma$. The conditions \eqref{cond: polar1}-\eqref{cond: polar5} thus determine the occurrence and structures of the solution curves of \eqref{cond: sigmab} depending on the parameter settings. For instance, changing the sign of certain expressions, but not their absolute values, merely rotates the structures by $\pi/2$.\newline

\subsubsection{Structure and values of $\sigma$}\label{s:SWE-sigma_structure}
\begin{figure}
\begin{center}
\subfigure[$\alpha_2=-3.5,\, m=-0.4$ $(\sigma=1.35)$]{
\includegraphics[trim=5.1cm 9.3cm 5.6cm 9.7cm, clip, width=0.32\linewidth]{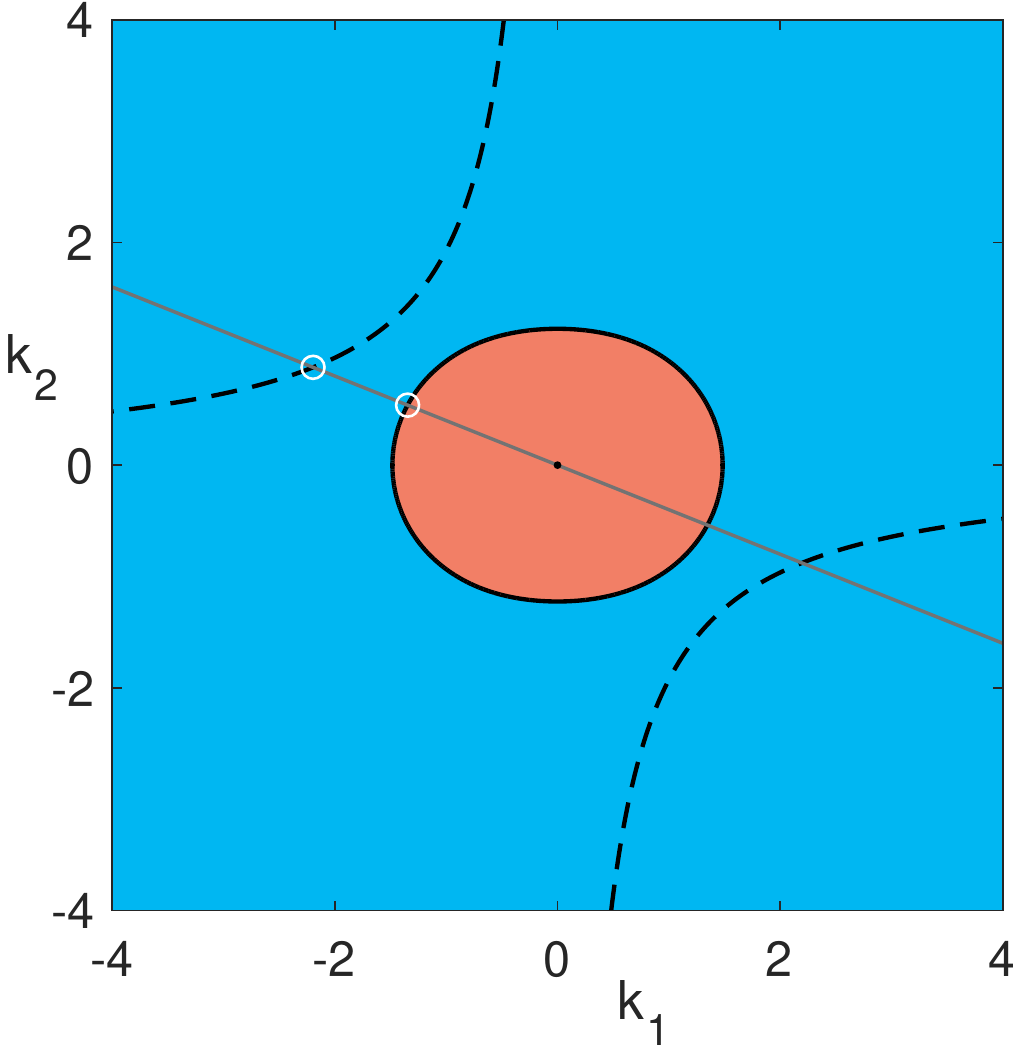}\label{Fig. 4a}}
\subfigure[$\alpha_2\approx-0.69,\, m=-0.4$ $(\sigma\approx0.51)$]{
\includegraphics[trim=5.1cm 9.3cm 5.6cm 9.7cm, clip, width=0.32\linewidth]{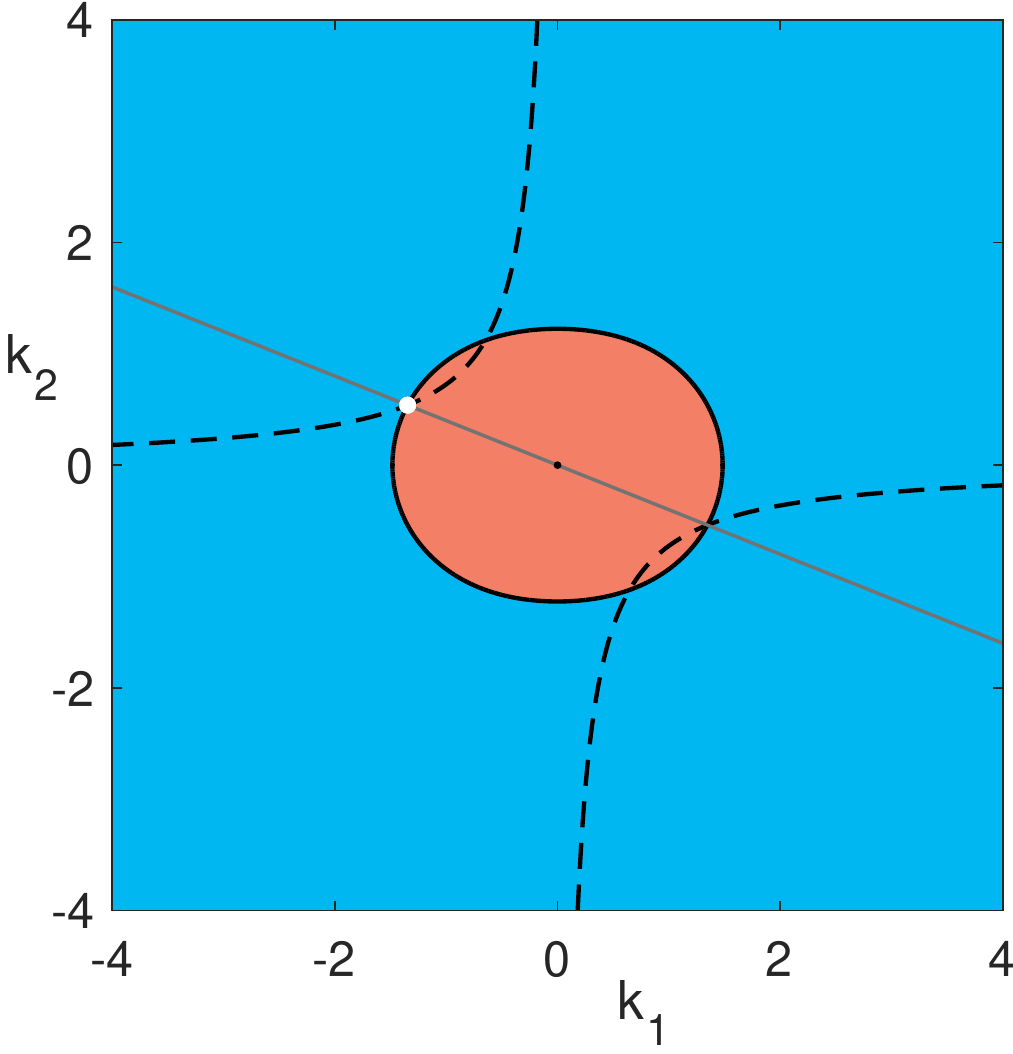}\label{Fig. 4b}}
\subfigure[$\alpha_2\approx-1.20,\, m\approx-0.84\qquad\qquad~$\hspace{10mm} $~\hspace{5mm}(\sigma=\sigma(m_{\pm})\approx0.66)$]{
\includegraphics[trim=5.1cm 9.3cm 5.6cm 9.7cm, clip, width=0.32\linewidth]{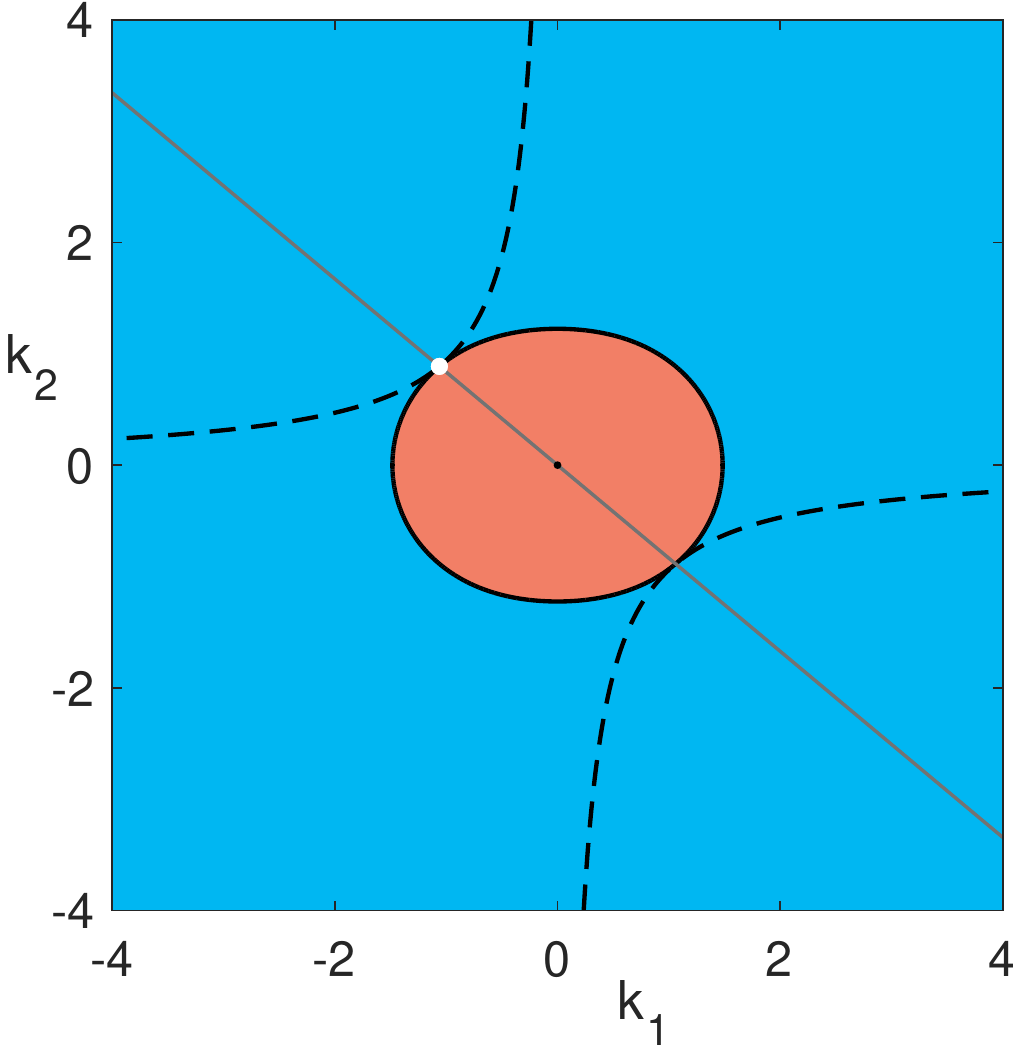}\label{Fig. 4c}}
\caption{\label{Fig. 4} Solution curves (dashed lines) of \eqref{cond: sigmab} in terms of $\sigma$, as well as their intersections with solution curve of \eqref{cond: sigmaa} with $\lambda=0$ and with gray line marking $k_2=mk_1$. Here we focus on the upper left quadrant. Denotations as in Figure \ref{Fig. 1}. In (a) there are no explicit solutions. In (b) and (c) steady solution occur at intersection point (white dot). Fixed parameters: $d_1=1.0,\, d_2=1.0,\, b_1=1.5,\, b_2=2.2,\, f=0.3,\, g = 9.8,\, H_0 = 0.1,\, \alpha_1=1.0$.}
\end{center}
\end{figure}

We next discuss the occurrence of explicit steady solutions of the form \eqref{sol: RSWB2} in the anisotropic case in more detail, in particular the values of $\sigma$, for which explicit steady solutions exist. These are the values of $\sigma$ for which the curves defined by \eqref{cond: sigmaa} with $\lambda=0$ and \eqref{cond: sigmab} intersect, see Figure \ref{Fig. 4}. Recall that time-dependent explicit solutions \eqref{sol: RSWB2} all have $\sigma=f$.\newline
To ease computations, we consider a line $k_2=mk_1$ with slope $m\in\R$, i.e. $m=\tan(\varphi)$ in the polar coordinates for wave vector used before. Inserting this into \eqref{cond: sigma} gives the values of $k_1$ for which line and curves intersect
\begin{subequations}\label{cond: intline}
\begin{align}
k_1^2 &= \frac{b_1m^2+b_2}{(d_1m^2+d_2)(1+m^2)}\qquad\text{and} \label{cond: intlinea} \\
k_1^2 &= \frac{\sigma}{m(b_1-b_2)} \qquad\text{for}\quad d_1=d_2\label{cond: intlineb}\\
k_1^2 &= \frac{b_1-b_2}{2(d_1-d_2)(1+m^2)}\pm\sqrt{\frac{m(b_1-b_2)^2-\sigma(d_1-d_2)(1+m^2)}{4m(d_1-d_2)^2(1+m^2)^2}} \qquad\text{for}\quad d_1\neq d_2\,.\label{cond: intlinec}
\end{align}
\end{subequations}
Here \eqref{cond: intlinea} is the intersection of the line with the curve defined by \eqref{cond: sigmaa}, while \eqref{cond: intlineb} and \eqref{cond: intlinec} the intersection of the line with the curve defined by \eqref{cond: sigmab} in the two cases (see Figure \ref{Fig. 4a}). 

We next choose $\sigma$ such that both intersection points are at the same position on the ray (see Figure \ref{Fig. 4b}). This occurs when the right hand side of \eqref{cond: intlinea} equals the right hand side of \eqref{cond: intlineb} or \eqref{cond: intlinec}. In both cases we find $\sigma=\sigma(m)$ is
\begin{align}\label{sigma} 
\sigma(m) = \frac{(b_1d_2-b_2d_1)(b_1m^2+b_2)m}{(d_1m^2+d_2)^2}\,, 
\end{align}
which satisfies $\sigma(-m)=-\sigma(m)$ and is zero for $b_1d_2-b_2d_1=0$, the aforementioned special anisotropic case $b_1/d_1=b_2/d_2$. In the remaining case $b_1d_2\neq b_2d_1$, we note that $\sigma(m)$ is differentiable and $\sigma(m)\rightarrow 0$ for $m\rightarrow \pm\infty$, so that it suffices to determine the extrema. The derivative of $\sigma(m)$ is given by
\begin{align}\label{derivative-sigma}
\sigma'(m) = \frac{b_1d_2-b_2d_1}{(d_1m^2+d_2)^3}\left(-b_1d_1m^4+3(b_1d_2-b_2d_1)m^2+b_2d_2\right)\,,
\end{align}
whose roots, and therefore the location of the extrema, are
\begin{align}\label{derivative-sigma}
m_{\pm} = \pm\sqrt{\frac{3(b_1d_2-b_2d_1)}{2b_1d_1}+\sqrt{\frac{9(b_1d_2-b_2d_1)^2}{4b_1^2d_1^2}+\frac{b_2d_2}{b_1d_1}}}\,.
\end{align}
Thus, steady explicit solutions of the form \eqref{sol: RSWB2} exist for $\sigma\in[\sigma(m_-),\sigma(m_+)]$, if $b_1d_2-b_2d_1>0$, and $\sigma\in[\sigma(m_+),\sigma(m_-)]$ for the case $b_1d_2-b_2d_1<0$. We may interpret the endpoints $\sigma(m_-)$ and $\sigma(m_+)$, where the solution curves of \eqref{cond: sigma} with $\lambda=0$ touch each other (see Figure \ref{Fig. 4c}), as bifurcation points of explicit steady solutions of the form \eqref{sol: RSWB2}.\newline

Using $m=\tan(\varphi)$ we equivalently obtain $\sigma$ as a function of the wave vector angle $\varphi$ that was used above. We plot an example of the resulting function $\sigma(\varphi)$ in Figure \ref{Fig. 7b}.

\subsection{Stability analysis of steady solutions}\label{Stability}
We study stability of a steady state $(\v_s,\eta_s)$ of \eqref{eq: RSWB} via the linear operator $\calL=\calL(\v_s,\eta_s)$, which results from linearising \eqref{eq: RSWB} in $(\v_s,\eta_s)$. A spectrum of $\calL$ with positive real part then implies that the steady solution $(\v_s,\eta_s)$ is linearly unstable. 
In the following we will show that the trivial steady state $(\v_s,\eta_s)\equiv(0,0,0)$ is linearly unstable, and in certain cases even unboundedly unstable. Afterwards, we focus on the stability of non-trivial steady solutions \eqref{sol: RSWB2}. We first analyse the unbounded instability of these in the full nonlinear equations \eqref{eq: RSWB} with respect to solutions of the form \eqref{sol: RSWB2}. Moreover, we study a certain long-wavelength instability in this case and briefly consider the energy of the explicit solutions. Finally, we investigate the linear stability of all steady solutions \eqref{sol: RSWB2} with small and large amplitudes.\newline

\subsubsection{Linear stability of trivial steady state}\label{s:SWE-lin_stability_trivial}
For the trivial homogeneous steady solution $(\v_s,\eta_s)\equiv(0,0,0)$ the linearisation of \eqref{eq: RSWB} is exactly \eqref{eq: RSWB}  without the nonlinear terms, and the corresponding linear operator $\calL$ is then the remaining right-hand side of \eqref{eq: RSWB}. The spectrum of $\calL$ in this case can be determined by the dispersion relation
\begin{align*}
d(\lambda,\k) := \det(\lambda\mathrm{Id} - \widehat\calL) = 0\,,
\end{align*}
with wave vectors $\k=(k_1,k_2)\transpose\in\R^2$, temporal rates $\lambda=\lambda(\k)\in\C$ and $\widehat\calL=\widehat\calL(\k)$ the Fourier transform of $\calL$ given by
\begin{align*}
\widehat\calL(\k) = 
\begin{pmatrix}
- d_1|\k|^4 + b_1|\k|^2 & f & -\rmi gk_1  \\
-f & - d_2|\k|^4 + b_2|\k|^2 & -\rmi gk_2 \\
-\rmi H_0k_1 & -\rmi H_0k_2 & 0
\end{pmatrix}\,.
\end{align*}
The dispersion relation is thus explicitly
\begin{align}\label{e:disp}
d(\lambda,\k) = \lambda^3 + c_2 \lambda^2 + c_1 \lambda + c_0 = 0\,,
\end{align}
with the coefficients
\begin{align*}
c_2 \ &:=\ (d_1 + d_2) |\k|^4 - (b_1 + b_2) |\k|^2 \,, \\
c_1 \ &:=\ (d_1 |\k|^4 - b_1 |\k|^2) (d_2 |\k|^4 - b_2 |\k|^2) + g H_0 |\k|^2 + f^2\,, \\
c_0 \ &:=\ - g H_0 |\k|^2 \left((b_1-d_1|\k|^2)k_2^2+(b_2-d_2|\k|^2)k_1^2  \right)\,.
\end{align*}

Recall that the explicit solutions \eqref{sol: RSWB2} with \eqref{cond: RSWB2} solve \textit{both} \eqref{eq: RSWB} with and without the nonlinear terms, since these terms vanish by construction of these solutions. Therefore, the wave vectors $\k$ and growth rates $\lambda$ of the explicit solutions are in fact real solutions of the dispersion relation \eqref{e:disp}. In other words, all these explicit solutions are real eigenmodes of $\calL$ and the values $\lambda$ defined by \eqref{cond: RSWB2a} are the corresponding real elements in the spectrum of $\calL$. 
Thus, the possible values for $\lambda$ of the explicit solutions \eqref{sol: RSWB2} with \eqref{cond: RSWB2} directly provide part of the spectrum of $\calL$, for instance all values of $\lambda$ on the white and black curves in Figure \ref{Fig. 1c}. In particular, the occurrence of positive growth rates $\lambda$ in \eqref{cond: RSWB2a} implies that the trivial steady solution $(\v_s,\eta_s)\equiv(0,0,0)$ is linearly unstable with respect to these exponentially growing explicit solutions. For instance, in Figure \ref{Fig. 1c} this happens for the wave vectors on the part of the white curves within the red region. More generally, even if the white curves do not intersect the red region, we next show that the red region is filled with unstable real modes of $\calL$. \newline

\begin{figure}
\begin{center}
\subfigure[~]{
\includegraphics[trim=2.1cm 0.0cm 2.5cm 0.4cm, clip, width=0.32\linewidth]{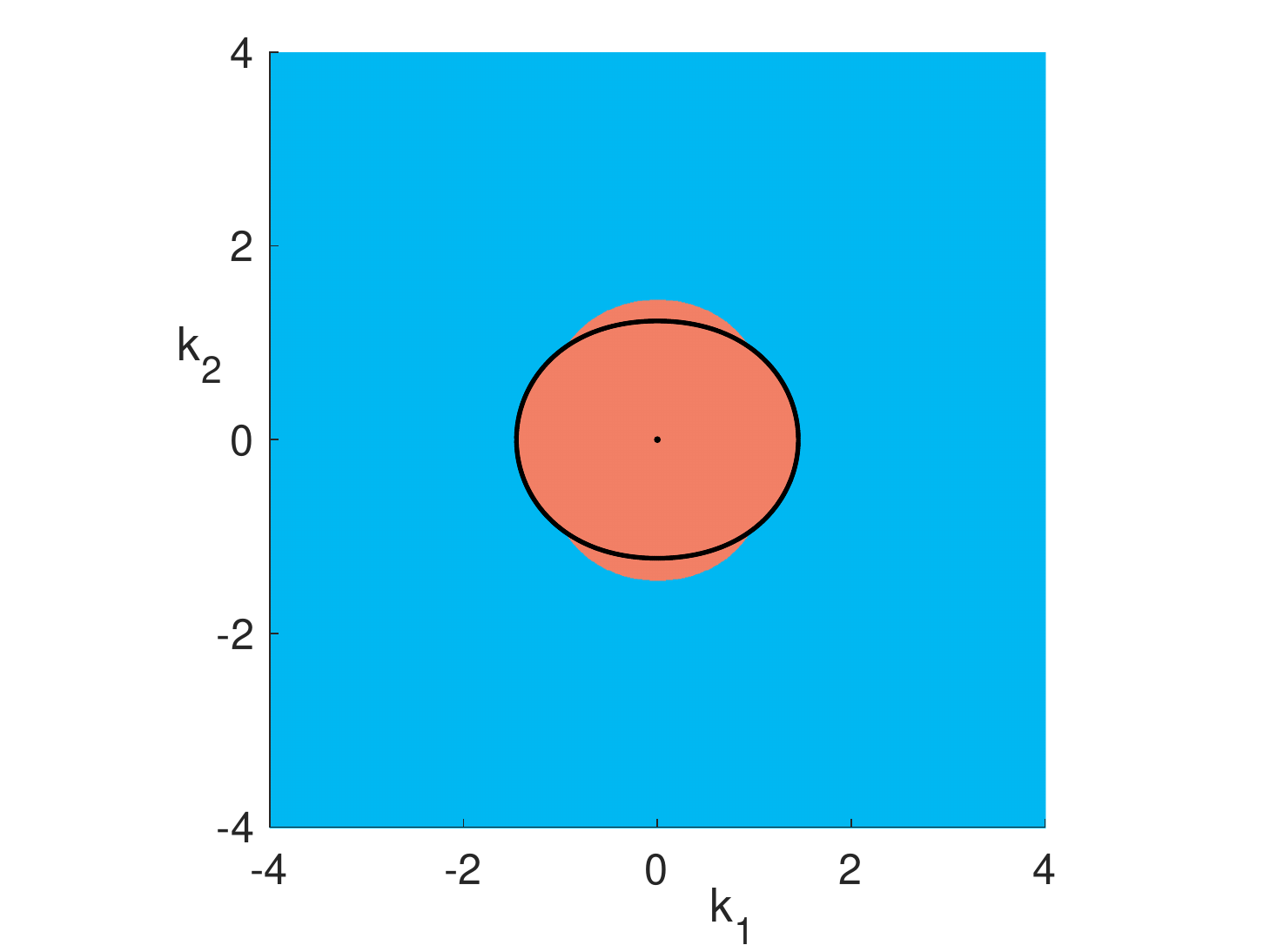}\label{Fig: SpecZeroStatea}}
\hspace{1cm}
\subfigure[~]{
\includegraphics[trim=3.4cm 9.3cm 4.3cm 9.6cm, clip, width=0.415\linewidth]{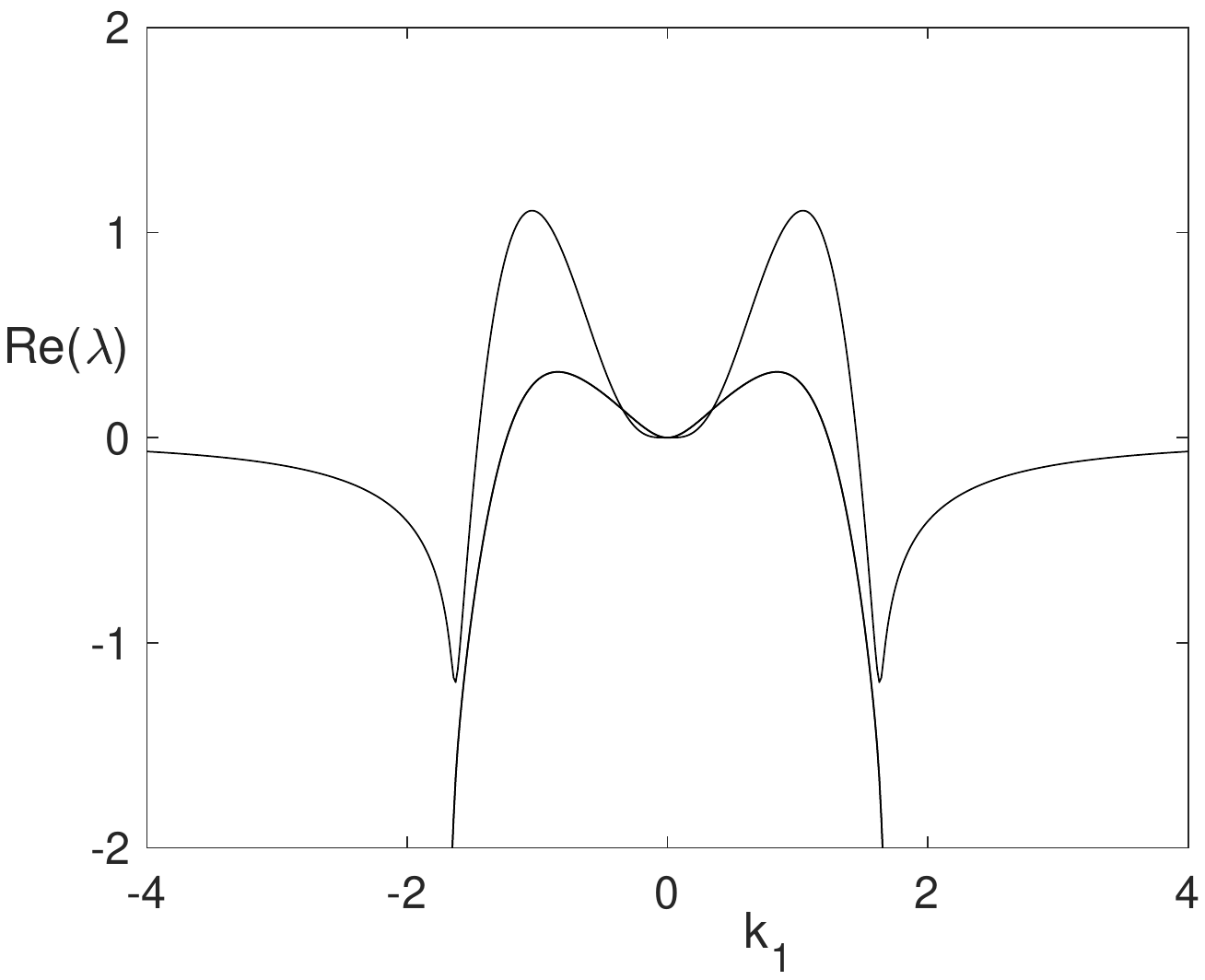}\label{Fig: SpecZeroStateb}}
\caption{\label{Fig: SpecZeroState} Information on the spectrum of the linearisation of \eqref{eq: RSWB} in the trivial steady solution $(\v_s,\eta_s)\equiv(0,0,0)$ for parameters $d_1=1.0,\, d_2=1.04,\, b_1=1.5,\, b_2=2.2,\, f=-0.3,\, g = 9.8,\, H_0 = 0.1$. (a): Signs of the most unstable real part of elements in the spectrum in terms of the wave vector $\k$ of the associated eigenmodes -- real part positive (red region); negative (blue region); black dot and curve correspond to steady states of \eqref{sol: RSWB2}. Note the unstable non-real spectrum in addition to, e.g. Figure \ref{Fig. 1}. (b): The real part of the spectrum for $k_2=0$, showing unstable spectrum in the vicinity of the origin.}
\end{center}
\end{figure}

We first note that in $\k=(0,0)\transpose$ the dispersion relation reduces to 
\begin{align*}
d(\lambda,(0,0)\transpose)=\lambda^3+f^2\lambda=0\,,
\end{align*}
which gives $\lambda=0$ and $\lambda= \pm\rmi f$, all having zero real part. A subset of the unstable spectrum can be determined by the sign of $c_0=-gH_0|\k|^2\beta(\k)$, where $\beta(\k)$ is the expression in brackets in the definition of $c_0$ above. The coefficient $c_0$ of the dispersion relation \eqref{e:disp} is zero if and only if $\k\in\R^2$ satisfies $\beta(\k)=0$, which means that $\lambda=0$ is in the spectrum of $\calL$ with corresponding eigenmodes having such wave vectors $\k$. Furthermore, $c_0$ is negative if and only if $\beta(\k)>0$, so according to the dispersion relation \eqref{e:disp} there is at least one positive real value $\lambda>0$ for each of these wave vectors $\k$.
We notice, that $\beta(\k)$ is also exactly the same expression as on the right-hand side of \eqref{cond: RSWB2a} or \eqref{cond: sigmaa}, whose sign we have already analysed above. In other words, the red regions plotted, e.g. in Figure \ref{Fig. 1c}, correspond to a part of the unstable spectrum of $\calL$ which are positive real. In particular, we conclude that $(\v_s,\eta_s)\equiv(0,0,0)$ is linearly unstable for any choice of parameters with horizontal backscatter. However, the spectrum of $\calL$ may also contain non-real unstable parts. We plot an example in Figure \ref{Fig: SpecZeroStatea}, where the unstable region extends into the blue region of Figure \ref{Fig. 1c}. This can be further studied based on the dispersion relation \eqref{e:disp}, but we will not do this here.\newline

The previous investigation regarding the instability of the trivial flow in fact shows that the explicit solutions \eqref{sol: RSWB2} of the full nonlinear equations \eqref{eq: RSWB} with $\lambda>0$ are real unstable eigenmodes. Here we see a specific case of what we refer to as unbounded instability: perturbations of the zero state by one such mode not only leads to infinitesimal or local growth, but to globally in time unbounded growth in the nonlinear system. \newline

\subsubsection{Unbounded and long-wavelength instability of non-trivial steady states}\label{Stability: unbounded growth}
\begin{figure}
\begin{center}
\subfigure[]{
\includegraphics[trim=5.1cm 9.3cm 5.6cm 9.7cm, clip, width=0.32\linewidth]{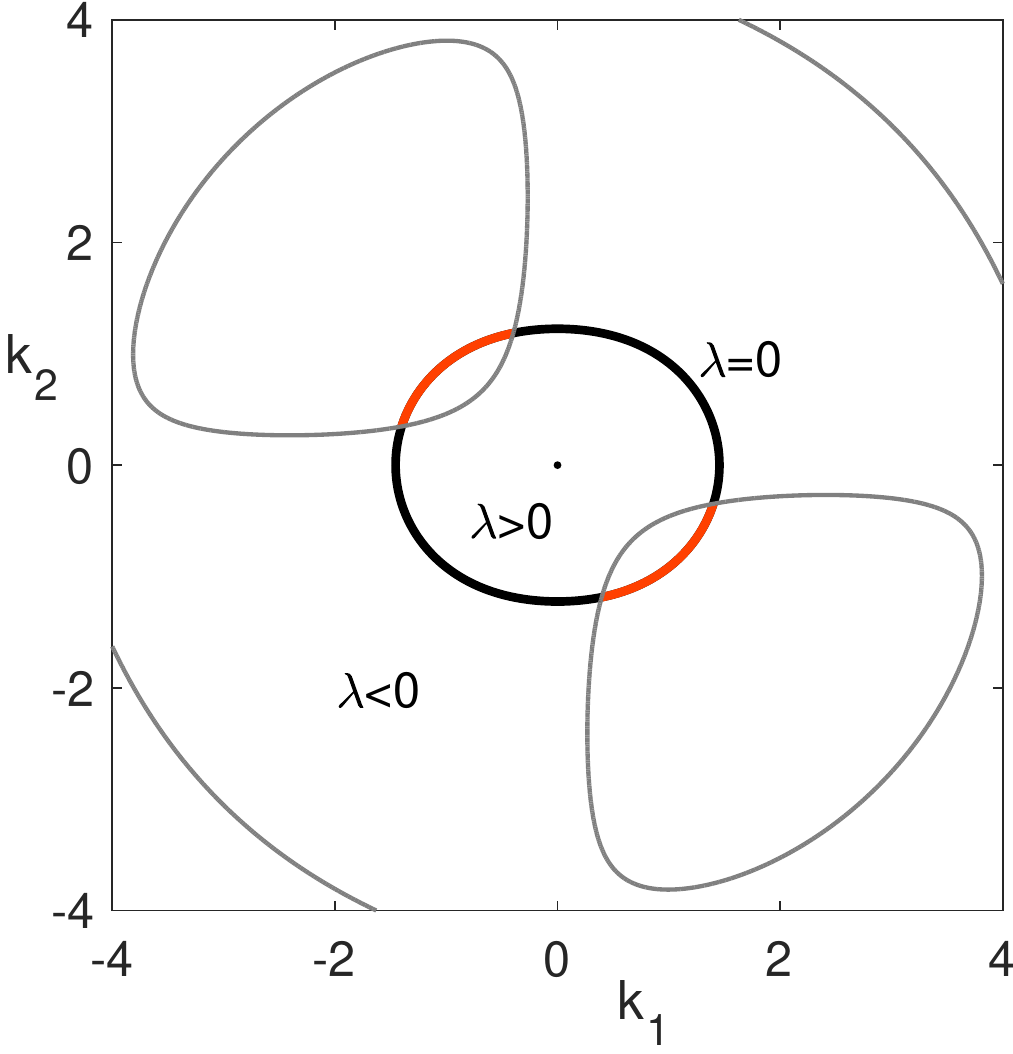}\label{Fig. 7a}}
\hspace{1cm}
\subfigure[]{
\includegraphics[trim=3.6cm 8.5cm 4.0cm 8.8cm, clip, width=0.3625\linewidth]{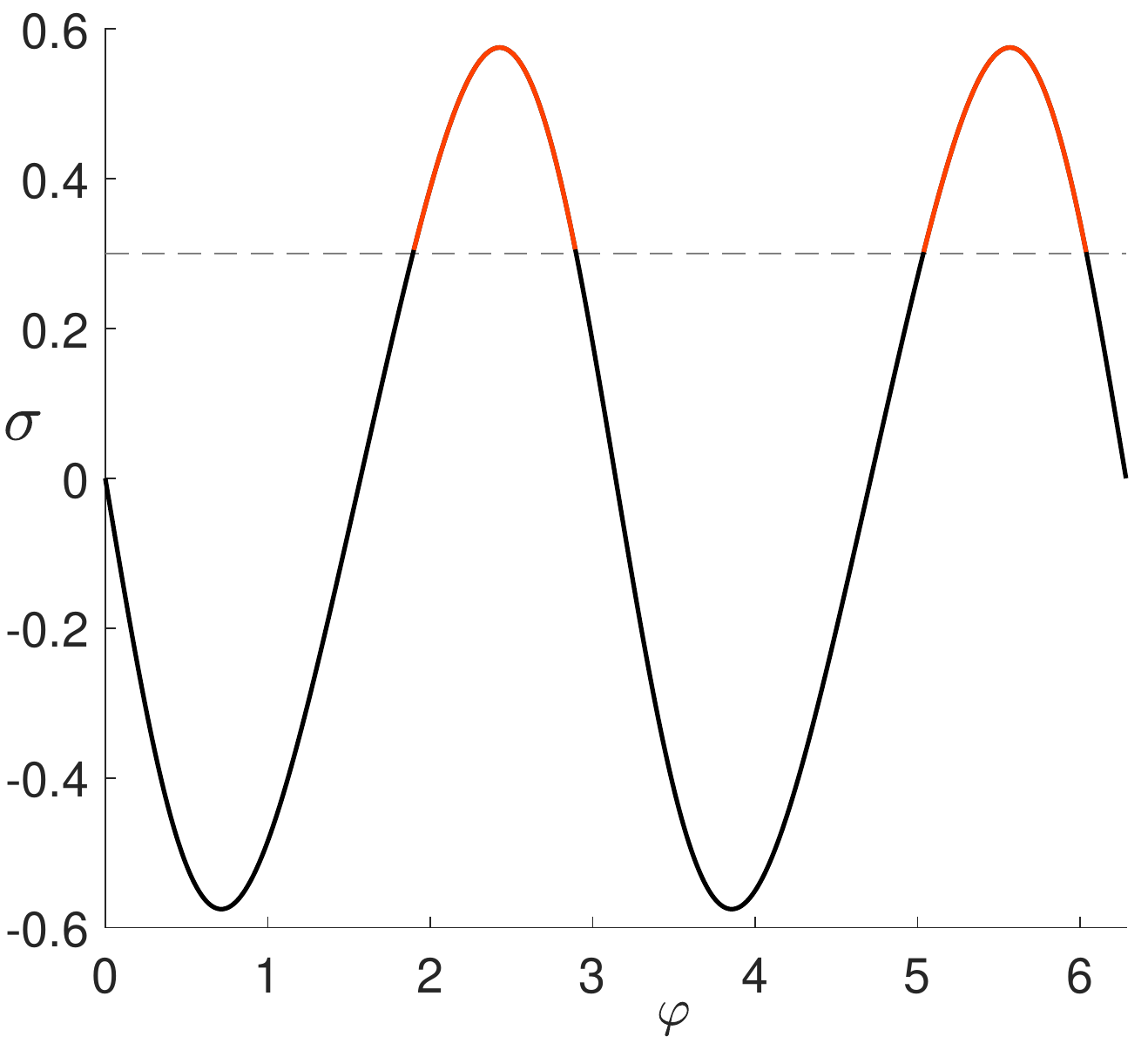}\label{Fig. 7b}}
\caption{\label{Fig. 7} Examples for unboundedly unstable explicit steady solutions \eqref{sol: RSWB2} for the same parameter values as in Fig.~\ref{Fig. 1c}. (a) Red arcs mark unboundedly unstable steady solutions. Gray curves mark explicit time-dependent solutions with $\alpha_2=0$. (b) Graph of $\sigma(\varphi)$. Red parts mark unboundedly unstable cases; dashed gray line marks the value of the Coriolis parameter $f$.}
\end{center}
\end{figure}

In the following we show that some of the steady solutions \eqref{sol: RSWB2} can be unboundedly unstable as well. We consider parameter values such that some time-dependent explicit solutions \eqref{sol: RSWB2} have positive growth rate $\lambda$, as in the example of Figure \ref{Fig. 1}. As already shown, steady solutions \eqref{sol: RSWB2} exist on the whole curve defined by \eqref{cond: RSWB2a} with $\lambda=0$ (see Figure \ref{Fig. 7a} as well as the black curve in Figure \ref{Fig. 1c}). Now superpositions of explicit solutions \eqref{sol: RSWB2}, which have the same wave vector direction (e.g. the intersections of white or black curves with the gray line in Figure \ref{Fig. 1c}), are also explicit solutions of \eqref{eq: RSWB}. In the case of Figure \ref{Fig. 1c} these are in particular a non-trivial steady solution $(\v_s,\eta_s)$ (white dot) and an exponentially growing solution $(\v_g,0)$ (white circle in red area). Any superposition $\alpha(\v_s,\eta_s)+\varepsilon(\v_g,0)$, with arbitrary $\alpha,\, \varepsilon\in\R$, is also an explicit solution of \eqref{eq: RSWB}; in particular, $\varepsilon$ can be arbitrarily close to zero. For any $\varepsilon\neq 0$, the resulting solution is exponentially and unboundedly growing. Thus, $(\v_s,\eta_s)$ is an unboundedly unstable steady solution. \newline

This implies the unbounded instability of the explicit steady solutions \eqref{sol: RSWB2} corresponding to wave vectors on the red arcs in Figure \ref{Fig. 7a}; in Figure \ref{Fig. 1c} these are between the intersections of black and white curves. These arcs connect intersection points of the curve defined by \eqref{cond: RSWB2a} for $\lambda=0$, with that for time-dependent explicit solutions defined by \eqref{cond: RSWB2b} with $\alpha_2=0$. 
The instability of the other explicit steady solutions (black regions in Figure \ref{Fig. 7a}) is not determined in this way; we discuss some cases later. However, the transition from the black to the red arcs can be associated with a long-wavelength instability (also called sideband or modulational instability). The numerical result plotted in Figure~\ref{fig:spec} shows that this instability should be expected on top of already unstable spectrum.

In order to study the long-wavelength instability, we consider the wave vector angle $\varphi$ and first discuss the values of $\sigma(\varphi)$, for which the corresponding explicit steady solutions \eqref{sol: RSWB2} are unboundedly unstable. Recall that since time-dependent explicit solutions \eqref{sol: RSWB2} require $\alpha_2=0$, according to \eqref{cond: RSWB2c}, these have $\sigma=f$. Thus, steady solutions with $\sigma(\varphi)=f$ lie at the intersections with the curve of time-dependent solutions and all steady solutions ``between'' those with $\sigma(\varphi)=f$ are unboundedly unstable, since those can be superposed with growing explicit solutions (cf. red regions in Figure \ref{Fig. 7a}). More precisely, there are at most four angles $\varphi_j\in[0,2\pi)$, ordered by size, so that $\sigma(\varphi_j)=f$ (cf.\ Figure \ref{Fig. 7b}), and steady solutions \eqref{sol: RSWB2} whose wave vectors have angles between $\varphi_1$ and $\varphi_2$, or $\varphi_3$ and $\varphi_4$, are unboundedly unstable. Hence, a steady solution \eqref{sol: RSWB2} is unboundedly unstable if and only if $\sgn(f)\sigma(\varphi)>|f|$, with its corresponding value $\sigma(\varphi)$  (see Figure \ref{Fig. 7b}). 

\begin{figure}
\begin{center}
\subfigure[]{\includegraphics[trim=3.4cm 9.2cm 4.4cm 9.6cm, clip, width=0.32\textwidth]{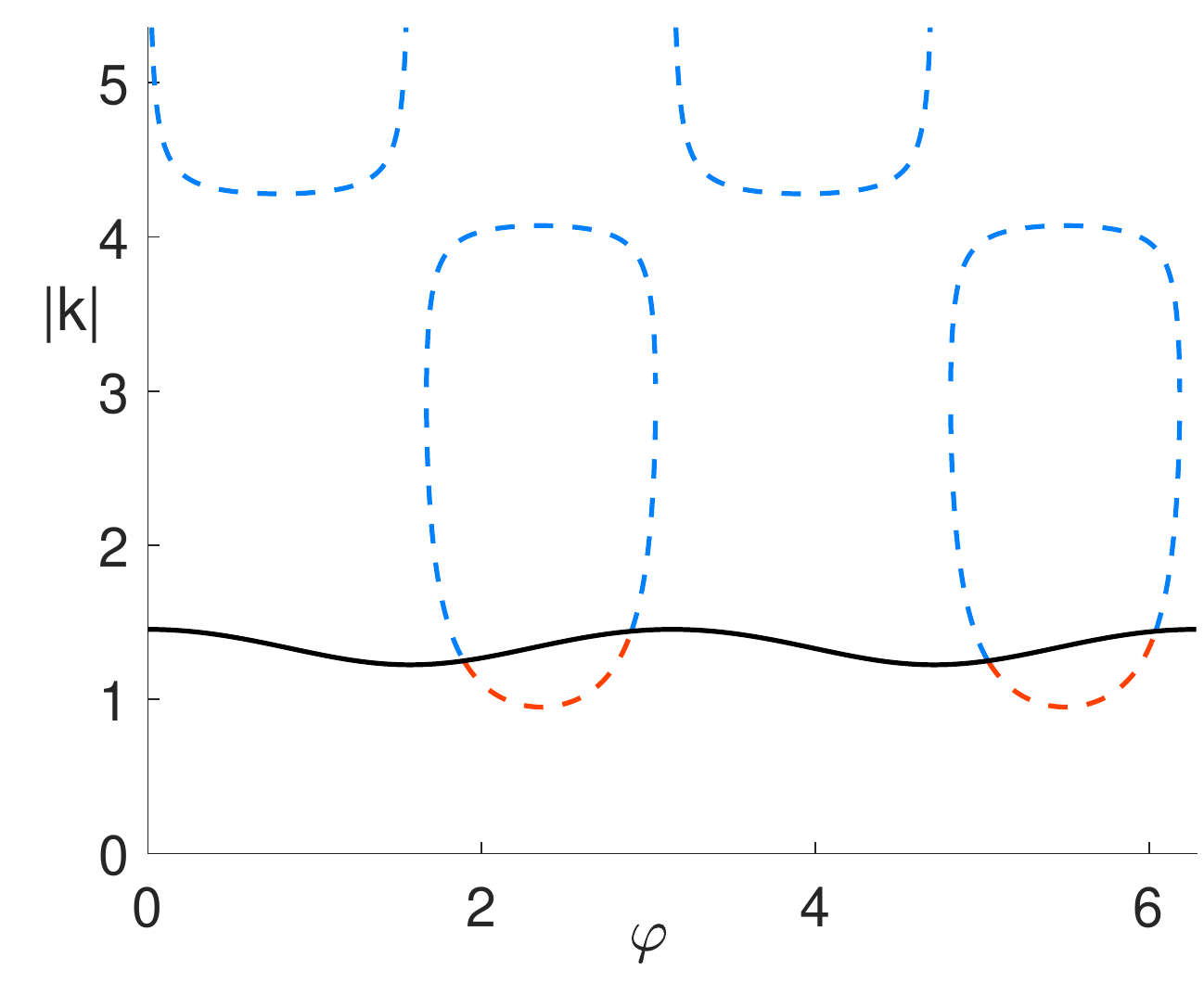}\label{Fig. 5a}}
\subfigure[]{\includegraphics[trim=3.4cm 9.2cm 4.4cm 9.6cm, clip, width=0.32\textwidth]{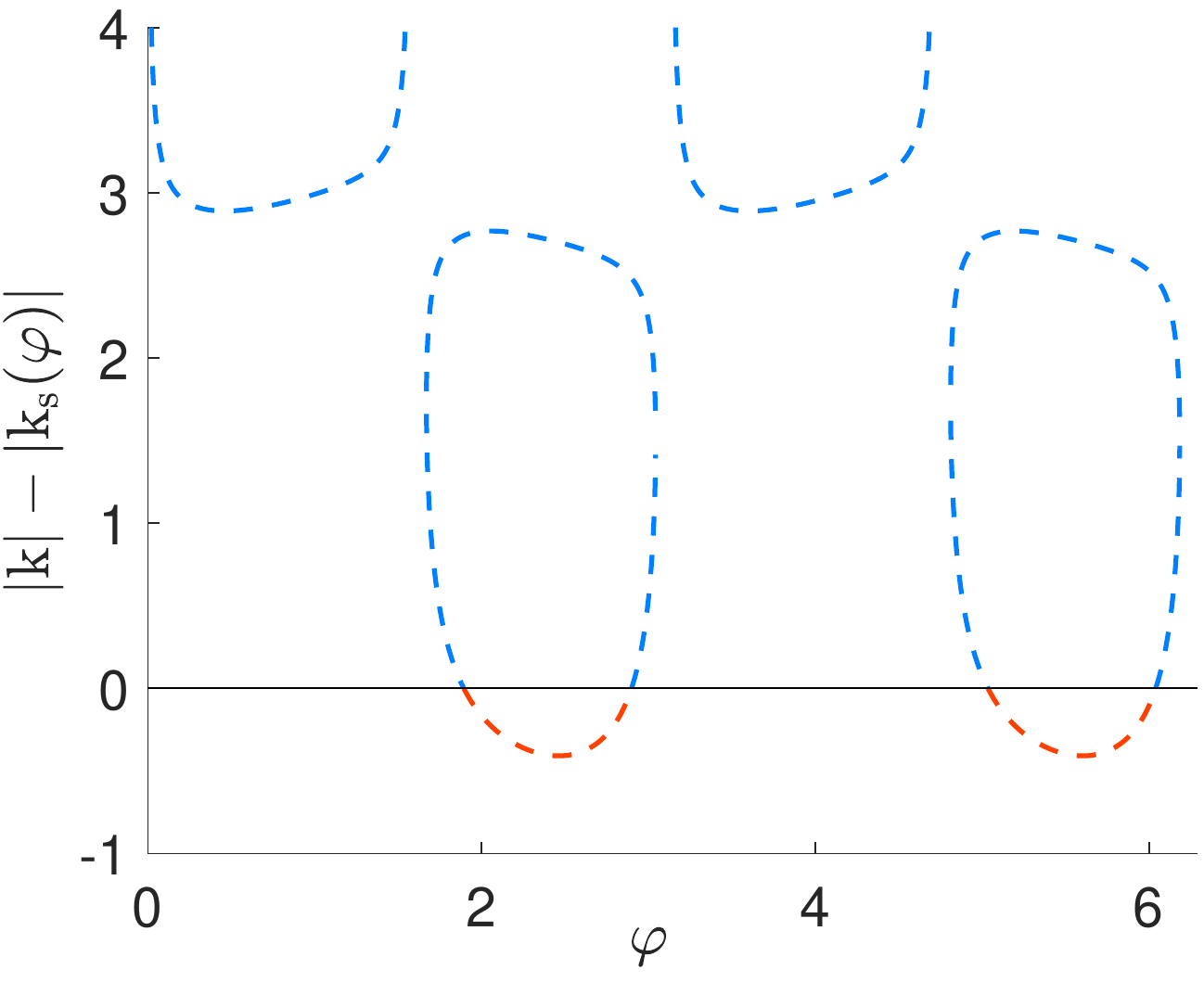}\label{Fig. 5b}}
\subfigure[]{\includegraphics[trim=3.9cm 9.6cm 4.4cm 9.8cm, clip, width=0.32\textwidth]{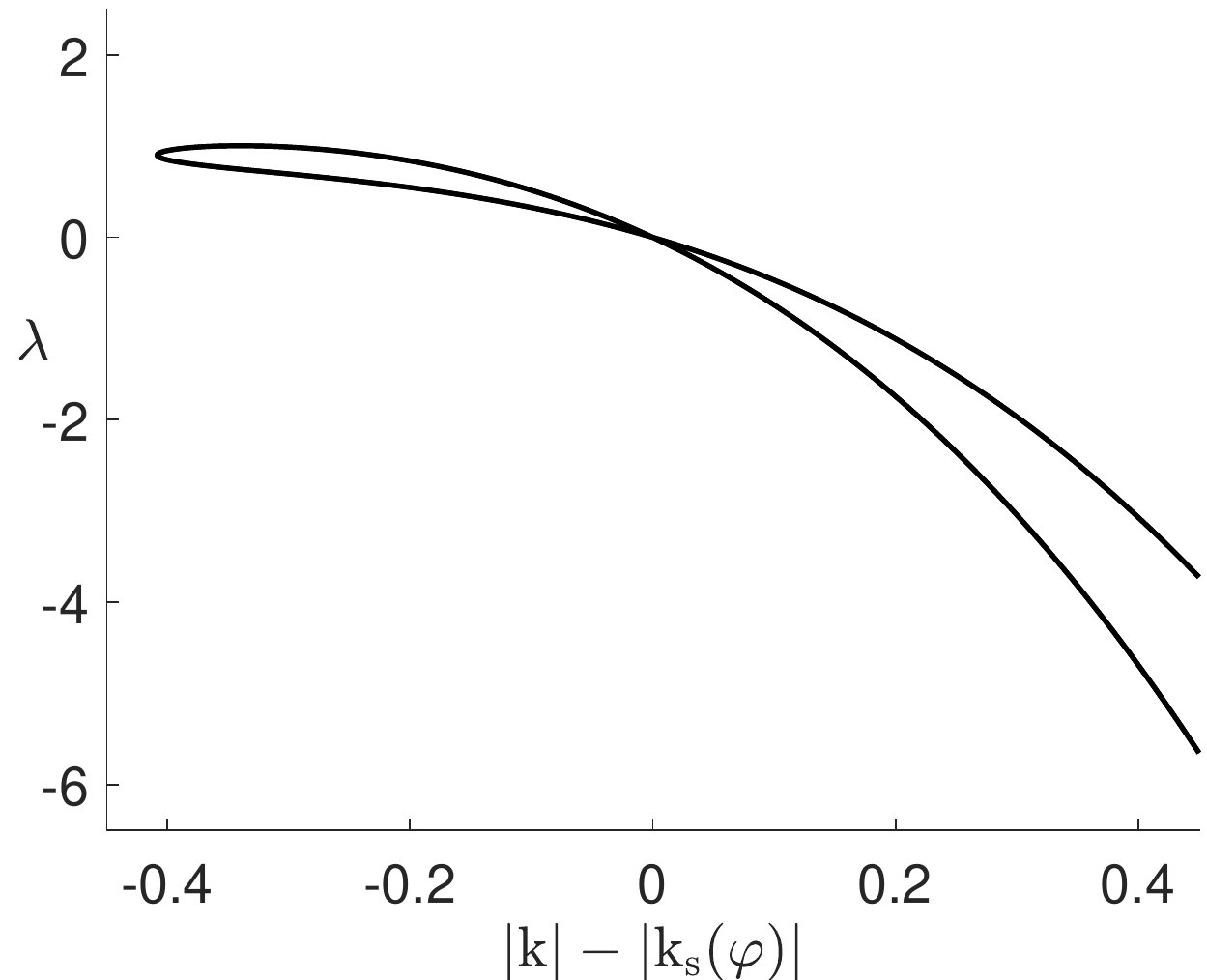}\label{Fig. 5c}}
\caption{\label{Fig. 5} Illustration of long-wavelength instabilities of explicit flows. (a)-(b): Wave vectors $\k=|\k|\bigl(\cos(\varphi),\sin(\varphi)\bigr)\transpose$ of explicit solutions \eqref{sol: RSWB2} as functions of $\varphi$. Black: steady solutions $\k_s(\varphi)$; blue: exponentially decaying; red: exponentially growing. (c): Growth rate $\lambda$ as a function of the difference of wave vector lengths from steady solution. Parameters for all three cases as in Figure \ref{Fig. 1c}.}
\end{center}
\end{figure}

Towards the long-wavelength instability, we parameterise the set of steady solutions by the angle $\varphi$ of their wave vectors. In Figure \ref{Fig. 5a} we plot for each $\varphi$ the wave vector lengths for which an explicit solution of the form \eqref{sol: RSWB2} exists and whether it is steady, exponentially decaying or growing. This also readily shows admissible superpositions of explicit solutions, since these must have the same angle $\varphi$; with respect to these exponentially growing explicit solutions, we thus have stable and unstable steady solutions. The stability change occurs at the intersections of the curves of the steady and time-dependent solutions, thus providing a long-wavelength instability character at these points, since the difference of wavelength between the steady solution and the unstable mode (the Floquet-Bloch parameter) crosses zero here (see Figure \ref{Fig. 5b}). Conversely, given any small Floquet-Bloch parameter one can find a value of $\varphi$, so that the corresponding steady solution is unstable with respect to it. See Figure \ref{Fig. 5c}, where the self-intersection point at the origin shows two such points along $\varphi$.\newline

We briefly consider some energetic aspects of the explicit solutions. The kinetic energy density of solutions to \eqref{eq: RSWB} is given by $\rm{KE}(\v,\eta)=\frac{1}{2}(H_0+\eta)|\v|^2$ and the potential energy density by $\rm{PE}(\v,\eta)=\frac{g}{2}(H_0+\eta)^2$. The superposed explicit solutions are generally of the form
\begin{align*}
\v&=\alpha_s\cos_s\k^{\perp}+\alpha_ne^{\lambda_nt}\cos_n\k^{\perp}+\alpha_pe^{\lambda_pt}\cos_p\k^{\perp}\,, \\
\eta&=\gamma\frac{f}{g}\sin(\kappa_s\k\cdot\x+ \tau_s)+c\,,
\end{align*}
with $|\k|=1$, wave shapes $\cos_a=\cos(\kappa_a\k\cdot\x+ \tau_a)$ for $a\in\{s,n,p\}$, growth rates $\lambda_n<0$ and $\lambda_p>0$, wave numbers $\kappa_s,\, \kappa_n,\, \kappa_p \in\R$ as well as arbitrary amplitudes $\alpha_a\in\R$ and shifts $\tau_a,\,c\in\R$ for any $a\in\{s,n,p\}$. The corresponding terms are steady, decaying and growing explicit solutions, determined by \eqref{sol: RSWB2} and \eqref{cond: RSWB2} (compare with steady solutions in red region in Figure \ref{Fig. 7a} and the possible superpositions with solutions which are decaying or growing in time). The energy densities of these explicit solutions explicitly read
\begin{align*}
\rm{KE}(\v,\eta)&=\frac{1}{2}\bigl(\gamma\frac{f}{g}\sin(\kappa_s\k\cdot\x+ \tau_s) +H_0+c\bigr)\bigl(\alpha_s\cos_s+\alpha_ne^{\lambda_nt}\cos_n+\alpha_pe^{\lambda_pt}\cos_p\bigr)^2\,,\\
\rm{PE}(\v,\eta)&=\frac{\gamma^2f^2}{2g}\sin^2(\kappa_s\k\cdot\x+ \tau_s)+\gamma f(H_0+c)\sin(\kappa_s\k\cdot\x+ \tau_s)+\frac{g(H_0+c)^2}{2}\,.
\end{align*}
Notably, being cubic in the sine/cosine terms, the kinetic energy in Fourier space features various diadic and triadic combinations of the wave vectors of the corresponding velocity components of the explicit solution.  On the temporal side, the squared linear combination of time-independent, decaying and growing parts yields doubling and adding of the individual rates. 
The potential energy is ignorant to the dynamic terms, but we note the constant and $2\kappa_s\k$ Fourier modes from the quadratic term.\newline

\subsubsection{Linear stability of non-trivial steady states with small and large amplitudes}\label{s:stabamp}
We now study stability properties of steady solutions \eqref{sol: RSWB2} for (asymptotically) small and large amplitudes, the natural asymptotic regimes for families of solutions with a free amplitude parameter. Since such linear spaces of solutions arise more broadly in incompressible fluid equations with transport nonlinearity, cf.~\citep{prugger2020explicit}, and for later use in \S\ref{Rotating Boussinesq}, we set up the notation for the more general setting of an evolution equation with linear term $\Lb$ and bilinear nonlinearity $\B$ given by
\[
\frac{\p}{\p t} \u =\Lb\u + \B(\u,\u) + \nabla p\,,
\] 
with a pressure $p$, that is trivial for the rotating shallow water equations \eqref{eq: RSWB} with $\u=(\v,\eta)$, and otherwise will derive from the incompressibility constraint $\nabla\cdot\v=0$.

We assume there exists a family of steady state solutions $\u=a\u_0$ with amplitude parameter $a\in\R$ and associated (possibly trivial) pressure $p=ap_0$. The spectral stability of the steady state $a\u_0$ is determined by the linearised right-hand side in $a\u_0$ and thus the solutions to the generalised eigenvalue problem
\[
\lambda \U = \LL_a\U+ \nabla P\,,\quad \LL_a:= \Lb + a\Lb_0 \,,\quad \Lb_0:= \B(\cdot ,\u_0) + \B(\u_0,\cdot)\,,
\] 
with eigenvalue parameter $\lambda\in\C$, eigenmode $\U=(\v,\eta)$ and $P$ either trivial or determined by the linearised constraint $\nabla\cdot \v=0$.

Since the resulting spectrum is locally uniformly continuous with respect to the parameter $a$, we immediately note that for $|a|\ll 1$ it is close to that for $a=0$ associated to $\Lb$. Since its spectrum is unstable for the backscatter setting, as shown in the linear stability analysis of the zero state above, it follows that all the discussed explicit flows for small amplitudes inherit unstable modes of the trivial state, more so for smaller amplitudes.

Regarding large amplitudes, $|a|\gg 1$, we consider eigenvalue parameters that scale with the amplitude, i.e. $\lambda = a\widetilde{\lambda}$, and set $P=a\widetilde{P}$. This gives the (generalised) eigenvalue problem 
\begin{align}\label{e:geneval}
\widetilde{\lambda} \U =(a^{-1}\Lb + \Lb_0)\U+ \nabla \widetilde{P}\,.
\end{align}
The operator of the limiting problem, as $|a|\to \infty$, is $\Lb_0$, and again by continuity of the spectrum, its stability properties partially predict those of $\LL_a$ for $|a|\gg 1$. In particular, an unstable eigenmode of $\Lb_0$ implies strongly unstable eigenmodes of $\LL_a$ for $|a|\gg 1$, for which the growth rate $\mathrm{Re}(\lambda)$ is proportional to the amplitude $a$ of the steady solution. However, eigenmodes of $\LL_a$ for which $\lambda$ is not proportional to $a$ will move to the origin in the scaled operator as $|a|\to\infty$, and thus contribute to the kernel of $\Lb_0$. In particular, $a\u_0$ may be unstable for all $a$ even though $\Lb_0$ does not possess unstable spectrum. Indeed, this turns out to be the case in the present setting. This is consistent with the unstable rates $\lambda$ of the explicit flows from the analysis of unbounded instability above, which are associated with unbounded growth, as these are constant with respect to $a$ so that in the scaling of $\LL_a$ satisfy $\widetilde{\lambda} = \lambda/a \to 0$ as $a\to\infty$.

Hence, we consider the limiting problem  
\[
\widetilde{\lambda} \U =\Lb_0\U+ \nabla \widetilde{P}\,,
\]
whose spectral properties do not seem to be known analytically for the explicit flows we are concerned with. Specifically, for \eqref{sol: RSWB2} we have $\u = (\v,\eta)$, the bilinear form is
\begin{align*}
\B\Big((\v_1,\eta_1), (\v_2,\eta_2)\Big) = \begin{pmatrix}-(\v_1\cdot\nabla)\v_2 \\-(\v_1\cdot\nabla)\eta_2-\eta_1 \nabla\cdot\v_2\end{pmatrix}\,,
\end{align*}
and the steady state family is generated by $\u_0 = (\v_0,\eta_0)\transpose$ from \eqref{sol: RSWB2}, i.e. with $\xi = \k\cdot\x$, 
\begin{align*}
\v_0 = \cos(\xi)\k^\perp\,,\quad
\eta_0 = \alpha_2\sin(\xi)+s\,,
\end{align*}
where $\alpha_2$ is chosen so that \eqref{cond: RSWB2b} holds. We are then interested in the spectrum of the operator $\Lb_0$ defined by
\begin{align*}
\Lb_0 \begin{pmatrix}\v\\ \eta\end{pmatrix}
&= -\begin{pmatrix}
(\v_0\cdot\nabla)\v+(\v\cdot\nabla)\v_0 \hfill \\
(\v_0\cdot\nabla)\eta+(\v\cdot\nabla)\eta_0 + \eta_0 \nabla\cdot\v + \eta \nabla\cdot\v_0\end{pmatrix}\\
&=-\begin{pmatrix}
\cos(\xi)(\k^\perp\cdot\nabla)\v 
-(\v\cdot\k)\sin(\xi)\k^\perp \hfill \\
\cos(\xi)(\k^\perp\cdot\nabla)\eta
+\alpha_2(\v\cdot\k)\cos(\xi)
+(\alpha_2\sin(\xi)+s)(\nabla\cdot\v)
\end{pmatrix}\,.
\end{align*}
We immediately note that the kernel of $\Lb_0$ is infinite dimensional: any perturbation $\v,\,\eta$ of the same form as the steady flow $\u_0$,  i.e. $\v(\x) = \phi_1(\xi)\k^\perp$, $\eta(\x)=\phi_2(\xi)$ with arbitrary $\phi_1$ and $\phi_2$, lies in the kernel, since $(\k^\perp\cdot \nabla)(\v,\eta)=0$ as well as $\v\cdot \k=0$ and $\nabla\cdot\v=0$ (in fact, one can show that here $\v$ can be an arbitrary function of $\xi$). Next, we show that the spectrum of $\Lb_0$ is purely imaginary.

First let $\alpha_2=s=0$, i.e. the steady state $\u_0$ is in the intersection of the sets of steady and time-dependent solutions as in Figure \ref{Fig. 1}, so that 
\begin{align*}
\Lb_0 \begin{pmatrix}\v\\ \eta\end{pmatrix}
&=-\begin{pmatrix}
\cos(\xi)(\k^\perp\cdot\nabla)\v 
-(\v\cdot\k)\sin(\xi)\k^\perp  \\
\cos(\xi)(\k^\perp\cdot\nabla)\eta\hfill
\end{pmatrix}\,.
\end{align*}
It is a diagonal operator where $\v$ and $\eta$ are decoupled. Let us change coordinates to $\xi = k_1 x+ k_2 y$, $\zeta = -k_2 x + k_1 y$. Then $\nabla$ becomes $(k_1\partial_\xi- k_2\partial_\zeta, k_2\partial_\xi+k_1\partial_\zeta)$ so that $\k^\perp\cdot\nabla$ turns into $K\partial_\zeta$, where $K:=k_1^2+ k_2^2$. We thus obtain the operator
\begin{align*}
\Lb_0 \begin{pmatrix}\v\\ \eta\end{pmatrix}
&=-\begin{pmatrix}
\cos(\xi)K \partial_\zeta\v 
-(\v\cdot\k)\sin(\xi)\k^\perp  \\
\cos(\xi)K \partial_\zeta\eta\hfill
\end{pmatrix}\,,
\end{align*}
whose Fourier transform with respect to $\zeta$ with wave number parameter $\vartheta$ read
\begin{align*}
\widehat\Lb_0 \begin{pmatrix}\hat\v\\ \hat\eta\end{pmatrix}
&=-\begin{pmatrix}
i \cos(\xi)K  \vartheta \hat\v 
-(\hat\v\cdot\k)\sin(\xi)\k^\perp  \\
i \cos(\xi)K \vartheta\hat\eta\hfill
\end{pmatrix} \\
&= -\begin{pmatrix}
i \cos(\xi)K \vartheta  \Id
-\sin(\xi)A(\k) & 0 \\
0 &i \cos(\xi)K \vartheta
\end{pmatrix} \begin{pmatrix}\hat\v\\ \hat\eta\end{pmatrix}\,,
\end{align*}
where $A(\k) = \begin{pmatrix}-k_1 k_2 & -k_2^2\\ k_1^2 & k_1 k_2\end{pmatrix}$.

The lower right entry, which corresponds to $\eta$, is a multiplication operator by $i \cos(\xi)K\vartheta$ and so its spectrum is the range of this function, which is $i K\vartheta [-1,1] \subset i\R$. Since this multiplication operator appears in the upper left entry as multiplying the identity, which commutes with any matrix, $A(\k)$ can be brought to normal form. This features a double zero eigenvalue so that the operator on the upper left block possesses purely imaginary spectrum. In particular, the spectrum is neutrally stable. 

\begin{figure}
\begin{center}
\includegraphics[height=6cm]{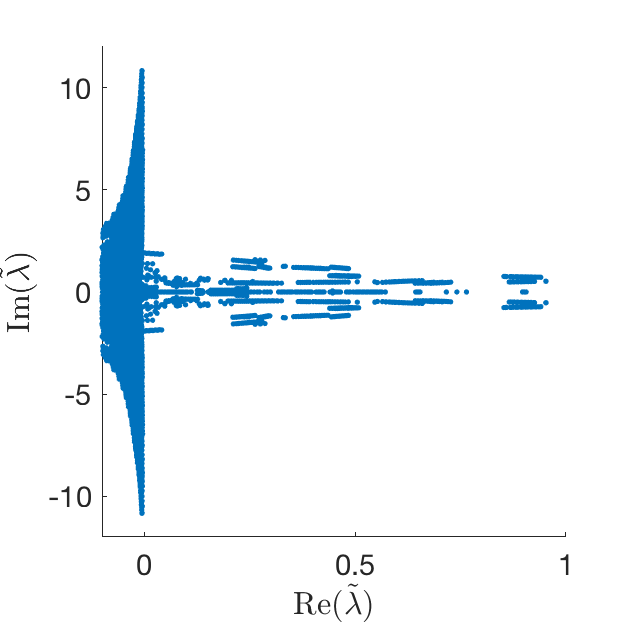} 
\caption{\label{fig:spec} Shown are the eigenvalues with real part larger than $-0.1$ of an approximation of $\calL_1$ with $N=10$ wave modes, i.e. $3(2N+1)^2$ Fourier modes on the periodic domain $[0,2\pi/k_1]\times [0,2\pi/k_2]$, and Bloch modes from the grid with distance $\pi/4$.
Parameters are as in Figure~\ref{Fig. 1c} and $s=0$. Amplitudes are $\alpha_1=1$ and $\alpha_2=0$, so $\sigma= f=0.3$ and the selected steady solution corresponds to the point between the red and black arcs in Figure~\ref{Fig. 7} with $\k\approx(-1.4,0.35)$. In particular, the solution is already unstable at marginal instability with respect to the explicit modes.}
\end{center}
\end{figure}

\medskip
For $\alpha_2,\, s\neq 0$ and writing $\v=(u,v)$ we analogously obtain the transformed operator
\begin{align*}
\widehat\Lb_0 \begin{pmatrix}\hat\v\\ \hat\eta\end{pmatrix}
&=-\begin{pmatrix}
i \cos(\xi)K  \vartheta \hat\v 
-(\hat\v\cdot\k)\sin(\xi)\k^\perp\hfill  \\
i \cos(\xi)K \vartheta\hat\eta + \calB\hat\v
\hfill
\end{pmatrix} = \begin{pmatrix}
\calA_1 & 0 \\
\calB & \calA_2
\end{pmatrix} \begin{pmatrix}\hat\v\\ \hat\eta\end{pmatrix}\,,
\end{align*}
with 
\begin{align*}
\calA_1 &:=-i \cos(\xi)K  \vartheta \Id  +\sin(\xi)A(\k)\\
\calA_2 &:= -i \cos(\xi)K \vartheta\\ 
\calB\hat\v &:=-\alpha_2(\hat\v\cdot\k)\cos(\xi) + (\alpha_2\sin(\xi)+s)(k_1\partial_\xi \hat u + k_2\partial_\xi \hat v + i\vartheta (k_1\hat v - k_2 \hat u))\,. 
\end{align*}
If $\lambda$ lies in the resolvent set of both operators on the diagonal $\calA_1$ and $\calA_2$ (by the above this includes any non-purely imaginary value), then $\lambda$ is also in the resolvent set of the present $\widehat\Lb_0$, since 
\begin{align*}
(\widehat\Lb_0-\lambda)^{-1} \begin{pmatrix}\hat\v\\ \hat\eta\end{pmatrix}
&= \begin{pmatrix}
(\calA_1-\lambda)^{-1} & 0 \\
 -(\calA_2-\lambda)^{-1} \calB(\calA_1-\lambda)^{-1}& (\calA_2-\lambda)^{-1}
\end{pmatrix} \begin{pmatrix}\hat\v\\ \hat\eta\end{pmatrix}\,.
\end{align*}
Hence, as claimed, the asymptotic operator possesses marginally stable spectrum and we cannot immediately infer in/stability information for large amplitudes. However, numerical computations based on truncated Fourier series suggest that the spectrum is in fact rather strongly unstable, cf.\ Figure~\ref{fig:spec}. 

\section{Rotating Boussinesq equations with backscatter}\label{Rotating Boussinesq}
We now turn to the study of various explicit solutions in the rotating Boussinesq equations augmented with backscatter \eqref{eq: introB}. To ease notation, we write these in the form 
\begin{subequations}\label{eq: B}
\begin{align}
\frac{\p\v}{\p t}+(\v\cdot\nabla)\v+f\3\times\v+\nabla p-\3 \b &= - \bigl(\diag(d_1,d_2,d_3) \Delta^2 + \diag(b_1,b_2,b_3) \Delta \bigr)\v\label{eq: Ba}\\
\nabla\cdot\v&= 0\label{eq: Bb}\\
\frac{\p \b}{\p t}+(\v\cdot\nabla)\b+N^2v_3 &= \mu\Delta \b\,,\label{eq: Bc}
\end{align}
\end{subequations}
where we focus on horizontal backscatter $d_1,\,d_2,\, b_1,\, b_2>0$ with usual viscosity vertically, $d_3=0,\, b_3=-\nu \leq0$, and stable stratification $N^2>0$. For comparison and illustration we also discuss briefly the usual horizontal viscous or inviscid cases $d_1=d_2=0$, $b_1,\,b_2\leq 0$, unstable stratification $N^2<0$, and artificial vertical backscatter $d_3,\,b_3>0$.
As in \S\ref{Shallow Water}, we are especially interested in the parameter relations and stability properties of steady solutions, in particular unbounded instability, as well as in unboundedly growing explicit solutions. We first investigate the horizontal flows, which are comparable with the explicit solutions of the shallow water equations in \S\ref{Shallow Water}, but are less restricted and have additional properties in this case here. Afterwards, we analyse other explicit solutions with vertical structure and coupled buoyancy.

\subsection{Horizontal flow and decoupled system}\label{Rotating Boussinesq decoupled}
In order to compare with the results of the rotating shallow water equations with backscatter, we consider here the barotropic case with horizontal velocity field. We therefore choose a velocity field $\v$ that is independent of the vertical coordinate $z$ and has $v_3\equiv 0$, as well as a horizontally independent buoyancy $\b=\b(t,z)$. This ansatz yields the reduced equations
\begin{subequations}\label{eq: rB}
\begin{align}
\frac{\p\v}{\p t}+(\v\cdot\nabla)\v+f\v^{\perp}+\nabla\tilde{p} &= - \mathrm{diag}\bigl(d_1\Delta+b_1,d_2\Delta+b_2\bigr)\Delta\v\label{eq: rBa}\\
\nabla\cdot\v&= 0\label{eq: rBb}\\
\frac{\p \b}{\p t} &= \mu\frac{\p^2}{\p z^2}\b\,,\label{eq: rBc}
\end{align}
\end{subequations}
with gradient and Laplacian for the horizontal directions $\x=(x,y)\transpose$,  $\v=\v(t,\x)\in\R^2$, $\b=\b(t,z)$ and $p=\tilde{p}+B(t,z)$, where $\frac{\p B(t,z)}{\p z}=\b(t,z)$. Here the buoyancy is decoupled from the velocity field and determined by the linear heat equation \eqref{eq: rBc}; on the idealized whole space this can be readily solved by Fourier transform. 

\medskip
Regarding the momentum equations, compared with the shallow water equations we may view the equation for the fluid depth \eqref{eq: RSWBb} to be replaced by the incompressibility condition \eqref{eq: rBb}. This is less restrictive and admits a larger set of explicit flow solutions as discussed in \citep{prugger2020explicit} for the setting without backscatter. In particular, the form \eqref{sol: RSWB2} for $\v$ satisfies \eqref{eq: rBb} and can readily be adjusted to solutions of \eqref{eq: rBa}. However, there are no additional a priori constraints for the pressure akin to condition \eqref{eq: RSWBb} or \eqref{cond: RSWB2c}, so that the resulting pressure can be exponentially decaying or growing along with the velocity field. Moreover, linear combinations of any of these solutions with the same wave vector direction but different wavelength (any wave vector on a whole ray like in Figure \ref{Fig. 1c}) also yield explicit solutions of \eqref{eq: rB}. 
 
The resulting set of explicit solutions of \eqref{eq: rBa} for which the nonlinear terms vanish can be identified by the following ansatz for wave shape $\psi$ and pressure profile $\phi$
\begin{align}\label{sol: B}
\v= \psi(t,\k\cdot\x)\k^{\perp}\,,\quad \tilde{p}=f\phi(t,\k\cdot\x)\,,
\end{align}
where $\k\in\R^2$ and without loss of generality $|\k|=1$ by the freedom in choosing $\psi$ and $\phi$. Substitution into \eqref{eq: rBa} gives the linear equations for $\psi$ and $\phi$, 
\begin{subequations}\label{cond: B}
\begin{align}
\frac{\p\psi}{\p t} &= -(d_1k_2^2+d_2k_1^2)\frac{\p^4\psi}{\p\xi^4}-(b_1k_2^2+b_2k_1^2)\frac{\p^2\psi}{\p\xi^2}\label{cond: Ba}\\
\frac{\p\phi}{\p\xi} &= \frac{k_1k_2}{f}\left((d_1-d_2)\frac{\p^4\psi}{\p\xi^4}+(b_1-b_2)\frac{\p^2\psi}{\p\xi^2}\right)+\psi\,.\label{cond: Bb}
\end{align}
\end{subequations}

For the Boussinesq equations with viscosity instead of the backscatter terms similar equations arise, cf.\ \citep{prugger2020explicit}. However, in that case the pressure gradient fully compensates the buoyancy and the Coriolis term in the equations, which makes the velocity field geostrophically balanced. In contrast, in the present case, equation \eqref{cond: Bb} for the pressure shape $\phi$ allows the pressure gradient to not only compensate the full Coriolis term, but also part of the backscatter terms. In particular, the velocity field \eqref{sol: B} with \eqref{cond: B} is in general not geostrophically balanced.\newline

\subsubsection{Superposition principles}\label{s:BE-superpositions}
The general wave shape $\psi$ in \eqref{sol: B} also contains the superpositions of arbitrary many sinusoidal waves in the \textit{same wave vector direction} $\k$ and \textit{any wave number} $|\k|$. This is possible in the Boussinesq equations, since  $\nabla p$ in the momentum equation \eqref{eq: Ba} is not further constrained, unlike \eqref{eq: RSWBb} for $\nabla \eta$. It is also possible to superpose by integrating over the wave numbers in the same wave vector direction.\newline

The structure of the Boussinesq equations admits superposing $\v$ in the form \eqref{sol: RSWB2} in another way, namely with \textit{different wave vector directions}, but the \textit{same wave number}, cf.\ \citep{prugger2020explicit}. For the decoupled momentum equation \eqref{eq: rBa}, and finite superposition of $N$ waves with arbitary $N\in\N$, the resulting superposed sinusoidal explicit solutions of a form similar to \eqref{sol: B} are given by
\begin{subequations}\label{sol: B-sup}
\begin{align}
\v(t,\x)&=\sum_{i=1}^{N}e^{\lambda_it}\psi_i \k^{\perp}_i \qquad\mbox{with}\quad \psi_i=\alpha_i\sin(\k_i\cdot\x+ \tau_i)\,,\quad 1\leq i\leq N\,,\label{sol: B-supa}\\
\tilde{p}(t,\x)&=-\sum_{i=1}^N\sum_{j=i+1}^Ne^{(\lambda_i+\lambda_j)t}\Bigl((\k_i\cdot\k_j)\psi_i\psi_j+\sc^2\frac{\p\psi_i}{\p\xi}\frac{\p\psi_j}{\p\xi}\Bigr) - f\sum_{i=1}^{N}\frac{\gamma_i}{\alpha_i}e^{\lambda_it}\frac{\p\psi_i}{\p\xi}\,,\label{sol: B-supb}
\end{align}
\end{subequations}
for any fixed $\sc>0$ and arbitrary $\alpha_i\in\R\setminus\{0\}$, $\tau_i\in\R$ and $\k_i=(k_{i,1},k_{i,2})\transpose\in\R^2$ with $|\k_i|=\sc$ for any $1\leq i\leq N$. Here, each $\lambda_i$ and $\gamma_i$ is defined by
\begin{subequations}\label{cond: B-sup}
\begin{align}
\lambda_i &= (b_1-d_1\sc^2)k_{i,2}^2+(b_2-d_2\sc^2)k_{i,1}^2\,,\label{cond: B-supa}\\
f\frac{\gamma_i-\alpha_i}{\alpha_i} &= \bigl((d_1-d_2)\sc^2+b_2-b_1\bigr)k_{i,1}k_{i,2}\,,\label{cond: B-supb}
\end{align}
\end{subequations}
in order to solve \eqref{eq: rBa}. Since each wave in \eqref{sol: B-supa} is divergence free, the whole superposed velocity $\v$ solves \eqref{eq: rBb} and thus is an explicit solution of \eqref{eq: rB}.\newline
It is also possible to superpose explicit solutions of the form \eqref{sol: B-sup} by integrating over the whole circle $S_{\sc}:=\{\k\in\R^2\mid |\k|=\sc\}$ for any fixed $\sc>0$. The exact form is then
\begin{subequations}\label{sol: B-intsup}
\begin{align}
\v(t,\x)&=\int\displaylimits_{S_{\sc}}\alpha(\k)e^{\lambda(\k)t}\sin\bigl(\k\cdot\x+ \tau(\k)\bigr)\k^{\perp} d\k \label{sol: B-intsupa}\\
\begin{split}
\tilde{p}(t,\x)&=-\sc^4\int_0^{2\pi}\int_{\varphi_1}^{2\pi}\Bigl(\cos\bigl(\xi_1\bigr)\cos\bigl(\xi_2\bigr)+\cos(\varphi_1-\varphi_2)\sin\bigl(\xi_1\bigr)\sin\bigl(\xi_2\bigr)\Bigr)\\
&\hspace{4.5mm}\cdot\alpha_1\alpha_2e^{(\lambda_1+\lambda_2)t}d\varphi_2d\varphi_1
-f\int\displaylimits_{S_{\sc}}\gamma(\k)e^{\lambda(\k)t}\cos\bigl(\k\cdot\x+ \tau(\k)\bigr)d\k  \label{sol: B-intsupb}\,,
\end{split}
\end{align}
\end{subequations}
where, for $i=1,\,2$, we set $\alpha_i:=\alpha(\k_i)$ with $\k_i:=\sc\bigl(\cos(\varphi_i),\sin(\varphi_i)\bigr)\transpose$, $\lambda_i:=\lambda(\k_i)$ and $\xi_i:= \k_i\cdot\x+ \tau(\k_i)$, for all $0\leq\varphi_i<2\pi$. Sufficient for the convergence of the integrals is $\alpha\in L^1(S_{\sc})$, $\tau\in L^{\infty}(S_{\sc})$ and for almost all $\k\in S_{\sc}$ we require 
\begin{subequations}\label{cond: B-intsup}
\begin{align}
\alpha(\k)\lambda(\k)&=\alpha(\k)\bigl((b_1-d_1\sc^2)k_2^2+(b_2-d_2\sc^2)k_1^2\bigr)\,,\label{cond: B-intsupa}\\
f\bigl(\gamma(\k)-\alpha(\k)\bigr)&=\alpha(\k)k_1k_2\bigl((d_1-d_2)\sc^2+b_2-b_1\bigr)\label{cond: B-intsupb}\,,
\end{align}
\end{subequations}
corresponding to \eqref{cond: B-sup} if $\alpha(\k)\neq0$. 

\medskip
The explicit solutions \eqref{sol: B-sup} with \eqref{cond: B-sup} differ from \eqref{sol: B} with \eqref{cond: B}, as well as the other explicit solutions before, not only by the structure of the superposition, but also by the resulting nonlinear terms, which are not vanishing. Due to the special structure of $\v$ in \eqref{sol: B-sup} and $|\k_i|=\sc$ for all $1\leq i\leq N$, the nonlinear terms form a gradient that can be fully compensated by the pressure gradient in the momentum equation; this gives the first sum in \eqref{sol: B-supb}. The same holds for the solutions \eqref{sol: B-intsup} with \eqref{cond: B-intsup} correspondingly. We refer to \citep{prugger2020explicit} for further discussion and literature references for explicit solutions with gradient nonlinearities without backscatter.\newline

Comparing the explicit solutions \eqref{sol: B-sup} with \eqref{cond: B-sup}, as well as \eqref{sol: B-intsup} with \eqref{cond: B-intsup}, with those without backscatter we notice two major differences: First, in the present case the amplitudes of the pressure can be different from those of the velocity; the conditions on the amplitudes are given in \eqref{cond: B-supb} and \eqref{cond: B-intsupb} respectively. The reason is that the pressure gradient in the momentum equation can additionally compensate a part of the backscatter terms as well. 
Second, the growth rates $\lambda_i$ and $\lambda(\k)$ can be different, so that each wave is decaying or growing differently. Both differences require anisotropy in the backscatter of the momentum equation, i.e. $d_1\neq d_2$ or $b_1\neq b_2$. Indeed, the usual viscosity is isotropic in this sense.\newline

In conclusion, the above constructions of explicit solutions can be viewed as superposition principles for the (nonlinear) Boussinesq equations in this setting: \eqref{sol: B} expresses a \textit{radial superposition principle} of flows in the same wave vector direction, and \eqref{sol: B-sup} an \textit{angular superposition principle} of flows on the same scale. Superpositions of plane waves with different wave vector directions and scales are, in general, not giving solutions.\newline

\subsubsection{Unbounded instability of steady states}\label{s:BE-unbounded_instability_steady}
Analogous to \S\ref{Stability: unbounded growth}, the possible superpositions of explicit solutions, as contained in \eqref{sol: B}, imply linear subspaces with linear dynamics, in particular unbounded exponential growth of perturbations. Compared with the rotating shallow water equations, restrictions on wave vectors are absent, and in this section we discuss implications for (in)stability of steady solutions, i.e. those with zero growth rate $\lambda_i$ in \eqref{cond: B-supa}.\newline

Since $b_1,\, b_2,\, d_1,\, d_2>0$ in \eqref{cond: B-supa} we have $\lambda_i>0$ for any sufficiently small wave numbers $\sc$ and $\lambda_i<0$ for any sufficiently large ones. In particular, the trivial flow with $\v\equiv0$  is unstable with exponential unbounded growth with respect to any small $\sc$. More importantly, the above radial superposition principle immediately implies that the same holds for every \textit{single-wave} steady solution \eqref{sol: B}: Such steady explicit solutions also arise from \eqref{sol: B-sup} with $N=1$ and $\lambda_1=0$ in \eqref{cond: B-supa}. Since the definition of $\lambda_i$ in \eqref{cond: B-supa} is exactly the same as in \eqref{cond: sigmaa}, we can use here the results from \S\ref{s:SWE-steady_solutions} about the growth rate $\lambda$ as well. Hence, for any $b_1,\, b_2,\, d_1,\, d_2>0$ the set of solutions with $\lambda_1=0$ forms a simple closed curve around the origin in the wave vector space that is symmetric with respect to axis reflections and whose interior is star shaped. Furthermore, $\lambda$ is positive in the interior of this closed curve, except $\lambda=0$ at the origin $\k=(0,0)\transpose$, and $\lambda$ is negative outside the closed curve. Thus, \textit{single-wave} steady solutions \eqref{sol: B} exist in any direction and can be radially superposed with explicit solutions with any smaller wave numbers $\sc$, which makes them unboundedly unstable.

\begin{figure}
\begin{center}
\subfigure[Case $\sc=\sc_\mathrm{max}=\sqrt{b_2/d_2}$ for steady solutions with largest possible wave number.]{
\includegraphics[trim=5.1cm 9.3cm 5.6cm 9.7cm, clip, width=0.35\linewidth]{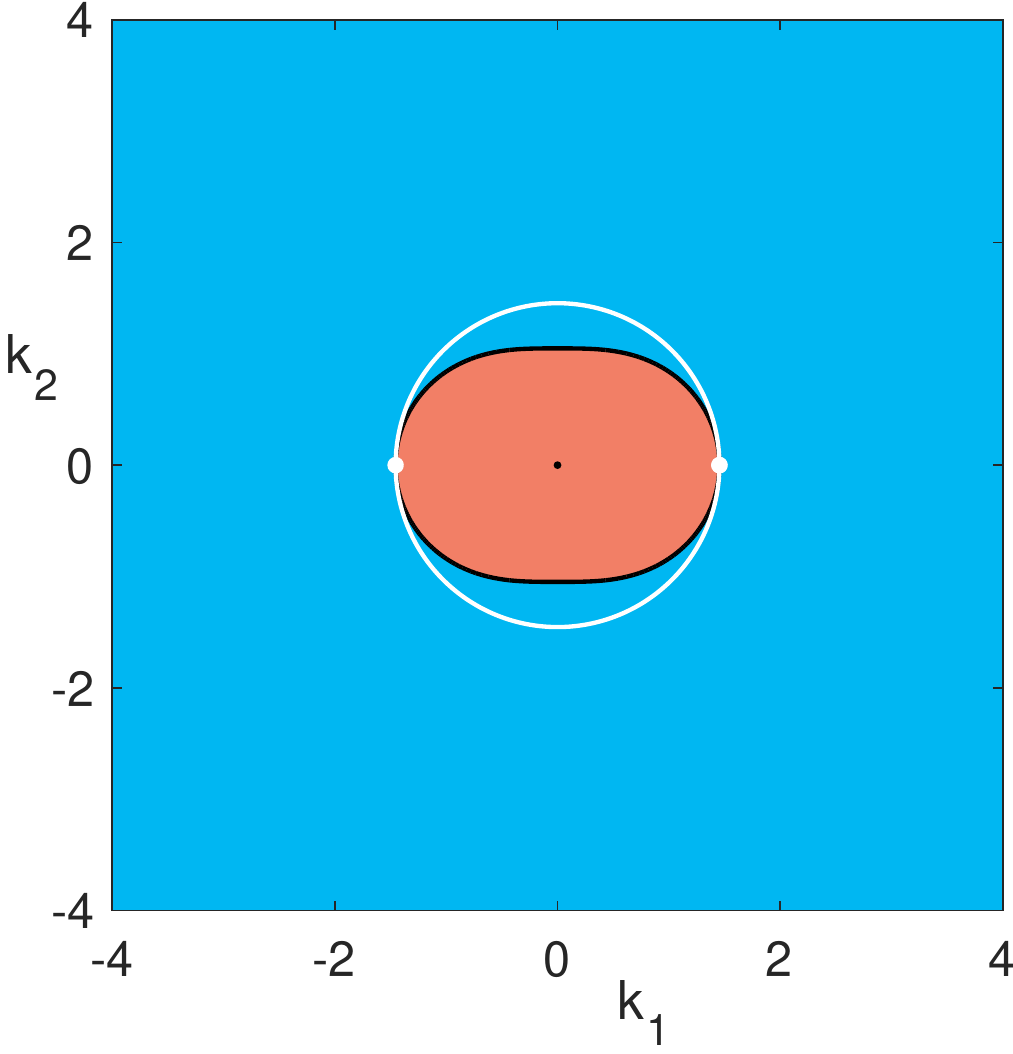}\label{Fig. RadSuperpositiona}}
\hspace{1cm}
\subfigure[Case $\sc\approx1.20<\sc_\mathrm{max}$ for steady solutions with different wave vector directions.]{
\includegraphics[trim=5.1cm 9.3cm 5.6cm 9.7cm, clip, width=0.35\linewidth]{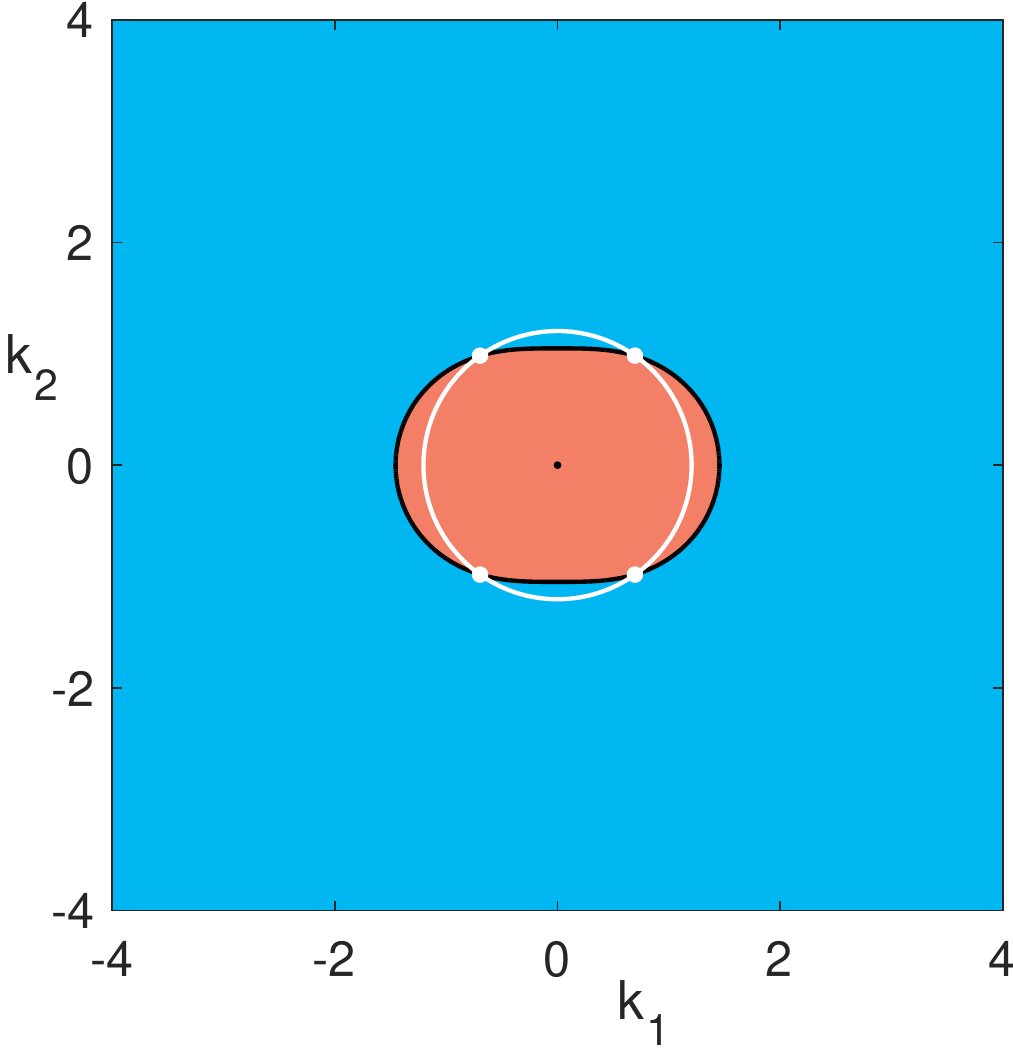}\label{Fig. RadSuperpositionb}}
\caption{\label{Fig. RadSuperposition} Signs of $\lambda$ as defined in \eqref{cond: B-supa} (red: $\lambda>0$, blue: $\lambda<0$, black: $\lambda=0$) and possible wave vectors for \eqref{sol: B-sup} for a fixed wave number $\sc>0$ (white circles). White dots mark wave vectors of corresponding steady solutions. Fixed parameters are $d_1=1.0,\, d_2=1.04,\, b_1=1.1,\, b_2=2.2$, i.e. $b_1/d_1\neq b_2/d_2$.}
\end{center}
\end{figure}

However, in the case $b_1/d_1\neq b_2/d_2$, for fixed $\sc$ there are up to four wave vectors for which $\lambda_i=0$; to see this note that for $\lambda_i=0$ \eqref{cond: B-supa} is linear in $\cos(\theta)^2$ for $\k_i=\sc (\cos (\theta), \sqrt{1-\cos^2 (\theta)})\transpose$ in the first quadrant. Thus, due to the reflection symmetry, there is at most one solution in each quadrant of the wave vector plane. Because of the symmetry, the steady states of the form \eqref{sol: B-sup} can consist of (at most) two different wave vector directions and for those we cannot infer instability by the radial superposition principle. In that case, if $b_1/d_1\neq b_2/d_2$ we next appeal to the angular superposition principle. 
First, we note that for $b_1/d_1\neq b_2/d_2$ there is a unique (up to reversing orientation) longest wave vector $\k_\mathrm{max}$ with length $\sc_\mathrm{max}$, so that \eqref{cond: B-supa} is satisfied with $\lambda_i=0$. See Figure \ref{Fig. RadSuperpositiona} for a typical example. Indeed, due to the aformentioned structures, $k_\mathrm{max}$ is along an axis, though the set of $\k$ with $\lambda=0$ need not be convex. Hence, any steady superposed solution \eqref{sol: B-sup} with $\sc=\sc_\mathrm{max}$ is built from $\pm\k_\mathrm{max}$, which lie on the same line in wave vector space. Thus, we can use the radial superposition principle with the exponentially growing explicit solutions of smaller wave numbers, which leads to unbounded growth with respect to modes on any larger scale. The same applies for steady solutions with minimal wave vector length $\sc_\mathrm{min}$. \newline
Second, we consider steady superposed solutions \eqref{sol: B-sup} with $\sc_\mathrm{min}<\sc<\sc_\mathrm{max}$ that can be built from  two different directions, cf.\ the white dots in Figure \ref{Fig. RadSuperpositionb}. Here, we apply the angular superposition principle and superpose with any explicit solution \eqref{sol: B-sup} on the same scale, i.e. whose wave vector has the same length $\sc$, cf.\ the white circle in Figure \ref{Fig. RadSuperpositionb}. Since for some wave vectors the corresponding $\lambda$ defined by \eqref{cond: B-supa} is positive for the length $\sc<\sc_\mathrm{max}$, at least for the wave vector $\sc/\sc_\mathrm{max}\k_\mathrm{max}$, we again have unbounded instability with respect to a range of modes, here on the same scale. We note that for $\sc=\sc_\mathrm{min}$ radial and angular superposition both give ranges of modes with unbounded growth. \newline

In case $b_1/d_1= b_2/d_2$ we cannot infer the instability of steady superposed solutions \eqref{sol: B-sup} using the above superposition principles, since the wave vectors of non-trivial steady solutions $\{\k\in\R^2\mid \lambda(\k)=0\}$, with $\lambda(\k)$ from \eqref{cond: B-supa}, form a circle with radius $\sqrt{b_1/d_1}$, which means they all have the same length. Thus, explicit solutions \eqref{sol: B-sup} with $\sc=\sqrt{b_1/d_1}$ consist only of steady solutions; there are also no steady solutions of the form \eqref{sol: B} for other wave numbers. However, in some cases unbounded instability still follows from angular superposition with unboundedly growing parallel flows, cf. \S\ref{s:parflow}.\newline

\subsubsection{Linear stability of steady states with small and large amplitudes}\label{s:BE-lin_stability_steady}
After the investigation of unbounded instability of steady states, we now turn to the linear stability of steady solutions with small and large amplitudes, analogous to \S\ref{s:stabamp}. Concerning small amplitudes, we are naturally led to linear and spectral stability of the zero state, as for the rotating shallow water equations. Instead of analysing the full spectrum of the linearisation of \eqref{eq: rB} in the trivial steady flow $\v\equiv0$, here we restrict attention to instability with respect to eigenmodes of the form of the horizontal flows \eqref{sol: B}, which means solutions to \eqref{cond: Ba}. The Fourier transform of \eqref{cond: Ba} yields the dispersion relation for perturbation wave vector $\k\in\R^2$ and temporal rate $\lambda\in\C$, 
\[
d_\psi(\lambda,\k) = -(d_1k_2^2+d_2k_1^2)|\k|^2+ b_1k_2^2+b_2k_1^2 - \lambda = 0\,,
\]
which is of course equivalent to \eqref{cond: B-supa} with $\lambda =\lambda_i$ and $\k=\k_i$. The above discussion for steady states built from single direction wave vectors $\pm\k$ implies that the spectrum of the linearisation in such horizontal flows is at least as unstable as that of the zero state in this wave vector direction, since the spectra contain the growth rates of the corresponding explicit solutions in this direction. Of course the result is much stronger in that these modes actually grow unboundedly in the nonlinear Boussinesq equations. In contrast, for the linearisation of \eqref{eq: rB} in steady multi-mode horizontal flow, similar comparison of its spectrum with that of the zero state holds, but with different wave vector directions and the same wave number $|\k|=\sc$.

However, analogous to \S\ref{s:stabamp}, any (superposed) steady horizontal flow inherit the instability of any unstable mode in the dispersion relation $d_\psi(\lambda,\k)=0$ of the zero state for sufficiently small amplitudes $0<|\alpha_i| \ll 1$, $1\leq i\leq N$, though the growth induced by such modes may be bounded. Recall that the explicit solutions of \eqref{eq: rB} also satisfy the full rotating Boussinesq equations \eqref{eq: B}, which admit modes that have vertical structures and are coupled with buoyancy. In \S\ref{Rotating Boussinesq coupled} we will discuss such explicit solutions of \eqref{eq: B}, which also satisfy these equations without the nonlinear terms. In particular, unboundedly growing flows of this type provide additional explicit unstable modes of the zero state $\v\equiv0$, which -- in contrast to the horizontal flows -- are also influenced by the Brunt-V\"ais\"al\"a frequency $N^2$, thermal diffusivity $\mu$ and the vertical viscosity. In addition, these imply linear instability modes of horizontal flows with sufficiently small amplitudes $0<|\alpha_i| \ll 1$, $1\leq i\leq N$.

\bigskip
As to large amplitude(s), where $1\ll |\alpha_i| $ for at least one $i$, we first note that, in the notation of \S\ref{s:stabamp} and with $\u=(\v,\b)$,  the bilinear form for the present case reads
\begin{align*}
\B\Big((\v_1,\b_1), (\v_2,\b_2)\Big) = -\begin{pmatrix}(\v_1\cdot\nabla)\v_2 \\(\v_1\cdot\nabla) \b_2\end{pmatrix}\,.
\end{align*}
The steady state family $\u=a\u_0$ in this case has $\u_0 = (\v_0,\b_0)\transpose$ and
\begin{align*}
\v_0 = \sum_{i=1}^N\alpha_i\sin(\k_i\cdot x + \tau_i)\k_i^\perp\,,
\end{align*}
where we consider in this section only horizontal (wave) vectors $\k:=(\hat{\k},0)\transpose$, $\k^\perp:=(\hat{\k}^\perp,0)\transpose$ for any $\hat{\k}\in\R^2$. Here the third component of $\v_0$ vanishes, and $\b_0$ is an arbitrary constant solving \eqref{eq: rBc}. 
In order to locate strongly unstable modes, whose growth rates $\lambda=a\widetilde{\lambda}$ are proportional to the amplitude parameter $a$, we are concerned with the generalised eigenvalue problem \eqref{e:geneval}, as $|a|\to \infty$, which here reads
\begin{align}\label{e:geneval2}
\widetilde{\lambda} \begin{pmatrix}\v\\ \b\end{pmatrix} = \Lb_0 \begin{pmatrix}\v\\ \b\end{pmatrix} + \begin{pmatrix}\nabla \bar{p}\\0\end{pmatrix}\,, \quad \nabla\cdot \v=0\,,
\end{align}
with $p=a\bar{p}$ and for the operator $\Lb_0$ defined by
\begin{align*}
\Lb_0 \begin{pmatrix}\v\\ \b\end{pmatrix}
&= -\begin{pmatrix}
(\v_0\cdot\nabla)\v+(\v\cdot\nabla)\v_0 \hfill \\
(\v_0\cdot\nabla)\b\end{pmatrix}\,.
\end{align*}
In order to simplify and illustrate the main finding, we investigate the stability of a certain superposed steady solution and reduce to only two modes, i.e. $\alpha_i=0$ for $i>2$, translate so that $\tau_1=\tau_2=0$, and set $\xi = \k\cdot\x$, $\zeta = \k^\perp\cdot\x$ so that with a certain wave vector $\k$
\begin{align*}
\v_0 = \alpha_1\sin(\xi)\k^\perp + \alpha_2\sin(\zeta)\k\,.
\end{align*}
We note, that even in the anisotropic case there is at least one wave vector $\k$, such that both $\k$ and $\k^\perp$ correspond to a steady mode. We omit the full proof, and instead explain the existence of such $\k$ based on Figure \ref{Fig. RadSuperposition}. We start with superposed steady solutions with wave number $s=s_{\max}$ as in Figure \ref{Fig. RadSuperpositiona}. Reducing $s$ towards the wave number $s_{\min}$ as in Figure \ref{Fig. RadSuperpositionb}, there is an intermediate value of $s$ such that a wave vector $\k$ with $|\k|=s$ exists, for which $\k$ and $\k^\perp$ each correspond to a single mode steady solution. This is ensured by the symmetry of the curve defined by \eqref{cond: B-sup} with $\lambda_i=0$. We then obtain 
\begin{align*}
\Lb_0 \begin{pmatrix}\v\\ \b\end{pmatrix}
&=-\begin{pmatrix}
\alpha_1\sin(\xi)(\k^\perp\cdot\nabla)\v + \alpha_2\sin(\zeta)(\k\cdot\nabla)\v + \alpha_1\cos(\xi)(\v\cdot\k)\k^\perp + \alpha_2\cos(\zeta)(\v\cdot\k^\perp)\k \hfill \\
\alpha_1\sin(\xi)(\k^\perp\cdot\nabla)\b + \alpha_2\sin(\zeta)(\k\cdot\nabla)\b
\end{pmatrix}\\
&=-\mathrm{diag}(L_{11}, L_{22},L_{22})\begin{pmatrix}\v\\ \b\end{pmatrix}\,,
\end{align*}
with block diagonal matrix operator in which $L_{11}$ is the 2-by-2 matrix operator 
\begin{align*}
L_{11} = L_{22}\Id + \alpha_1\cos(\xi)A(\k) - \alpha_2\cos(\zeta)A\transpose(\k)\,,
\end{align*} 
with $A(\k)$ as in \S\ref{s:stabamp} (for which $A\transpose(\k)=-A(\k^\perp)$) and
\begin{align*}
L_{22} &= \alpha_1\sin(\xi)(\k^\perp\cdot\nabla) + \alpha_2\sin(\zeta)(\k\cdot\nabla)\,.
\end{align*}
Taking the divergence of \eqref{e:geneval2} gives, using $\nabla\cdot \v=0$, the linear pressure Poisson equation $\Delta \bar{p} = \nabla\cdot (L_1 \v)$, with block matrix operator $L_1:=\mathrm{diag}(L_{11},L_{22})$. We denote the solution as $\bar{p}=\Delta^{-1}\nabla\cdot (L_1 \v)$ and substitution into \eqref{e:geneval2} yields the eigenvalue problem
\begin{align}\label{e:eval2}
\widetilde{\lambda} \begin{pmatrix}\v\\ \b\end{pmatrix} = \left(\Lb_0 +  \begin{pmatrix}\nabla \Delta^{-1}\nabla\cdot (L_1 \circ) & 0\\0& 0\end{pmatrix}\right)\begin{pmatrix}\v\\ \b\end{pmatrix}\,,
\end{align}
in which $\circ$ denotes the slot for $\v$. The resulting operator on the right-hand side features a diagonal block structure such that the spectrum is the union of the spectra of $-L_{22}$ and $-\widetilde{L}_1$ defined by
\[
\widetilde{L}_1:= L_1 - \nabla \Delta^{-1}\nabla\cdot (L_1 \circ)\,.
\] 
Since $L_{22}$ is skew-adjoint ($\rmi L_{22}$ is self-adjoint on suitable spaces), its spectrum is purely imaginary. In case the steady state is a single mode flow, i.e. $\alpha_i=0$ for $i> 1$, we readily infer as in \S\ref{s:stabamp} that the spectrum of $-\widetilde{L}_1$ is also purely imaginary, so that the spectrum of $\Lb_0$ is purely imaginary. 

\begin{figure}[htbp]
\begin{center}
\includegraphics[height=6cm]{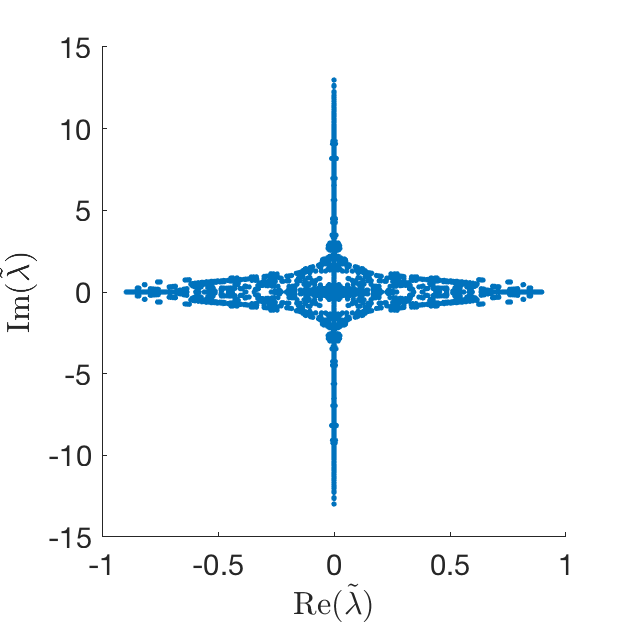}
\caption{\label{fig:spec2} Shown is an approximation of part of the spectrum of $\widetilde{L}_1$. Using \eqref{e:evalFour} we reduced to two-dimensional wave vectors by fixing the third component of $\k_m$ at zero. Here $\k=(1,1)$, $\alpha_1=0.1,\, \alpha_2=1$. As in Figure~\ref{fig:spec}  we use $N=10$ wave modes, i.e. $3(2N+1)^2$ Fourier modes on the periodic domain $[0,2\pi/k_1]\times [0,2\pi/k_2]$, and Bloch modes in the first component from the grid with distance $\pi/8$. In particular, the spectrum is unstable, so that large amplitude solutions are linear unstable with growth rates proportional to the amplitude.}
\end{center}
\end{figure}

Otherwise, if $\alpha_1\alpha_2\neq 0$, it appears difficult to determine the spectrum of $\widetilde{L}_1$ analytically and we resort to numerical computations. For this let $(\cdot)_m$ denote the projection onto the mode $\exp(\rmi \k_m\cdot \x)$. Then 
\begin{align}\label{e:evalFour}
(\widetilde{L}_1 \v)_m = (L_1 \v)_m - \frac{\k_m\cdot (L_1\v)_m}{|\k_m|^2} \k_m = (\Id - B(\k_m))(L_1\v)_m\,,
\end{align}
with suitable matrix $B(\k_m)$. This admits straightforward numerical computation of spectra on truncated Fourier series. We plot results for an example in Figure~\ref{fig:spec2}, which gives unstable spectrum and thus strong evidence for unstable spectrum of $\Lb_0$. 

Notably, this means that for steady states that are mixed mode flows \eqref{sol: B-sup} it is possible that linear growth rates are proportional to the amplitude parameter $a$. In contrast, such modes do not exist for steady single mode flows since in this case the spectrum of $\Lb_0$ is purely imaginary, as in \S\ref{s:stabamp}. Again we remark that we expect the growth induced by these modes in the nonlinear system is bounded.

\subsection{Flows with vertical structure and coupled buoyancy}\label{Rotating Boussinesq coupled}
The rotating Boussinesq equations with backscatter \eqref{eq: B} also admit explicit solutions of different form in which the velocity and the buoyancy are coupled, and in which the vertical dependence and velocity component is non-trivial. Here we investigate parallel flows, Kolmogorov flows and monochromatic inertia gravity waves. As before, we are particularly interested in the occurrence of unboundedly growing explicit solutions as well as the existence of such steady solutions and their stability properties.\newline

\subsubsection{Parallel flow}\label{s:parflow}
This class of explicit flows is well-known in the inviscid and viscous case, e.g. \citep{Wang90}. It possesses only a vertical velocity component and is thus different from the horizontal flows, and admits more general dependence on the horizontal space variables. Specifically, 
\begin{align}\label{sol: Parallel}
\v(t,\x)=w(t,x,y)\3 \,,\quad \b(t,\x)=\tb(t,x,y) \,,\quad p(t,\x)=\tilde{p}(t)z \,,
\end{align}
where $w$ and $\tb$ satisfy (with horizontal Laplacian and bi-Laplacian)
\begin{subequations}\label{cond: Parallel}
\begin{align}
\frac{\p w}{\p t} + \left(d_3\Delta^2+b_3\Delta\right)w+\tilde{p}&=\tb\label{cond: Parallela}\\
\frac{\p\tb}{\p t} - \mu\Delta\tb&=-N^2w\,.\label{cond: Parallelb}
\end{align}
\end{subequations}
Recall that kinetic energy backscatter, which has $d_3=0$ and $b_3\leq0$, has no vertical impact so that parallel flows are in fact independent of backscatter. Plane wave parallel flows can be superposed with the horizontal flows \eqref{sol: B} that have zero buoyancy, if their wave vector directions $\k$ are the same. In this case the orthogonality conditions for wave vectors and wave directions are satisfied and the nonlinear terms vanish, so that the superposition of both solutions is also an explicit solution due to the remaining linear system. Thus, a priori, any parallel flow of this form is unboundedly unstable concerning perturbations \eqref{sol: B} with the same wave vector $\k$ and small enough wave number $|\k|$. 
 
Existence and dynamics of parallel flows can be inferred from the dispersion relation of the linear equations \eqref{cond: Parallel}. By Fourier transformation with wave vector $\k$ and growth rate $\lambda$ this is given by
\begin{align*}
\det(\lambda\Id - \widehat \Lb) = 0\,,\quad \widehat \Lb:= \begin{pmatrix} -d_3|\k|^4 + b_3|\k|^2 & 1 \\ -N^2 & -\mu|\k|^2 \end{pmatrix}\,,
\end{align*}
or equivalently as the characteristic polynomial
\begin{align}\label{eq: parallelpoly}
\lambda^2 + c_1 \lambda + c_0= 0\,,
\end{align}
where $c_1 := d_3K^2 + (\mu - b_3)K$ and $c_0 := \mu(d_3K - b_3)K^2 + N^2$ with $K:=|\k|^2\geq0$. 
Steady solutions to \eqref{cond: Parallel} require constant $\tilde p$ and consist of Fourier modes with $\k\neq 0$ that solve \eqref{eq: parallelpoly} with $\lambda=0$, i.e. $c_0=0$. Growing spatially non-constant solutions to \eqref{cond: Parallel} exist if and only if \eqref{eq: parallelpoly} possesses a root with positive real part and $\k\neq 0$, and then do so exponentially and unboundedly.  Note that both roots have negative real parts only for $c_1,\, c_0>0$ and complex conjugate solutions can be superposed to form a real parallel flow solution. 

With vertical viscosity, $d_3=0$ and $b_3<0$, and focusing on non-constant solutions $\k\neq0$, we have $c_1>0$ and $c_0=-\mu b_3 K^2 + N^2$. So steady states require $N^2<0$ and then $K^2=N^2/(\mu b_3)$, or $\mu=N^2=0$ and any $K$. For $\mu>0$ also the unstable case $c_0<0$ requires unstable stratification $N^2<0$, and then $c_0<0$ occurs on a disc of wave vectors. 

Regarding small amplitudes, analogous to \S\ref{s:BE-lin_stability_steady}, any steady (or decaying) parallel flow with small amplitude is unstable, though typically not unboundedly, with respect to unstable modes of the trivial steady state that are exhausted for decreasing amplitude. 
In the large amplitude scaling, the resulting operator $\Lb_0$ for steady parallel flow $w_0$ is a lower triangular matrix operator with diagonal entries $L_1:=w_0(x,y)\partial_z$. Hence, as for the spectrum of $\Lb_0$ in \S\ref{s:BE-lin_stability_steady}, the spectrum is given by the diagonal entries. For any (smooth) $w_0$ the operator $L_1$ is skew self-adjoint, similar to $L_{22}$ in \S\ref{s:BE-lin_stability_steady}, and thus the spectrum of $\Lb_0$ is purely imaginary. Hence, no real parts of the spectrum of the steady parallel flow are proportional to its amplitude $a$.

\medskip
Finally, in order to illustrate the abstract structure and in preparation of the flows discussed below, next we briefly consider the artificial case $d_3,\, b_3>0$. \newline
Without thermal diffusion ($\mu=0$), we have $c_0=N^2$ and $c_1=\delta_3(K):=(d_3K-b_3)K$. For stable stratification, $N^2>0$, growing Fourier modes occur if and only if $c_1<0$ and $c_1^2\geq 4c_0$, which is equivalent to $K<b_3/d_3$ and $N^2\leq\delta_3^2(K)/4$, respectively. For $K\in(0,b_3/d_3)$ the global maximum of $\delta_3^2(K)/4$ is $b_3^4/(64d_3^2)$ at $K=K_1:=b_3/(2d_3)$;  its global minimum is zero at $K=0$ and $K=b_3/d_3$. Specifically, if $N^2\leq b_3^4/(64d_3^2)$, i.e. the stability of the stratification is sufficiently weak compared with the backscatter destabilisation, then $c_1^2\geq 4c_0$ for $K$ in a positive interval $I_1\subset(0,b_3/d_3)$; in particular, $I_1=\{K_1\}$ if $N^2= b_3^4/(64d_3^2)$. Hence, a parallel flow \eqref{sol: Parallel} grows exponentially and unboundedly if it contains a Fourier mode with wave vector $\k$ in the annulus $\{\k\in\R^2\,|\,|\k|^2\in I_1\}$.\newline
In the presence of thermal diffusion ($\mu>0$), we first note that $c_0$ is a cubic polynomial in $K$, so $c_0$ has a local maximum at $K=0$ and there is a global minimum at some $K>0$. Specifically, if $0<N^2<4\mu b_3^3/(27d_3^2)$, then $c_0<0$ in a positive interval $I_2\subset(0,b_3/d_3)$; if $N^2<0$, then $c_0<0$ in an interval $I_3:=[0,K_2)$ for some $K_2>b_3/d_3$. Hence, in this case, a parallel flow \eqref{sol: Parallel} grows exponentially and unboundedly if it contains a Fourier mode with wave vector $\k$ in the annulus $\{\k\in\R^2\,|\,|\k|^2\in I_2\}$ for (not too strongly) stable stratification, or in the disc $\{\k\in\R^2\,|\,|\k|^2\in I_3\}$ for unstable stratification, and its Fourier coefficient vector is not an eigenvector for a possible negative root. Other modes that have $c_0\geq0$ also yield such growth if $c_1<0$ and $c_1^2\geq 4c_0$. We omit details but note that $c_1<0$ occurs for $K<(b_3-\mu)/d_3$, possibly containing $I_2$.\newline

\subsubsection{Kolmogorov flow}\label{Kolmogorov flow}
Another well-known class of explicit solutions for the Boussinesq equations in absence of backscatter are the so-called Kolmogorov flows, see e.g. \citep{BalmforthYoung2005}, with wave vectors of the form $\k=(k,0,-m)\transpose$, where $k,\,m\in\R$. Here we study their occurrence in the case of backscatter and start with the ansatz 
\begin{align}\label{sol: Kolmogorov}
\v(t,\x)= e^{\lambda t}\cos(\k\cdot \x)\a \,,\quad \b(t,\x)=ce^{\lambda t}\cos(\k\cdot \x) \,,\quad
p(t,\x)=\gamma e^{\lambda t}\sin(\k\cdot \x) \,,
\end{align}
and the flow direction $\a=\alpha(0,1,0)\transpose +\beta(m,0,k)\transpose.$ 

Compared with the Kolmogorov flows without backscatter and rotation from \citep{prugger2020explicit}, here we have rotation ($f\neq0$), time dependence ($\lambda\neq0$) and a nonzero second component of the velocity direction ($\alpha\neq0$). A superposition of these Kolmogorov flows with the horizontal flow solutions \eqref{sol: B} is not possible, since the orthogonality conditions of wave vectors and velocity directions are not satisfied, thus leading to non-gradient terms from the nonlinearity. However, superposition of different Kolmogorov flows is possible, as long as all wave vectors $\k$ have the same direction, as is the superposition with such monochromatic inertia gravity waves as discussed in \S\ref{s:MGW}. \newline 

We next determine the necessary relations for the coefficients of \eqref{sol: Kolmogorov} in order to solve the Boussinesq equations. For better readability we define the following terms resulting from the backscatter and thermal diffusion:
\begin{align*}
\delta_{\mu}(k,m)=\mu|\k|^2\,,\quad \delta_i(k,m)=d_i|\k|^4-b_i|\k|^2 \,\mbox{ for any }\, 1\leq i\leq3\,,
\end{align*}
where $|\k|^2=k^2+m^2$. Upon inserting \eqref{sol: Kolmogorov} into \eqref{eq: B} we find that the coefficients have to satisfy 
\begin{align}\label{cond: coeff}
\begin{pmatrix}
-f & m(\lambda+\delta_1) & 0 & k\\
\lambda+\delta_2 & fm & 0 & 0\\
0 & k(\lambda+\delta_3) & -1 & -m\\
0 & N^2k & \lambda+\delta_\mu & 0
\end{pmatrix}
\begin{pmatrix}
\alpha \\ \beta \\ c \\ \gamma
\end{pmatrix}
=0\,.
\end{align}
For $(k,m)=(0,0)$ these require $\alpha=0$ and $c=0$, which is the zero state. From the second row of the 4-by-4 matrix in \eqref{cond: coeff} and $f\neq0$ we immediately find that $\alpha=0$ implies $\beta m=0$. In case $\beta=0$, the Kolmogorov flow \eqref{sol: Kolmogorov} is the trivial zero solution, and in case $m=0$, \eqref{sol: Kolmogorov} is a parallel flow. Hence, we may assume $\alpha\neq 0$. Since \eqref{cond: coeff} is a homogeneous linear system in $(\alpha,\beta,c,\gamma)$, non-trivial solutions require a kernel of the associated matrix. Hence, either there is no non-trivial Kolmogorov flow or a linear space of these, which requires vanishing  determinant of this matrix. Assuming $(k,m)\neq(0,0)$ and dividing by $-|\k|^2$, this gives 
\begin{align}\label{eq: poly}
\lambda^3 + c_2 \lambda^2 + c_1\lambda + c_0 = 0\,,
\end{align}
with coefficients 
\begin{align*}
c_2 &:= (\delta_2+\delta_{\mu}) + |\k|^{-2}(\delta_3 k^2 + \delta_1 m^2)\,,\\
c_1 &:= \delta_2\delta_{\mu} + |\k|^{-2}[(\delta_2+\delta_{\mu})(\delta_3 k^2 + \delta_1 m^2) + N^2k^2+f^2m^2]\,,\\
c_0 &:= |\k|^{-2}[\delta_2\delta_{\mu}(\delta_3 k^2 + \delta_1 m^2) + \delta_2 N^2k^2 + \delta_{\mu} f^2m^2]\,.
\end{align*}

\begin{figure}[t]
\begin{center}
\subfigure[$\Lambda=\{-4,0,0,0,0,0,0,0\}$]{
\includegraphics[width=0.28\linewidth]{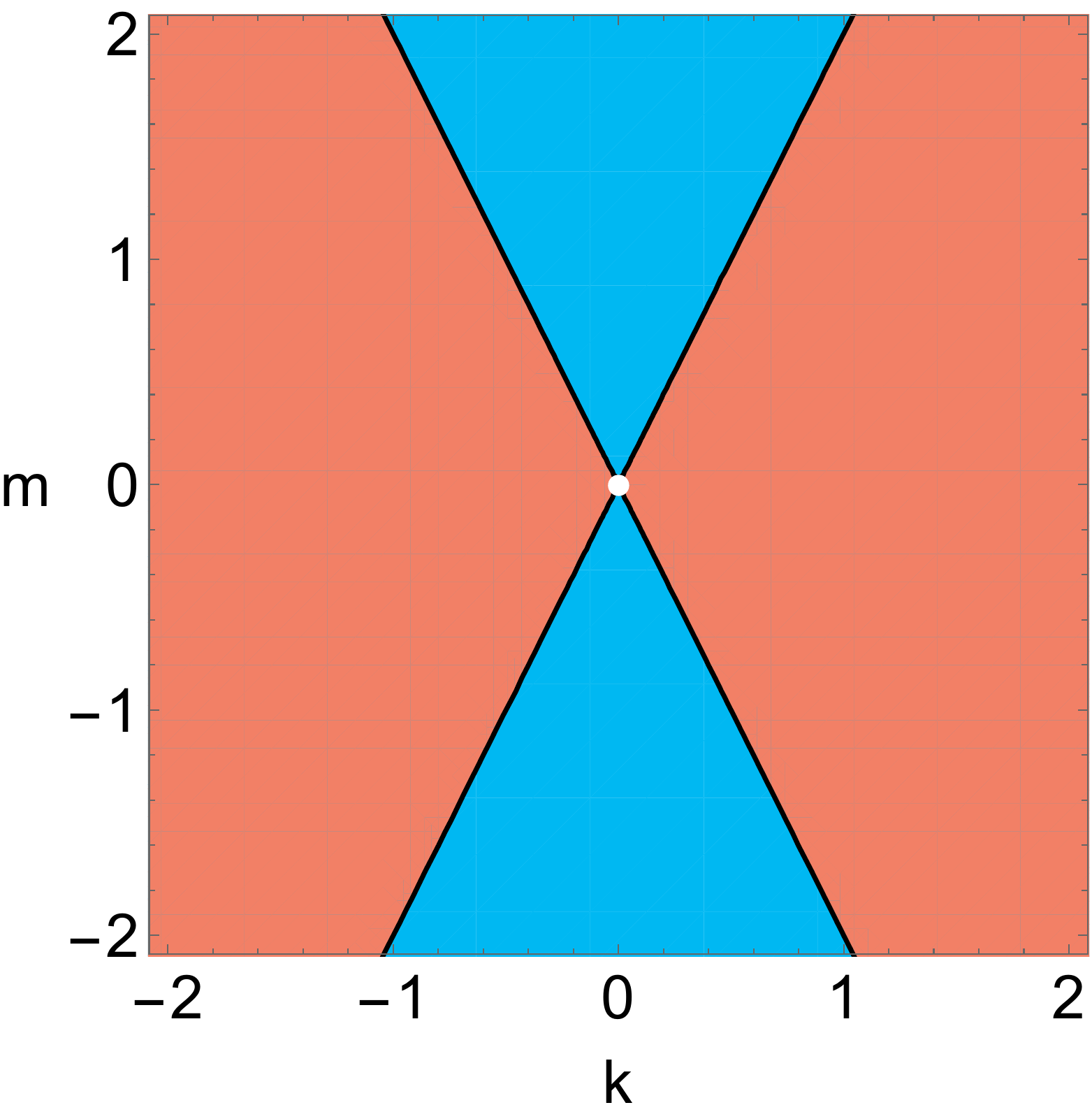}
\label{fig: Kolmogorov_f_1_N2_m4_d1_0_b1_0_d2_0_b2_0_d3_0_b3_0_d4_0_b4_0}}
\hfil
\subfigure[$\Lambda=\{-4,1,2,0.5,1,0,-1,0\}$]{
\includegraphics[width=0.28\linewidth]{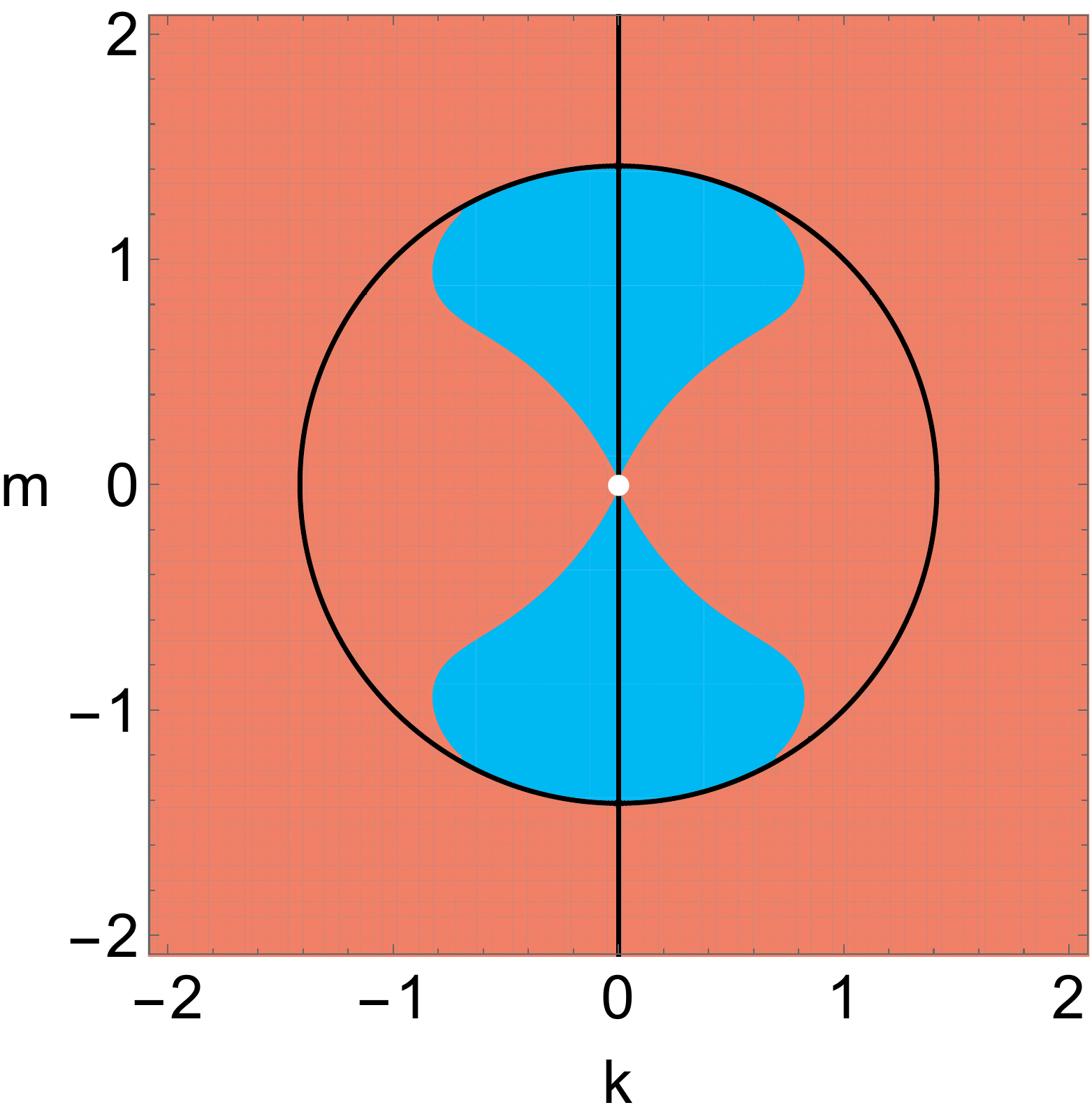}
\label{fig: Kolmogorov_f_1_N2_m4_d1_1_b1_2_d2_0p5_b2_1_d3_0_b3_m1_d4_0_b4_0}}
\hfil
\subfigure[$\Lambda=\{-4,1,2,0.5,1,0,-1,3\}$]{
\includegraphics[width=0.28\linewidth]{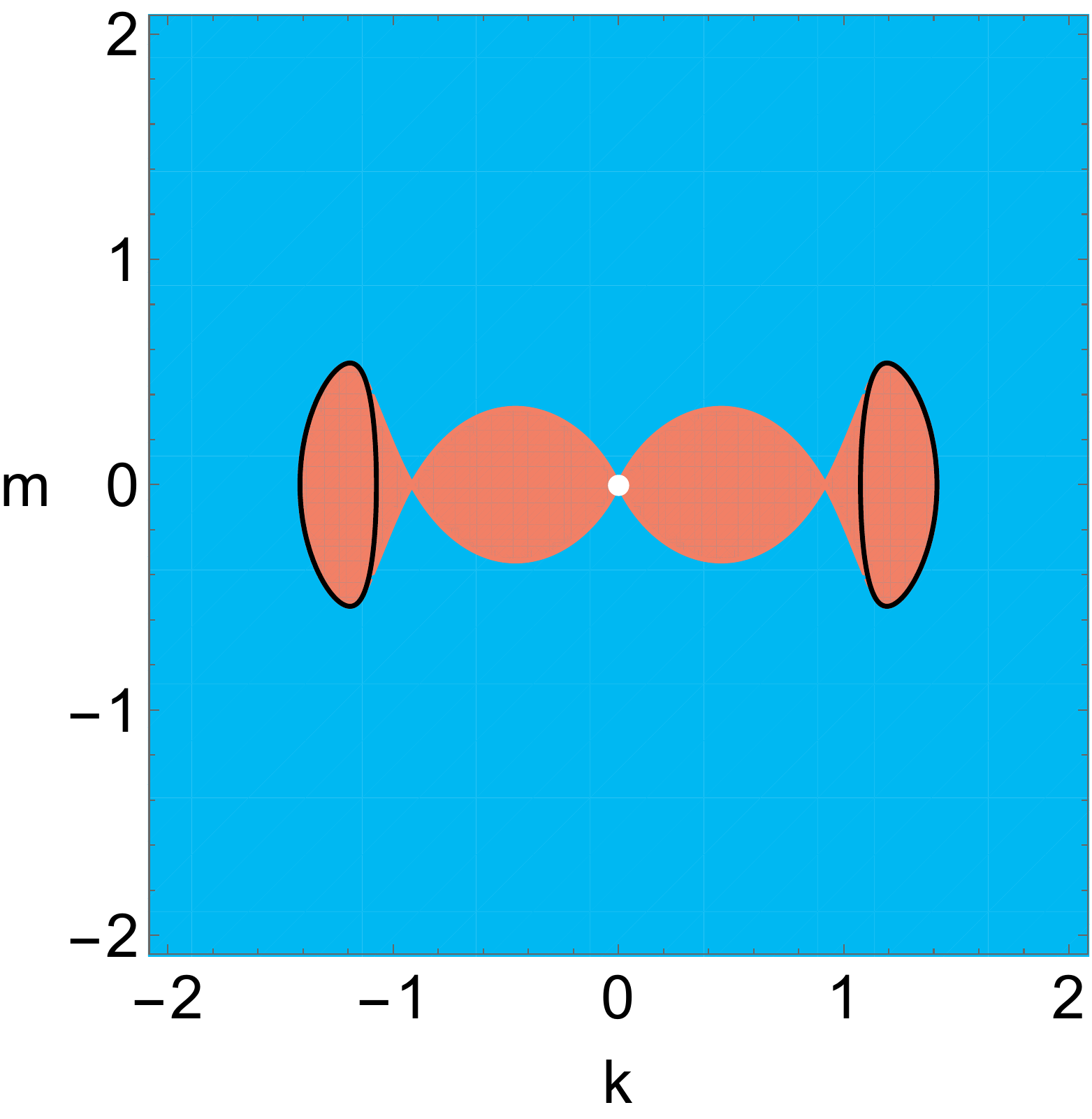}
\label{fig: Kolmogorov_f_1_N2_m4_d1_1_b1_2_d2_0p5_b2_1_d3_0_b3_m1_d4_0_b4_m3}}

\vspace{0.5cm}
\subfigure[$\Lambda=\{4,1,2,0.5,1,0,-1,0\}$]{
\includegraphics[width=0.28\linewidth]{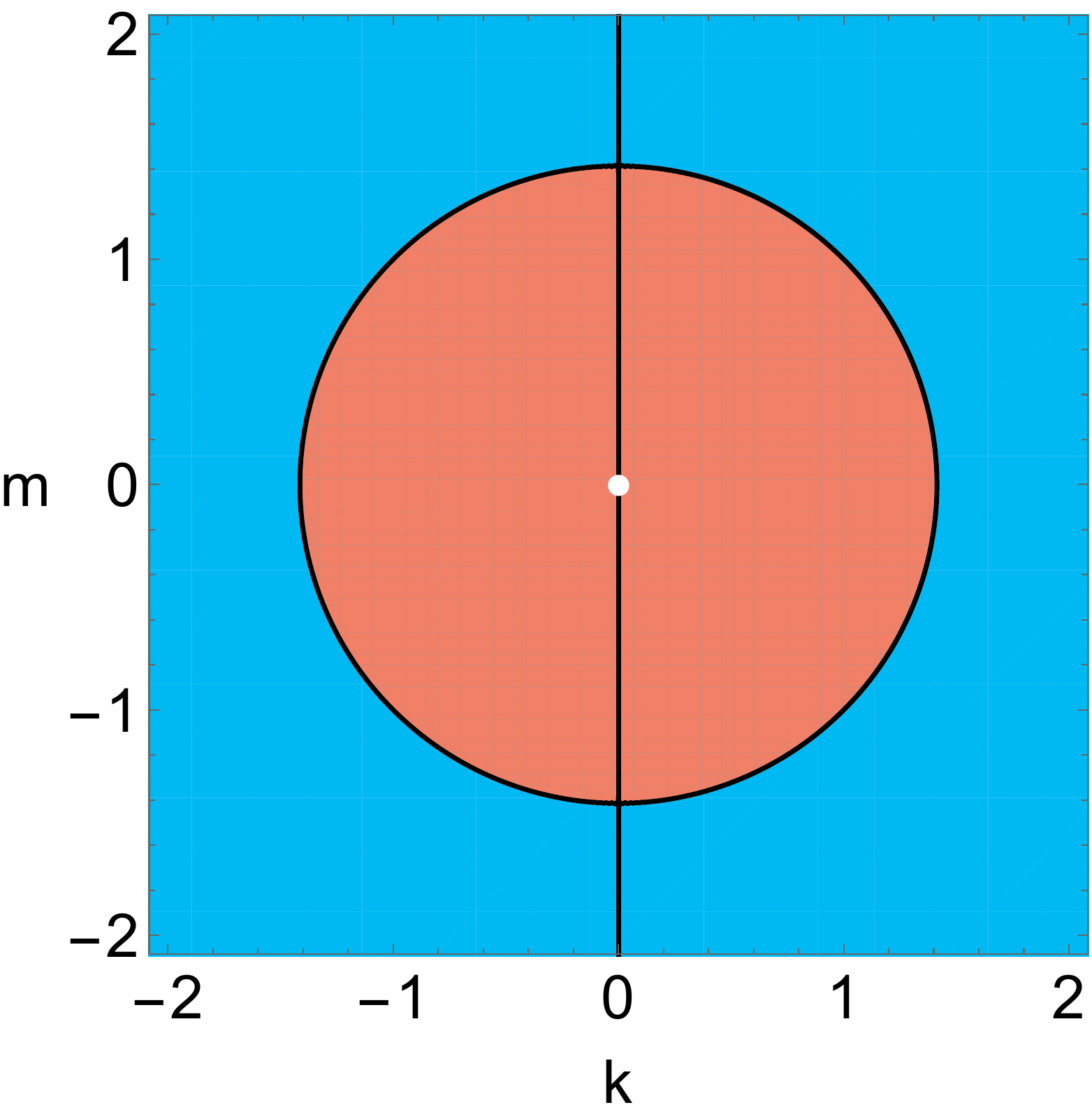}
\label{fig: Kolmogorov_f_1_N2_4_d1_1_b1_2_d2_0p5_b2_1_d3_0_b3_m1_d4_0_b4_0}}
\hfil
\subfigure[$\Lambda=\{4,1,2,0.5,1,0,-1,3\}$]{
\includegraphics[width=0.28\linewidth]{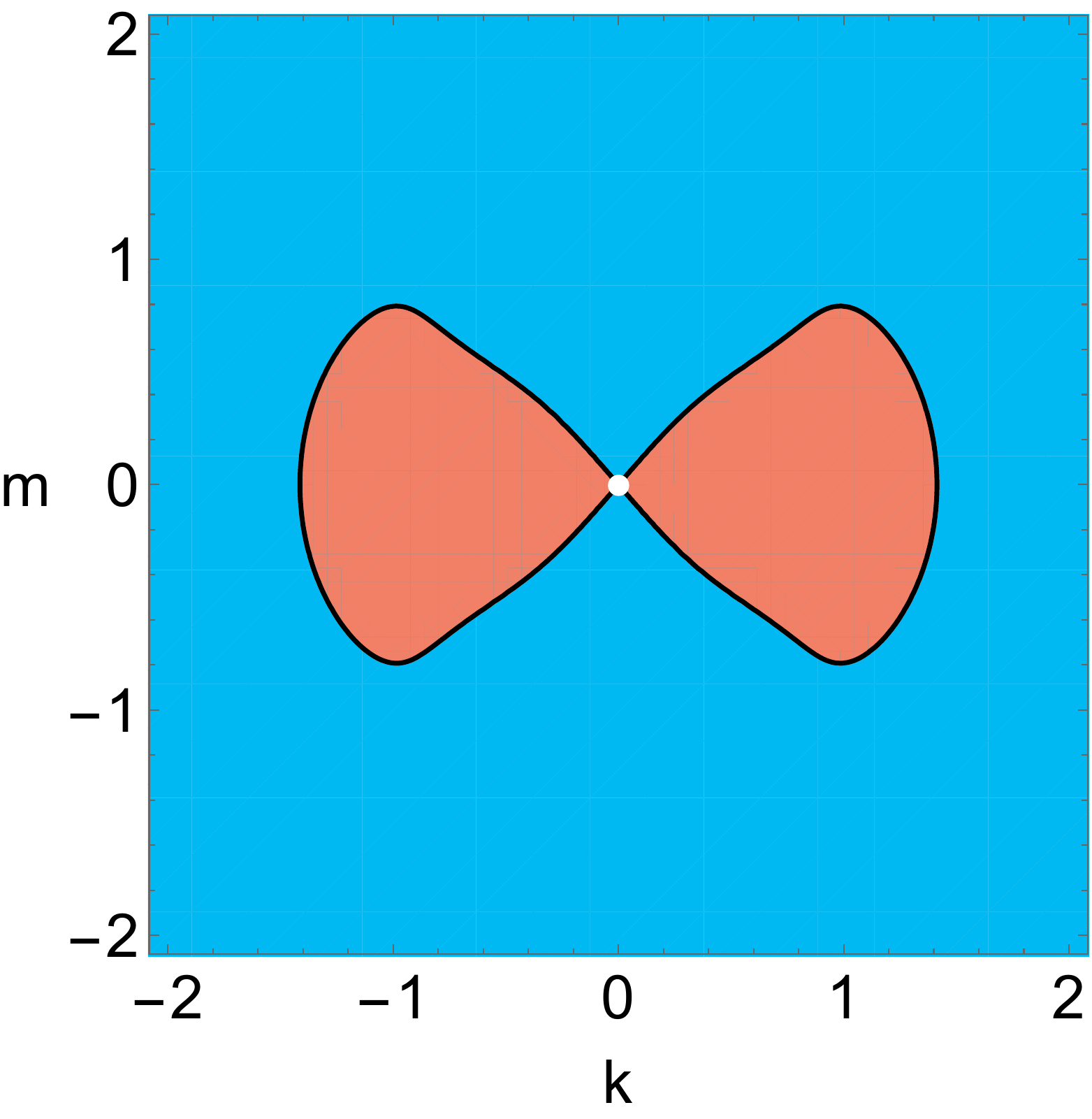}
\label{fig: Kolmogorov_f_1_N2_4_d1_1_b1_2_d2_0p5_b2_1_d3_0_b3_m1_d4_0_b4_m3}}

\caption{\label{fig: KFgrow} Unboundedly growing Kolmogorov flows occur for wave vectors $(k,m)$ in the red regions for parameters sets $\Lambda:=\{N^2,d_1,b_1,d_2,b_2,d_3,b_3,\mu\}$, where the Coriolis parameter is fixed at $f=1$. Red regions: one of conditions (1), (2) and (3) is satisfied; blue regions: none of conditions (1), (2) and (3) is satisfied; black curves in (a): loci of $c_0=c_1=c_2=0$; black curves in (b-e): loci of $c_0=0$; white dots: the zero state at $(k,m)=(0,0)$.}
\end{center}
\end{figure}

\pparagraph{Steady Kolmogorov flows and linear stability.} 
The condition $\lambda=0$ for steady Kolmogorov flow reduces \eqref{eq: poly} to
\begin{align}\label{eq: zerolam}
\delta_2\delta_{\mu}(\delta_3 k^2 + \delta_1 m^2) + \delta_2 N^2k^2 + \delta_{\mu} f^2m^2 = 0\,.
\end{align}
For comparison, note that the left-hand side identically equals to zero in the absence of backscatter and viscosity ($d_j,\, b_j=0$ for $j=1,\,2,\,3$) and thermal diffusion ($\mu=0$) so that in this case steady and non-trivial Kolmogorov flow exists for all $(k,m)\in\R^2\setminus\{(0,0)\}$.
In contrast, in the presence of backscatter $d_2,\, b_2>0$, but still without thermal diffusion ($\mu=0$), only $\delta_2N^2k^2 = 0$ remains so that either $k=0$ or $\delta_2 = 0$. The latter means $|\k|^2=b_2/d_2$, so that non-trivial and steady Kolmogorov flow occurs on the $m$-axis and the circle in the $(k,m)$-plane with radius $\sqrt{b_2/d_2}$, cf. Fig~\ref{fig: Kolmogorov_f_1_N2_m4_d1_1_b1_2_d2_0p5_b2_1_d3_0_b3_m1_d4_0_b4_0}, \ref{fig: Kolmogorov_f_1_N2_4_d1_1_b1_2_d2_0p5_b2_1_d3_0_b3_m1_d4_0_b4_0}. Conversely, we can create steady Kolmogorov flows for any wave vector $(k,m)\neq (0,0)$ by suitable choice of $d_2,\,b_2$ such that  $\delta_2=0$.\newline

Concerning stability, analogous to \S\ref{s:BE-lin_stability_steady}, small amplitude steady Kolmogorov flows are unstable, though typically not unboundedly, due to the instability of the zero state under backscatter. In the large amplitude scaling, the resulting operator $\Lb_0$ is a triangular block matrix operator, similar to \S\ref{s:parflow}, with skew-adjoint parts that imply purely imaginary spectrum. Hence, there are again no growth rates that are proportional to the amplitude of the steady Kolmogorov flow. Regarding stability of Kolmogorov flows without backscatter we refer to \citep{BalmforthYoung2005}.

A source of unbounded instability of steady Kolmogorov flows are possible superpositions with monochromatic inertia gravity waves discussed in \S\ref{s:MGW} below. In the following we will examine the existence of exponentially and unboundedly growing Kolmogorov flows, which then also proves unbounded instability of steady Kolmogorov flows due to possible superpositions.\newline

\pparagraph{Unboundedly growing Kolmogorov flows.}
Such flows with $\beta,\, k\neq 0$ transfer the horizontal backscatter to growing vertical velocity component. They correspond to positive roots of \eqref{eq: poly}, which occur as follows in terms of the sign of $c_0$:
\begin{itemize}
\item[(1)]  If $c_0<0$, then \eqref{eq: poly} has a positive root.
\item[(2)] For $c_0=0$, \eqref{eq: poly} has a positive root if and only if $-c_2+\sqrt{c_2^2 - 4c_1}>0$.
\item[(3)] For $c_0>0$, \eqref{eq: poly} has a positive root if and only if $-c_2+\sqrt{c_2^2-3c_1}>0$ and $2c_2^3-9c_1c_2+27c_0 + (6c_1-2c_2^2)\sqrt{c_2^2-3c_1}\leq0$.
\end{itemize}
The conditions in (3) imply that the local minimum of the cubic polynomial in \eqref{eq: poly} lies at a positive value and the value on the local minimum is non-positive.

\medskip
For comparison we start with the common situation without backscatter, viscosity and thermal diffusion ($\mu,\,b_j,\,d_j=0$ for $j=1,\,2,\,3$), where a growing Kolmogorov flow \eqref{sol: Kolmogorov} requires unstable stratification $N^2<0$. Indeed, in this case \eqref{eq: poly} has $c_0=c_2=0$ and $c_1=|\k|^{-2}(N^2k^2+f^2m^2)$, so that a positive root occurs if and only if $c_1<0$, which requires $N^2<0$, and thus growing solutions occur for $m^2/k^2<-N^2/f^2$, cf. Fig.~\ref{fig: Kolmogorov_f_1_N2_m4_d1_0_b1_0_d2_0_b2_0_d3_0_b3_0_d4_0_b4_0}. More precisely, for such wave vectors a steady flow co-exists with a growing and a decaying flow (on the red regions in Fig.~\ref{fig: Kolmogorov_f_1_N2_m4_d1_0_b1_0_d2_0_b2_0_d3_0_b3_0_d4_0_b4_0}), which turn into a triple steady flow on the boundary, where $c_0=c_1=c_2=0$ (black curves).

\medskip
In the presence of (horizontal) backscatters ($b_j,\,d_j>0$ for $j=1,\,2$) growth of Kolmogorov flows \eqref{sol: Kolmogorov} is possible also for stable stratification $N^2>0$. We next focus on case (1) with negative $c_0$ and omit details of cases (2) and (3) with non-negative $c_0$. Some examples are plotted in Fig.~\ref{fig: KFgrow}. \newline
First, we note that, for $N^2>0$, the coefficient $c_0$ is negative for sufficiently small $k$ and $m$. Its sign is that of the left-hand side of \eqref{eq: zerolam}, whose leading order term as $(k,m)\to (0,0)$ is $|\k|^2(-b_2N^2k^2 + \mu f^2m^2)$ and can be negative only if $N^2>0$ and $k\neq 0$. Then, to leading order, growing Kolmogorov flows occur for $m^2/k^2<b_2N^2/(\mu f^2)$ with $\mu>0$ (cf. Fig~\ref{fig: Kolmogorov_f_1_N2_4_d1_1_b1_2_d2_0p5_b2_1_d3_0_b3_m1_d4_0_b4_m3}) and increase of $N^2$ or $b_2$ enlarges this region in the $(k,m)$-plane near the origin, while increase of $\mu$ or $f$ shrinks it. Rewriting the condition as $N^2>\mu f^2m^2/(b_2k^2)$, for a given wave vector, then on the one hand increase of $\mu$ or $f$ requires sufficiently stable stratification, and on the other hand increase of $b_2$ allows for less stable stratification. The leading order term of the left-hand side of \eqref{eq: zerolam} is always positive if $N^2<0$, so $c_0>0$ near the origin, which refers to case (3) above; we omit details and just plot an example in Fig.~\ref{fig: Kolmogorov_f_1_N2_m4_d1_1_b1_2_d2_0p5_b2_1_d3_0_b3_m1_d4_0_b4_m3}.\newline
Second, for general $(k,m)$ and vanishing thermal diffusion ($\mu=0$), only $\delta_2 N^2 k^2$ remains. Hence, $c_0<0$ for $|\k|^2<b_2/d_2$ if $N^2>0$ and $k\neq0$ (cf. Fig.~\ref{fig: Kolmogorov_f_1_N2_4_d1_1_b1_2_d2_0p5_b2_1_d3_0_b3_m1_d4_0_b4_0}), and $c_0<0$ for $|\k|^2>b_2/d_2$ if $N^2<0$ and $k\neq0$ (cf. Fig.~\ref{fig: Kolmogorov_f_1_N2_m4_d1_1_b1_2_d2_0p5_b2_1_d3_0_b3_m1_d4_0_b4_0}). For $k=0$, $c_0$ is zero so we refer to the case (2) above for the growing Kolmogorov flows. The numerical computations show that in the examples of Fig.~\ref{fig: Kolmogorov_f_1_N2_m4_d1_1_b1_2_d2_0p5_b2_1_d3_0_b3_m1_d4_0_b4_0} and~\ref{fig: Kolmogorov_f_1_N2_4_d1_1_b1_2_d2_0p5_b2_1_d3_0_b3_m1_d4_0_b4_0} the condition in case (2) is not satisfied for $k=0$ so that here a Kolmogorov flow \eqref{sol: Kolmogorov} is not growing. Near $\k=0$, the examples of Fig.~\ref{fig: Kolmogorov_f_1_N2_m4_d1_1_b1_2_d2_0p5_b2_1_d3_0_b3_m1_d4_0_b4_0} and \ref{fig: Kolmogorov_f_1_N2_m4_d1_1_b1_2_d2_0p5_b2_1_d3_0_b3_m1_d4_0_b4_m3} can be regarded as perturbations of the example of Fig.~\ref{fig: Kolmogorov_f_1_N2_m4_d1_0_b1_0_d2_0_b2_0_d3_0_b3_0_d4_0_b4_0}. As before, the red region indicates existence of growing flows; note that decaying or oscillating ones may co-exist. Notably, in Fig.~\ref{fig: Kolmogorov_f_1_N2_m4_d1_1_b1_2_d2_0p5_b2_1_d3_0_b3_m1_d4_0_b4_0} and \ref{fig: Kolmogorov_f_1_N2_m4_d1_1_b1_2_d2_0p5_b2_1_d3_0_b3_m1_d4_0_b4_m3} the part of the red region's boundary  without black marking stems from a bifurcation of saddle-node-type in terms of the wave vector, where \eqref{eq: poly} possesses a positive double root. In contrast, the black curves mark loci of steady flows, where $c_0=0$.

\medskip
Similar to parallel flow, we briefly consider artificial vertical backscatter since isotropic backscatter $d_j=d>0,\,b_j=b>0$, $j=1,\,2,\,3$ admits an analytical consideration. In this case $\delta:=d|\k|^4-b|\k|^2=\delta_j$ for $j=1,\,2,\,3$. With thermal diffusion ($\mu>0$), we then have $c_0<0$ for small wave number $|\k|^2<b/d$ (i.e. $\delta<0$) if stratification is sufficiently stable $N^2>-\mu|\k|^2(f^2m^2+\delta^2|\k|^2)/(k^2\delta)>0$, and for large wave number $|\k|^2>b/d$ (i.e. $\delta>0$) only for sufficiently unstable stratification $N^2<-\mu|\k|^2(f^2m^2+\delta^2|\k|^2)/(k^2\delta)<0$. Unlike the growing parallel flow, which requires sufficiently weak stable stratification, here the increase of positive $N^2$ increases the set of wave vectors -- within the small wave number region -- for growing Kolmogorov flow. Another observation is that for the fixed stratification -- stable or unstable -- larger $\mu$ leads to smaller region in wave vector space of growing solutions.

The latter can also be observed (numerically) without vertical backscatter $d_3=0,\, b_3<0$ as shown in Figs.~\ref{fig: Kolmogorov_f_1_N2_4_d1_1_b1_2_d2_0p5_b2_1_d3_0_b3_m1_d4_0_b4_0}, \ref{fig: Kolmogorov_f_1_N2_4_d1_1_b1_2_d2_0p5_b2_1_d3_0_b3_m1_d4_0_b4_m3} for $N^2>0$, and Figs.~\ref{fig: Kolmogorov_f_1_N2_m4_d1_1_b1_2_d2_0p5_b2_1_d3_0_b3_m1_d4_0_b4_0}, \ref{fig: Kolmogorov_f_1_N2_m4_d1_1_b1_2_d2_0p5_b2_1_d3_0_b3_m1_d4_0_b4_m3} for $N^2<0$.\newline

\subsubsection{Monochromatic inertia gravity waves}\label{s:MGW}
The last kind of explicit plane wave-type solutions we are aware of are the so-called monochromatic inertia gravity waves (MGWs), which are for example discussed in \citep{Achatz06} for the inviscid Boussinesq equations. Here we study the occurrence in the rotating Boussinesq equations with backscatter. These solutions again form an invariant subspace of linear dynamics since the nonlinear terms vanish, but structurally differ from the aforementioned flows. In particular, the wave profile of a MGW is a time-dependent travelling wave with phase variable $\xi=kx+mz-\omega t$ and takes the form
\begin{subequations}\label{sol: MGW2}
\begin{align}
\v(t,\x)&= \alpha_1 e^{\lambda t}\sin(\xi)(0,1,0)\transpose + \alpha_2\omega e^{\lambda t}\cos(\xi)(-m,0,k)\transpose \,,\label{sol: MGW2a}\\
\b(t,\x)&= \beta_1 e^{\lambda t}\sin(\xi) + \beta_2\omega e^{\lambda t}\cos(\xi) \,,\label{sol: MGW2b}\\
p(t,\x)&= \gamma_1 e^{\lambda t}\cos(\xi) + \gamma_2\omega e^{\lambda t}\sin(\xi)\,.\label{sol: MGW2c}
\end{align}
\end{subequations}
The conditions for these to be (non-trivial) explicit solutions of \eqref{eq: B}, with $\k=(k,m)\transpose\in\R^2\backslash\{(0,0)\transpose\}$, in particular depend on $\omega$. In fact, for $\omega=0$ these are Kolmogorov flows from \S\ref{Kolmogorov flow} with $\beta=0$, and the existence conditions read, after inserting \eqref{sol: MGW2} into \eqref{eq: B},
\begin{align}\label{cond: MGWzero}
\begin{pmatrix}
\lambda+\delta_2 & 0 & 0\\
mf & -k & 0\\
kf & m & |\k|^2\\
0 & \lambda+\delta_{\mu} & 0
\end{pmatrix}
\begin{pmatrix}
\alpha_1 \\ \beta_1 \\ \gamma_1
\end{pmatrix}
=0\,,
\end{align}
with $\delta_j=\delta_j(\k):=d_j|\k|^4-b_j|\k|^2$ for any $1\leq j\leq 3$ and $\delta_{\mu}=\delta_{\mu}(\k):=\mu|\k|^2$ as above.\newline
For $\omega\neq0$ the existence conditions are at first the following eight:
\begin{align}\label{cond: MGWnonzero}
\begin{pmatrix}
\lambda+\delta_2 & 0 &  0 &0 & 0 & 0\\
mf & -k & 0 & \omega^2|\k|^2 & 0 & 0\\
kf & m & |\k|^2 & 0 & 0 & 0\\
0 & \lambda+\delta_{\mu} & 0 & 0 & \omega^2 & 0\\
1 & 0 & 0 & mf & 0 & 0\\
0 & -1 & 0 & kN^2 & \lambda+\delta_{\mu} & 0\\
0 & 0 & 0 & \delta_3k^2+\delta_1m^2+\lambda|\k|^2 & -k & 0\\
0 & 0 & 0 & (\delta_3-\delta_1)km & -m & |\k|^2
\end{pmatrix}
\begin{pmatrix}
\alpha_1 \\ \beta_1 \\ \gamma_1 \\ \alpha_2 \\ \beta_2 \\ \gamma_2
\end{pmatrix}
=0\,.
\end{align}
However, most of these can be readily solved directly in terms of the coefficients. In the following we explicitly determine all non-trivial solutions \eqref{sol: MGW2}. Afterwards we shortly discuss the stability of MGWs and the possible superpositions with other types of explicit solutions. \newline

\pparagraph{MGWs with steady phase ($\omega=0$).}
We start with $\omega=0$, which gives a certain type of Kolmogorov flows. There is no propagation of the travelling wave profile and the second terms of $\v$, $\b$ and $p$ in \eqref{sol: MGW2} vanish. Solutions with 
\begin{align*}
k=0\,,\quad \alpha_1=0\,,\quad \beta_1=-\gamma_1m\,,\quad \lambda=-\delta_{\mu}
\end{align*}
have $\v\equiv0$, the pressure depends on the buoyancy only, and there are no further conditions on the wave vector $\k=(0,m)\transpose$.\newline

Other solutions with $\omega=0$ satisfy 
\begin{align*}
k\neq0\,,\quad \beta_1=\alpha_1\frac{m}{k}f\,,\quad \gamma_1=-\alpha_1\frac{f}{k}\,,\quad \lambda=-\delta_{\mu}\,,\quad \delta_2=\delta_{\mu}\,,
\end{align*}
so $\v\neq 0$ for $\alpha_1\neq0$. The last of these equations gives a condition on the wave vector $\k$, which is equivalent to $d_2|\k|^2-b_2-\mu=0$ and constrains the wave number to $|\k|^2=\frac{b_2+\mu}{d_2}$.\newline

In both of these cases the growth rate is defined by the thermal diffusion as $\lambda=-\mu|\k|^2$, so that all these MGWs are exponentially decaying for $\mu>0$ and steady for $\mu=0$. Furthermore, both solutions only depend on the parameter of the second momentum or the buoyancy equation, but are independent of the Brunt-V\"ais\"al\"a frequency $N^2$.\newline

\pparagraph{MGWs with oscillating phase ($\omega\neq0$).}
We turn to non-trivial MGW solutions \eqref{sol: MGW2} with $\omega\neq0$. The simplest class are `vertically varying' MGWs with 
\begin{align*}
k=0\,,\quad \alpha_1=-\alpha_2mf\,,\quad \beta_1=\beta_2=\gamma_1=\gamma_2=0\,,\quad \lambda=-\delta_2\,,\quad \omega=\pm f\,,\quad \delta_1=\delta_2\,,
\end{align*}
so that $\b\equiv p\equiv0$, while $\v\neq 0$ -- in contrast to the case $k=\omega=0$ above. Notably, these solutions depend on the parameters of the horizontal momentum equations only, while the solutions with $k=\omega=0$ only depend on those from the buoyancy equation. The last of the above equations is a condition on the wave vector $\k=(0,m)\transpose$, which is equivalent to $(d_1-d_2)m^2+b_2-b_1=0$. Thus, these solutions exist for all $m\neq0$ in the isotropic case, while for $d_1\neq d_2$ they are restricted to $m^2=\frac{b_1-b_2}{d_1-d_2}>0$.\newline
Since $\lambda=-\delta_2$, these MGWs are exponentially and unboundedly growing for wave numbers $m^2<b_2/d_2$, and steady for $m^2=b_2/d_2$, thus transferring the horizontal backscatter to growing vertical dependence. In particular, the thermal diffusion $\mu$ and Brunt-V\"ais\"al\"a frequency $N^2$ have no impact, since these solutions trivially satisfy the buoyancy equation by $v_3\equiv\b\equiv0$.\newline

Another type are `zonally varying' MGWs with the coefficients satisfying
\begin{align*}
&\alpha_1=m=0\,,\quad \beta_1=\alpha_2\Bigl(N^2-\frac{(\delta_3-\delta_{\mu})^2}{4}\Bigr)k\,,\quad \beta_2=\alpha_2\frac{\delta_3-\delta_{\mu}}{2}k\,,\\
&\gamma_1=\gamma_2=0\,,\quad \lambda=-\frac{\delta_3+\delta_{\mu}}{2}\,,\quad \omega^2=N^2-\frac{(\delta_3-\delta_{\mu})^2}{4}\,.
\end{align*}
These have vanishing pressure and depend only on the parameters of the buoyancy and vertical momentum equations. In contrast to the MGW solutions before, these here do also depend on the Brunt-V\"ais\"al\"a frequency $N^2$. Furthermore, different from the solutions before, the growth rate $\lambda$ and phase frequency $\omega$ depend on both, the vertical term $\delta_3$ and the thermal diffusion. Since $\omega^2>0$, the last equation is a condition on the wave vector $\k=(k,0)\transpose$, which requires stable stratification $N^2>0$ satisfying $(\delta_3-\delta_{\mu})^2<4N^2$. In case $N^2>0$ these MGW solutions exist at least for sufficiently small $k^2$, since $\delta_3,\,\delta_{\mu}\rightarrow0$ as $|\k|\rightarrow0$.
Due to the equation for the growth rate $\lambda$, these MGWs are exponentially decaying for kinetic energy backscatter, where $\delta_3= -b_3|\k|^2>0$. Hence, this kind of MGWs are steady or exhibit growth only in the artificial case $d_3,\, b_3>0$ for wave numbers $k^2\leq \frac{b_3-\mu}{d_3}$, i.e. only in the case $b_3>\mu$.\newline

\begin{figure}
\begin{center}
\subfigure[$\mu=0.0$]{
\includegraphics[trim=4.5cm 9cm 4.5cm 9cm,clip,width=0.42\linewidth]{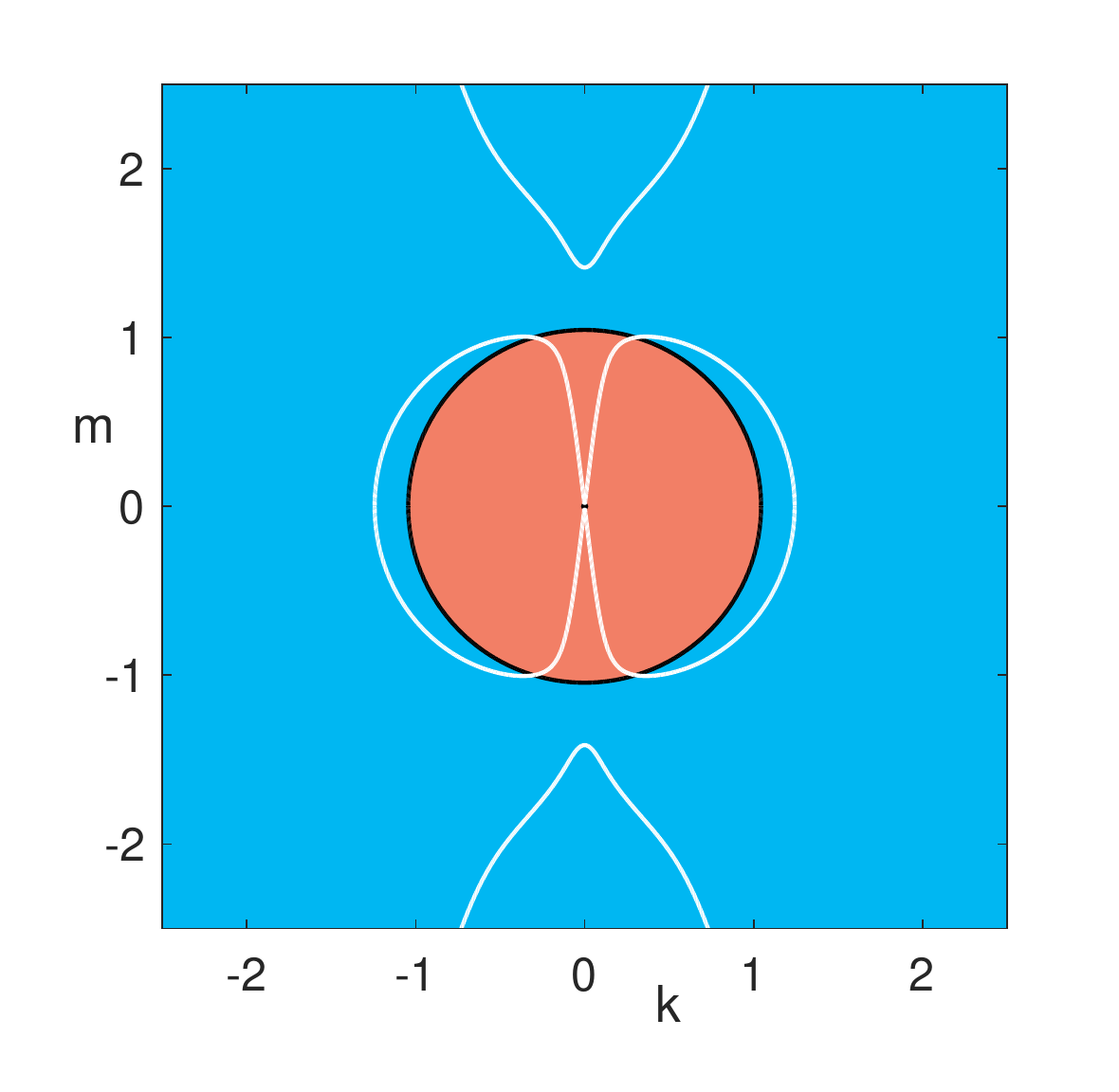}\label{Fig. MGWa}}
\subfigure[$\mu=0.12$]{
\includegraphics[trim=4.5cm 9cm 4.5cm 9cm,clip,width=0.42\linewidth]{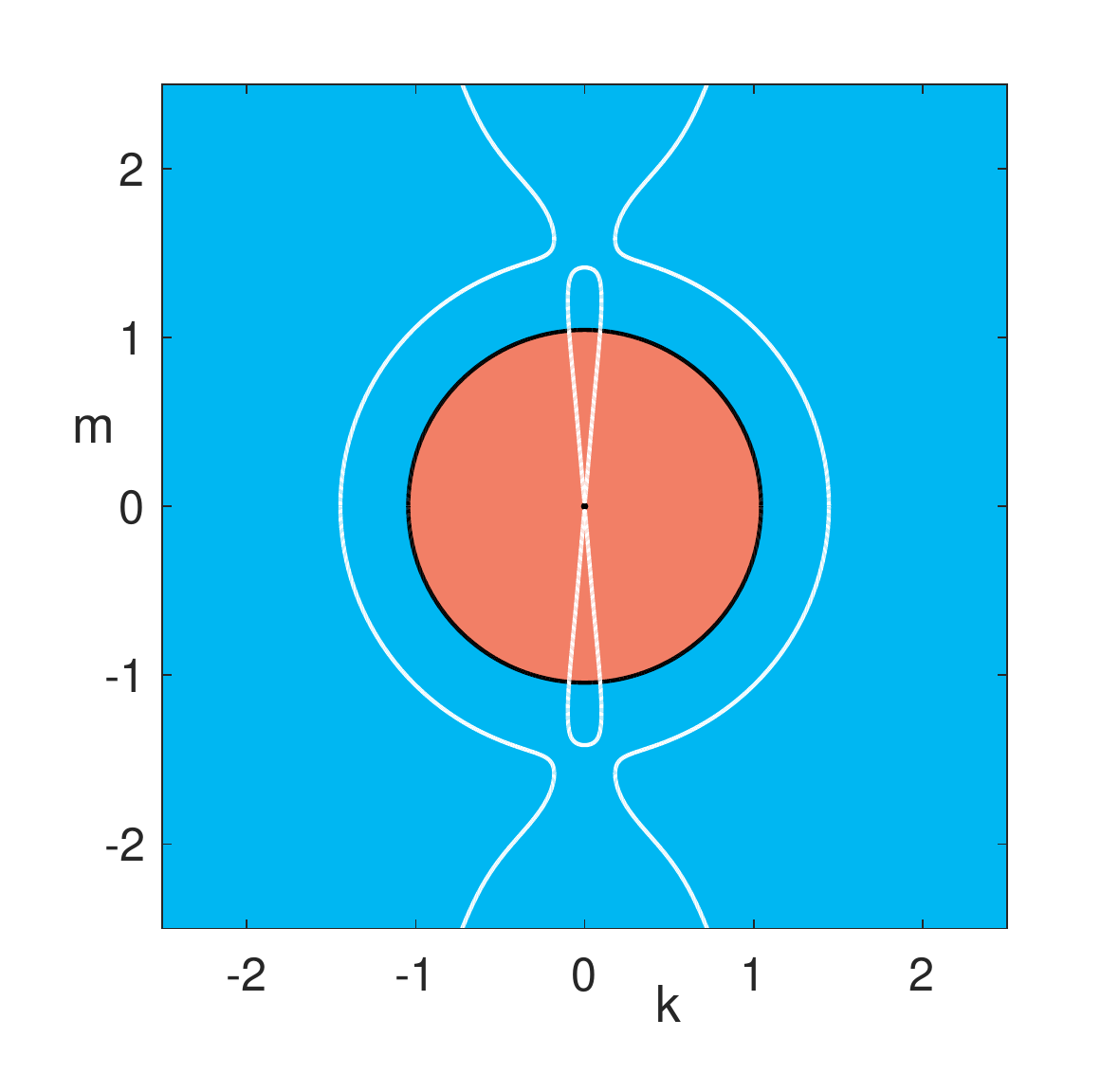}\label{Fig. MGWb}}
\caption{\label{Fig. MGW} Existence of MGW solutions on the wave vector plane $(k,m)$ in the case $\omega, \alpha_1\neq0$ (white curves) with growth rate $\lambda$ positive (red), zero (black) or negative (blue). Fixed parameter: $d_1=0.12,\, d_2=0.11,\, d_3=0.0,\, b_1=0.14,\, b_2=0.12,\, b_3=-0.1,\, f=0.3,\, N^2=1$.}
\end{center}
\end{figure}

The existence analysis for the remaining solutions to \eqref{sol: MGW2} with $\omega,\, k,\, \alpha_1 \neq0$ is more involved. These solutions have the coefficients
\begin{align*}
&\alpha_1=-\alpha_2mf\,,\quad \beta_1=\alpha_1\frac{m}{k}f+\alpha_2\frac{\omega^2}{k}|\k|^2\,,\quad \beta_2=\frac{\alpha_2}{k}(\delta_3k^2+\delta_1m^2-\delta_2|\k|^2)\,,\\
&\gamma_2=\alpha_2(\delta_1-\delta_3)\frac{km}{|\k|^2}+\beta_2\frac{m}{|\k|^2}\,,\quad \gamma_1=-\alpha_1\frac{kf}{|\k|^2}-\beta_1\frac{m}{|\k|^2}\,,\quad \lambda=-\delta_2\,,
\end{align*}
with additional conditions, that also define the phase frequency $\omega$, given by
\begin{align*}
&\omega^2=\Bigl((\delta_3-\delta_2)(\delta_{\mu}-\delta_2)+N^2\Bigr)\frac{k^2}{|\k|^2}+\Bigl((\delta_1-\delta_2)(\delta_{\mu}-\delta_2)+f^2\Bigr)\frac{m^2}{|\k|^2}>0\,,\\
&\Bigl((\delta_{\mu}-\delta_2)N^2+(\delta_3-\delta_2)\bigl((\delta_{\mu}-\delta_2)^2+\omega^2\bigr)\Bigr)k^2+(\delta_1-\delta_2)\bigl((\delta_{\mu}-\delta_2)^2+\omega^2\bigr)m^2=0\,.
\end{align*}
The first condition has solutions $(k,m)$ for any parameter: For stable stratification $N^2>0$ the factors of $k^2$ and $m^2$ are both positive for $|\k|$ sufficiently small, since $\delta_j,\,\delta_{\mu}\rightarrow0$ for $|\k|\rightarrow0$ for all $1\leq j\leq 3$. For unstable stratification $N^2<0$ one can first choose $|\k|$ sufficiently small, so that the factor of $m^2$ is positive. Then, for a fixed $|\k|^2=\sc^2$ so that $\delta_j$ and $\delta_{\mu}$ are constant for these wave vectors, one can choose $k^2=\sc^2-m^2$ small enough, so that the term with $k^2$ does not make the whole expression negative.\newline
We omit a complete analysis of the more complicated second condition here. Both conditions together generate a set of solutions with rather complex structure, as plotted in Figure \ref{Fig. MGW} for the cases $\mu=0$ and $\mu>0$.\newline
Due to the equation for the growth rate $\lambda=-\delta_2$, these MGWs are exponentially and unboundedly growing for wave numbers $|\k|^2<b_2/d_2$ and steady states for $|\k|^2=b_2/d_2$ (see Figure \ref{Fig. MGW}). Notably, since $k\neq 0$, these flows can transfer the horizontal backscatter (in fact only the meridional component) to growing vertical velocity. \newline

\pparagraph{Superpositions and stability.}
It is possible to superpose MGWs and Kolmogorov flows as a solution to \eqref{eq: B} if these have the same direction of wave vectors $\k$, and this can also be in the form of an integral. Indeed, the similar structure of wave vector and velocity direction yields vanishing nonlinear terms and a remaining system of linear equations, which each solution satisfies. Note that Kolmogorov flows exist on the whole wave vector space $(k,m)\transpose\in\R^2$, while MGW in general not. This means, that superposition in one wave vector direction is possible for arbitrary wave number of the Kolmogorov flow, but the wave number of the MGW is in general restricted. Depending on the wave vectors, the Kolmogorov flows and MGWs in such a superposition can be steady, exponentially growing or decaying.\newline

With this superposition we can prove in certain cases the unbounded instability of steady MGW solutions due to perturbations with exponentially growing Kolmogorov flows, and vice versa. In \S\ref{Kolmogorov flow} we found that growing Kolmogorov flows in case $\mu>0$ may occur only in certain directions. In this case, steady MGW solutions are unboundedly unstable a priori only in these certain wave vector directions (see investigation of $c_0$ in \S\ref{Kolmogorov flow}). Without thermal diffusion ($\mu=0$) there are always growing Kolmogorov flows in any direction with $k\neq0$, for stable stratification $N^2>0$ at least for all $|\k|^2<b_2/d_2$, and for unstable stratification $N^2<0$ for $|\k|^2>b_2/d_2$. Thus, in these cases, the steady MGWs with $k\neq0$ are always unboundedly unstable with respect to certain Kolmogorov flows. However, steady MGWs with $k=0$, as well as Kolmogorov flows, are also unboundedly unstable, since there are exponentially and unboundedly growing MGWs with $k=0$ and $m^2<b_2/d_2$, with which they can be superposed (see the case of `vertically varying' MGWs with $\omega\neq0$ and $k=0$). \newline

As in \S\ref{s:BE-lin_stability_steady}, steady small amplitude MGWs are unstable due to the unstable zero state under backscatter. In the large amplitude scaling, the resulting operator $\Lb_0$ is a triangular block matrix operator with skew-adjoint parts that imply purely imaginary spectrum as for Kolmogorov flows. Hence, there are again no unstable eigenvalues that scale with the amplitude of the steady MGW.

\section{Discussion}
Motivated by the numerical backscatter scheme \citep{Jansen2014,M3,juricke2020kinematic,Perezhogin20}, we have studied the impact of simplified kinetic energy backscatter via classes of explicit flows in the shallow water and rotating Boussinesq equations on the whole space. Here we have found that backscatter induces unbounded instability of the zero state, as well as of certain non-trivial steady solutions, in the sense of unboundedly and exponentially growing flows, also with vertical structure and also for stable stratification. This highlights the possibility of  concentration of energy due to backscatter, and is in contrast to the desired energy redistribution. Since all flows we have considered simultaneously solve the linear equations from dropping the transport nonlinearity and the full nonlinear equations, these are directly linked to spectrum and linear stability of the trivial flow. Indeed, in wave vector space, the condition to solve the nonlinear equations emposes additional constraints, which are stronger in the shallow water case than in the purely horizontal Boussinesq equations. We have discussed the corresponding linear and nonlinear constraints on coefficients for these and also for flows with vertical structure and coupled buoyancy that relate to parallel flows, Kolmogorov flows and monochromatic inertia gravity waves. 

We have identified superposition principles of these flows in the nonlinear equations and have discussed the resulting unbounded instability of the flows themselves, in particular of steady flows induced by the backscatter. Due to the linear-nonlinear structure, these steady flows come as a family with an amplitude scaling parameter and we have discussed linear stability in the regimes of small and large amplitudes. Flows of small amplitudes inherit the instability of the zero state, but the treatment of large amplitudes is more subtle. Here we have considered a renormalised eigenvalue problem and have found that the resulting spectrum is purely imaginary in all except one class of the aforementioned flows. Based on numerical computations, in the exceptional case of steady multi mode horizontal flows unstable rates are proportional to the amplitude of the steady flow, thus leading to arbitrarily strong growth rates. \newline

The approach that we have presented can be applied to operators with other constant coefficient linear or derivative terms. An interesting case that we will pursue further is the inclusion of bottom drag, which admits an onset of instability with finite wave number. Also forcing of plane wave form can be treated, similar to \citep{prugger2020explicit}. \newline

It would be interesting to study analytically and numerically in what way the undesired growth occurs in numerical discretisations. For fixed backscatter coefficients under decreasing grid size, the unbounded instability in the limit is expected to readily imply arbitrarily large growth that may however be bounded for any fixed small grid size. The actual backscatter scheme scales the strength of backscatter with the grid size so that it would be interesting to identify, for a given discretisation scheme, the relation between the growth of flows and the grid-scaling of backscatter coefficients. This may provide another approach to energetic consistency of backscatter schemes.

\subsubsection*{Acknowledgments}
\vspace{-3mm}The authors thank Marcel Oliver, Gualtiero Badin, Stephan Juricke and Ulrich Achatz for fruitful discussions.

\subsubsection*{Disclosure statement}
\vspace{-3mm}No potential conflict of interest was reported by the author(s).

\subsubsection*{Funding}
\vspace{-3mm}This paper is a contribution to the project M2 (Systematic multi-scale modelling and analysis for geophysical flow) of the Collaborative Research Centre TRR 181 ``Energy Transfers in Atmosphere and Ocean" funded by the Deutsche Forschungsgemeinschaft (DFG, German Research Foundation) under project number 274762653.

\bibliographystyle{apa}
\bibliography{References}

\begin{thebibliography}{}

\bibitem[\protect\astroncite{Achatz}{2006}]{Achatz06}
Achatz, U. (2006).
\newblock {\em Gravity-Wave Breakdown in a Rotating Boussinesq Fluid: Linear
  and Nonlinear Dynamics}.
\newblock Habilitation Thesis. University of Rostock.

\bibitem[\protect\astroncite{Balmforth and Young}{2005}]{BalmforthYoung2005}
Balmforth, N.~J. and Young, Y.-N. (2005).
\newblock Stratified kolmogorov flow. ii.
\newblock {\em J. Fluid Mech.}, 528:23--42.

\bibitem[\protect\astroncite{Chai et~al.}{2020}]{Chai20}
Chai, J., Wu, T., and Fang, L. (2020).
\newblock Single-scale two-dimensional-three-component
  generalized-beltrami-flow solutions of incompressible navier-stokes
  equations.
\newblock {\em Physics Letters A}, 384(34):126857.

\bibitem[\protect\astroncite{Danilov et~al.}{2019}]{M3}
Danilov, S., Juricke, S., Kutsenko, A., and Oliver, M. (2019).
\newblock Toward consistent subgrid momentum closures in ocean models.
\newblock In Eden, C. and Iske, A., editors, {\em Energy Transfers in
  Atmosphere and Ocean}, pages 145--192. Springer-Verlag, Cham.

\bibitem[\protect\astroncite{Drazin and Riley}{2006}]{Drazin06}
Drazin, P.~G. and Riley, N. (2006).
\newblock {\em The {N}avier-{S}tokes equations: a classification of flows and
  exact solutions}, volume 334 of {\em London Mathematical Society Lecture Note
  Series}.
\newblock Cambridge University Press, Cambridge.

\bibitem[\protect\astroncite{Dyck and Straatman}{2019}]{Dyck19}
Dyck, N.~J. and Straatman, A.~G. (2019).
\newblock Exact solutions to the three-dimensional navier–stokes equations
  using the extended beltrami method.
\newblock {\em Journal of Applied Mechanics}, 87(1).
\newblock 011004.

\bibitem[\protect\astroncite{Ghaemsaidi and
  Mathur}{2019}]{ghaemsaidi_mathur_2019}
Ghaemsaidi, S.~J. and Mathur, M. (2019).
\newblock Three-dimensional small-scale instabilities of plane internal gravity
  waves.
\newblock {\em Journal of Fluid Mechanics}, 863:702--729.

\bibitem[\protect\astroncite{Jansen et~al.}{2019}]{JansenEtAl2019}
Jansen, M.~F., Adcroft, A., Khani, S., and Kong, H. (2019).
\newblock Toward an energetically consistent, resolution aware parameterization
  of ocean mesoscale eddies.
\newblock {\em Journal of Advances in Modeling Earth Systems},
  11(8):2844--2860.

\bibitem[\protect\astroncite{Jansen and Held}{2014}]{Jansen2014}
Jansen, M.~F. and Held, I.~M. (2014).
\newblock Parameterizing subgrid-scale eddy effects using energetically
  consistent backscatter.
\newblock {\em Ocean Modelling}, 80:36--48.

\bibitem[\protect\astroncite{Juricke et~al.}{2020}]{juricke2020kinematic}
Juricke, S., Danilov, S., Koldunov, N., Oliver, M., Sein, D., Sidorenko, D.,
  and Wang, Q. (2020).
\newblock A kinematic kinetic energy backscatter parametrization: From
  implementation to global ocean simulations.
\newblock {\em Journal of Advances in Modeling Earth Systems},
  12(12):e2020MS002175.

\bibitem[\protect\astroncite{Kalogirou et~al.}{2015}]{Kalogirou15}
Kalogirou, A., Keaveny, E.~E., and Papageorgiou, D.~T. (2015).
\newblock An in-depth numerical study of the two-dimensional
  {K}uramoto-{S}ivashinsky equation.
\newblock {\em Proc. R. Soc. A.}, 471(2179):20140932, 20.

\bibitem[\protect\astroncite{Lelong and Dunkerton}{1998}]{LeLongDunkerton1998}
Lelong, M.-P. and Dunkerton, T.~J. (1998).
\newblock Inertia–gravity wave breaking in three dimensions. part i:
  Convectively stable waves.
\newblock {\em Journal of the Atmospheric Sciences}, 55(15):2473 -- 2488.

\bibitem[\protect\astroncite{Majda}{2003}]{Majda03}
Majda, A. (2003).
\newblock {\em Introduction to {PDE}s and waves for the atmosphere and ocean},
  volume~9 of {\em Courant Lecture Notes in Mathematics}.
\newblock New York University, Courant Institute of Mathematical Sciences, New
  York; American Mathematical Society, Providence, RI.

\bibitem[\protect\astroncite{Majda and Wang}{2006}]{MajdaWang06}
Majda, A.~J. and Wang, X. (2006).
\newblock {\em Non-linear dynamics and statistical theories for basic
  geophysical flows}.
\newblock Cambridge University Press, Cambridge.

\bibitem[\protect\astroncite{Nicolaenko et~al.}{1985}]{NST85}
Nicolaenko, B., Scheurer, B., and Temam, R. (1985).
\newblock Some global dynamical properties of the kuramoto-sivashinsky
  equations: Nonlinear stability and attractors.
\newblock {\em Physica D: Nonlinear Phenomena}, 16(2):155--183.

\bibitem[\protect\astroncite{Onuki et~al.}{2021}]{onuki_joubaud_dauxois_2021}
Onuki, Y., Joubaud, S., and Dauxois, T. (2021).
\newblock Simulating turbulent mixing caused by local instability of internal
  gravity waves.
\newblock {\em Journal of Fluid Mechanics}, 915:A77.

\bibitem[\protect\astroncite{Perezhogin}{2020}]{Perezhogin20}
Perezhogin, P.~A. (2020).
\newblock Testing of kinetic energy backscatter parameterizations in the nemo
  ocean model.
\newblock {\em Russian Journal of Numerical Analysis and Mathematical
  Modelling}, 35(2):69--82.

\bibitem[\protect\astroncite{Prugger and
  Rademacher}{2021}]{prugger2020explicit}
Prugger, A. and Rademacher, J. D.~M. (2021).
\newblock {Explicit superposed and forced plane wave generalized Beltrami
  flows}.
\newblock {\em IMA Journal of Applied Mathematics}.
\newblock hxab015.

\bibitem[\protect\astroncite{Smyrlis and Papageorgiou}{1991}]{Smyrlis91}
Smyrlis, Y. and Papageorgiou, D. (1991).
\newblock Predicting chaos for infinite dimensional dynamical systems: the
  kuramoto-sivashinsky equation, a case study.
\newblock {\em Proceedings of the National Academy of Sciences of the United
  States of America}, 88(24):11129—11132.

\bibitem[\protect\astroncite{Wang}{1990}]{Wang90}
Wang, C.~Y. (1990).
\newblock Exact solutions of the {N}avier-{S}tokes equations---the generalized
  {B}eltrami flows, review and extension.
\newblock {\em Acta Mech.}, 81(1-2):69--74.

\bibitem[\protect\astroncite{Wei}{2006}]{Wei06}
Wei, H.-H. (2006).
\newblock Shear-flow and thermocapillary interfacial instabilities in a
  two-layer viscous flow.
\newblock {\em Physics of Fluids}, 18(6):064109.

\bibitem[\protect\astroncite{Weinbaum and O'Brien}{1967}]{Weinbaum67}
Weinbaum, S. and O'Brien, V. (1967).
\newblock Exact navier‐stokes solutions including swirl and cross flow.
\newblock {\em The Physics of Fluids}, 10(7):1438--1447.

\bibitem[\protect\astroncite{Yau et~al.}{2004}]{Yau04}
Yau, K.-H., Klaassen, G.~P., and Sonmor, L.~J. (2004).
\newblock Principal instabilities of large amplitude inertio-gravity waves.
\newblock {\em Physics of Fluids}, 16(4):936--951.

\bibitem[\protect\astroncite{Zurita-Gotor et~al.}{2015}]{ZuritaEtAl2015}
Zurita-Gotor, P., Held, I.~M., and Jansen, M.~F. (2015).
\newblock Kinetic energy-conserving hyperdiffusion can improve low resolution
  atmospheric models.
\newblock {\em Journal of Advances in Modeling Earth Systems}, 7(3):1117--1135.

\end{thebibliography}

\end{document}